\documentclass[sn-basic]{sn-jnl}

\usepackage{graphicx}%
\usepackage{multirow}%
\usepackage{amsmath,amssymb,amsfonts}%
\usepackage{amsthm}%
\usepackage{mathrsfs}%
\usepackage[title]{appendix}%
\usepackage{xcolor}%
\usepackage{textcomp}%
\usepackage{manyfoot}%
\usepackage{booktabs}%
\usepackage{algorithm}%
\usepackage{algorithmicx}%
\usepackage{algpseudocode}%
\usepackage{listings}%
\usepackage{subcaption}

\usepackage{mdframed}
\usepackage{xcolor}
\newmdenv[
  leftline=true,
  topline=false,
  bottomline=false,
  rightline=false,
  linecolor=blue,
  linewidth=2pt,
  backgroundcolor=white,
  skipabove=10pt,
  skipbelow=10pt,
  innerleftmargin=6pt,
  innertopmargin=4pt,
  innerbottommargin=4pt,
  nobreak=true
]{InterviewQuote}
\newcommand{\interviewquote}[2]{%
  \begin{InterviewQuote}
    {\small ``\emph{#1}'' — \texttt{#2}}
  \end{InterviewQuote}
}

\lstdefinelanguage{Java}{
  morekeywords={abstract, assert, boolean, break, byte, case, catch, char, class, const,
    continue, default, do, double, else, enum, extends, final, finally, float, for, if,
    goto, implements, import, instanceof, int, interface, long, native, new, null,
    package, private, protected, public, return, short, static, strictfp, super, switch,
    synchronized, this, throw, throws, transient, try, void, volatile, while},
  sensitive=true,
  morecomment=[l]{//},
  morecomment=[s]{/*}{*/},
  morestring=[b]",
  moredelim=**[is][\textcolor{red}{\rule[0.3ex]{0.8em}{0.25pt}\hspace{0.4em}}]{@r@}{@}
}

\definecolor{lightblue}{rgb}{0.8, 0.9, 1.0}

\usepackage{tcolorbox}
\tcbuselibrary{skins, breakable}

\usepackage{enumitem}

\theoremstyle{thmstyleone}%
\theoremstyle{thmstyletwo}%

\theoremstyle{thmstylethree}%

\begin{document}

\title[Echoes of AI]{Echoes of AI: Investigating the Downstream Effects of AI Assistants on Software Maintainability}


\author*[1,3]{\fnm{Markus} \sur{Borg}}\email{markus.borg@codescene.com}

\author[2]{\fnm{Dave} \sur{Hewett}}\email{dave.hewett@equalexperts.com}

\author[3]{\fnm{Nadim} \sur{Hagatulah}}\email{nadim.hagatulah@cs.lth.se}

\author[3]{\fnm{Noric} \sur{Couderc}}\email{noric.couderc@cs.lth.se}

\author[3]{\fnm{Emma} \sur{Söderberg}}\email{emma.soderberg@cs.lth.se}

\author[4]{\fnm{Donald} \sur{Graham}}\email{donald.graham@equalexperts.com}

\author[5]{\fnm{Uttam} \sur{Kini}}\email{uttam.kini@equalexperts.com}

\author[6]{\fnm{Dave} \sur{Farley}}\email{info@continuous-delivery.co.uk}

\affil*[1]{\orgname{CodeScene}, \orgaddress{\city{Malmö}, \country{Sweden}}}

\affil[2]{\orgname{Equal Experts}, \orgaddress{\city{London}, \country{UK}}}

\affil[3]{\orgdiv{Dept. of Computer Science}, \orgname{Lund University}, \orgaddress{\city{Lund}, \country{Sweden}}}

\affil[4]{\orgname{Equal Experts}, \orgaddress{\city{Cape Town}, \country{South Africa}}}

\affil[5]{\orgname{Equal Experts}, \orgaddress{\city{Bengaluru}, \country{India}}}

\affil[6]{\orgname{Continuous Delivery}, \orgaddress{\city{London}, \country{UK}}}

\abstract{
[Context] AI assistants, like GitHub Copilot and Cursor, are transforming software engineering. While several studies highlight productivity improvements, their impact on maintainability requires further investigation. 
[Objective] This study investigates whether co-development with AI assistants affects software maintainability, specifically how easily other developers can evolve the resulting source code. 
[Method] We conducted a two-phase, preregistered controlled experiment involving 151 participants, 95\% of whom were professional developers. In Phase~1, participants added a new feature to a Java web application, with or without AI assistance. In Phase~2, a randomized controlled trial, new participants evolved these solutions without AI assistance. 
[Results] Phase~2 revealed no significant differences in subsequent evolution with respect to completion time or code quality. Bayesian analysis suggests that any speed or quality improvements from AI use were at most small and highly uncertain. Observational results from Phase~1 corroborate prior research: using an AI assistant yielded a 30.7\% median reduction in completion time, and habitual AI users showed an estimated 55.9\% speedup.
[Conclusions] Overall, we did not detect systematic maintainability advantages or disadvantages when other developers evolved code co-developed with AI assistants. Within the scope of our tasks and measures, we observed no consistent warning signs of degraded code-level maintainability. Future work should examine risks such as code bloat from excessive code generation and cognitive debt as developers offload more mental effort to assistants.
}



\maketitle

\section{Introduction} \label{sec:intro}
Generative AI is rapidly transforming software development, disrupting the discipline as we know it. Tools based on Large Language Models (LLMs), such as GitHub Copilot and ChatGPT, have seen widespread adoption among developers~\citep{jetbrains_state_2024}. The former exemplifies an IDE-integrated code completion assistant, while ChatGPT represents a general-purpose tool that supports chat-based programming. The appeal of AI assistants for code synthesis is clear and, as we will review in Section~\ref{sec:rw}, several empirical studies, in fact, suggest that working with them can lead to significant productivity gains.

For many developers, AI assistants are now a natural part of the development context. The first secondary studies on their use have now been published, e.g., reviews by \citet{ani_recent_2024} and \citet{husein_large_2025}, which summarize the risks of LLM-based code synthesis identified in primary studies. Risks include the generation of incorrect code, the introduction of security vulnerabilities, unnecessarily complex code completions, and reduced code maintainability. In this study, we focus on the last aspect: how AI-assisted development influences maintenance and evolution.

As many organizations move into a hybrid world where human- and machine-generated code coexist, we urgently need to understand how the use of AI assistants affects maintainability. Recent public announcements from Microsoft, Meta, and Google state that roughly a third of their new code is already AI-generated~\citep{prosser_worried_2025}, demonstrating that the change is happening now. Thus, we argue that maintainability has never been more important since the sheer volume of code will increase rapidly from this point on. Furthermore, we posit that in the foreseeable future, human developers must remain able to manually evolve source code no matter its provenance.


We conduct a preregistered two-phase controlled experiment~\citep{borg_does_2024}. Our starting point is that maintainable code should be easy to reason about and modify by someone other than the original author. In Phase~1, participants extend a Java web application with or without the help of an AI assistant. Phase~2 is a Randomized Control Trial (RCT), in which new participants are randomly assigned to evolve a solution from Phase~1 \textit{without using AI assistants}. Accordingly, we primarily assess maintainability by the ease with which a new developer can add features to existing code. To complement this perspective, we include CodeScene's CodeHealth~\citep{tornhill_code_2022} and test coverage measurements to broaden the assessment. Finally, we measure perceived productivity guided by the SPACE framework~\citep{forsgren_space_2021} and collect rich free-text data through exit questionnaires.

For an experiment relying on volunteers from a wide range of organizations, our study is unique in both the number of participants and their level of seniority. Our Phase~2 RCT provides no clear evidence that code co-developed with AI assistants is more efficient to evolve manually, and any speed advantage in the manual evolution task was small and statistically unreliable. For code quality, we found no significant differences in CodeHealth between treatment and control, although Bayesian analysis suggests a small CodeHealth improvement when the original solution was co-developed with AI by a habitual AI user. Although not the primary focus of our study, the Phase~1 results corroborate previous findings: AI assistants significantly reduced task completion time, with a median improvement of 30.7\%. Moreover, the posterior mean effect for habitual AI users among the participants was 55.9\%. 

Coding agents, introduced as the third-generation AI assistants in Section~\ref{sec:ai-gens}, are now on the rise, offering autonomous code synthesis based on a developer's intent. Our study was conducted in late 2024, so its empirical results predate this trend. Instead, our study might be one of the last to include a large sample of developers who had not yet been fundamentally affected by AI assistants. There is a clear need for additional empirical studies to investigate these agentic trends, both controlled experiments like ours and longitudinal case studies, to explore direct and indirect effects on maintainability.

The rest of the manuscript is organized as follows:
\begin{itemize}
    \item Section~\ref{sec:bg_rw} introduces fundamental constructs, presents different generations of AI assistants for code, and reviews related empirical studies.
    \item Section~\ref{sec:method} describes the research method, including data collection and analysis.
    \item Section~\ref{sec:res} presents the results from the RCT in Phase~2 and observational findings from Phase~1.
    \item Section~\ref{sec:disc} discusses the results in light of previous work and shares our interpretation of implications for practice.
    \item Section~\ref{sec:threats} discusses the limitations of our study and its threats to validity.
    \item Section~\ref{sec:conc} concludes the paper and outlines directions for future work.
\end{itemize}

\section{Background and Related Work} \label{sec:bg_rw}
This section first introduces background on software maintainability and productivity. Second, we present the disruptive trend of AI-assisted development. Finally, we review empirical studies on the impact of AI assistants.

\subsection{Software Maintainability and Productivity} \label{sec:maint}
The two constructs maintainability and productivity are cornerstones of this research. As both are complex and multi-faceted, this section describes how we rely on state-of-the-art approaches to measure them.

The ISO~25010 quality model defines maintainability as \textit{``the degree of effectiveness and efficiency with which a product or system can be modified to improve it, correct it or adapt it to changes in environment, and in requirements''}~\citep{international_organization_for_standardization_systems_2011}. To acknowledge the different constituents of this quality, ISO~25010 further defines five sub-qualities: 1) modularity, 2) reusability, 3) analyzability, 4) modifiability, and 5) testability. We use the top-level maintainability definition in this work, and argue that completion time on an adaptation task is a useful proxy for maintainability, since more maintainable code should require less effort and time to evolve.

A related concept is Technical Debt (TD), defined as: \textit{``In software-intensive systems, TD is a collection of design or implementation constructs that are expedient in the short term but set up a technical context that can make future changes more costly or impossible. Technical Debt presents an actual or contingent liability whose impact is limited to internal system qualities, primarily maintainability and evolvability''}~\citep{avgeriou_managing_2016} Note that this definition includes the term evolvability, which is not explicitly mentioned in ISO~25010. We rely on its definition by~\citet{cook_software_2003} as \textit{``the
capability of software products to be evolved to continue to serve its customer in a cost-effective way''} and consider it subsumed by the ISO~25010 maintainability definition.

The presence of code smells is widely recognized as a factor that degrades maintainability. Code smells, a term (like TD) coined by the agile community, refer to problematic code that developers should preferably refactor -- typically to make it easier to understand and extend~\citep{mantyla_subjective_2006}. The academic literature on code smells is extensive, to the point that even a tertiary study has been published~\citep{lacerda_code_2020}. Thanks to extensive lists of code smells, and capable detection tools, maintainability can pragmatically be treated as a negative quality, i.e., the degree to which code does not contain smells. 

CodeScene measures maintainability using the CodeHealth\textsuperscript{TM} metric. The file-level score penalizes the presence of 25+ code smells\footnote{Some are language- or paradigm-dependent, thus not an exact number.} known to increase cognitive load on developers. Our previous work shows that CodeHealth correlates with increased maintenance costs and defect risks~\citep{tornhill_code_2022,borg_increasing_2024}, as well as slower onboarding~\citep{borg_u_2023}. Furthermore, CodeHealth outperforms competing metrics~\citep{borg_ghost_2024}, including SonarQube's Maintainability Rating and Microsoft's Maintainability Index on the Maintainability Dataset -- a benchmark developed by~\citet{schnappinger_defining_2020}.

CodeHealth is a numeric value between 1 and 10. The metric aligns with \citet{fenton_software_1994}'s seminal work on software measurement, particularly the philosophy that the best way to assess code complexity is by identifying and quantifying specific complexity-inducing attributes. A file with no detected code smells receives a perfect score of 10. For each code smell that is detected, the file gets a decreased CodeHealth score -- infinitesimally approaching 1 for the very worst files. When aggregating CodeHealth across multiple files, CodeScene reports a file-size-weighted average of the file-level values. In our study, we complement task completion time with CodeHealth to better measure the maintainability construct. Note that we customize the CodeHealth scoring for increased sensitivity in this study, as described in Section~\ref{sec:vars}.

Developer productivity is notoriously difficult to define. Numerous publications have addressed the topic since the early days of software engineering~\citep{sadowski_rethinking_2019}, and there is broad agreement that productivity is inherently multi-dimensional~\citep{jaspan_no_2019}. Yet, many naïve myths and simplistic metrics continue to circulate in the software industry. In this study, we rely on the SPACE framework by \citet{forsgren_space_2021} to discuss productivity in the context of software maintenance and evolution. Rather than providing an overarching definition of productivity, SPACE decomposes the construct into five dimensions:  

\begin{itemize}
    \item[\textbf{S}]\textbf{atisfaction and wellbeing.} How fulfilled developers feel with their work, team, tools, and culture. Also, how their work impacts their happiness and health.
    \item[\textbf{P}]\textbf{erformance.} The outcome of a process. Note that outcome goes beyond mere output, as in ``Did the contributed code reliably do what it was supposed to do?''
    \item[\textbf{A}]\textbf{ctivity.} The volume of actions or outputs completed while performing work, such as commits or issues closed.
    \item[\textbf{C}]\textbf{ommunication and collaboration.} How developers exchange information and coordinate work, including perceptions of effectiveness in information seeking.
    \item[\textbf{E}]\textbf{fficiency and flow.} The ability to complete work or make progress on it with minimal interruptions or delays.
\end{itemize}

\begin{figure}
    \centering
    \includegraphics[width=0.8\linewidth]{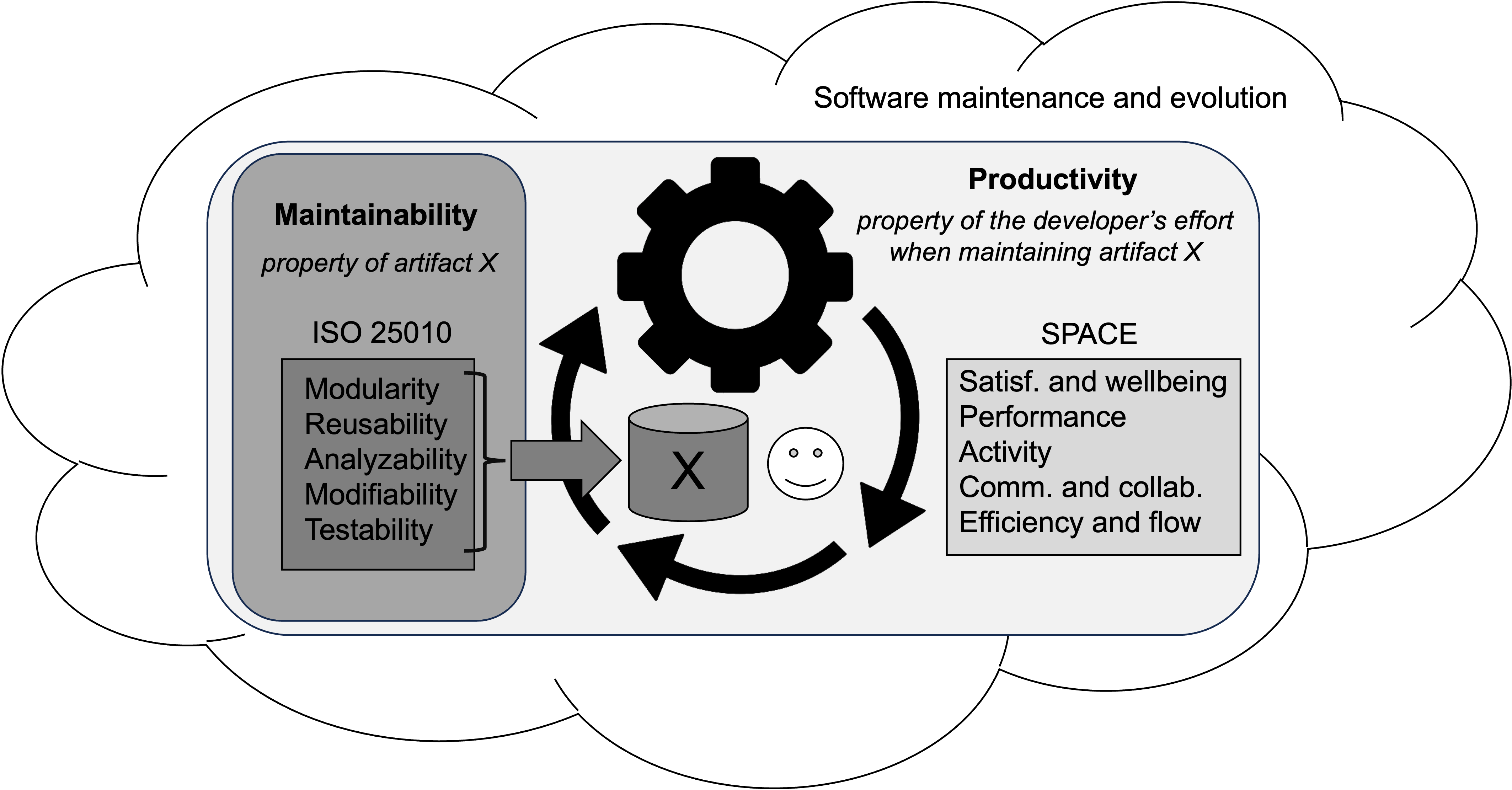}
    \caption{Conceptual relationship between maintainability (artifact-centric) and productivity (developer-centric). ISO~25010 frames maintainability, SPACE frames productivity, and artifact~X links the two through the maintenance activity.}
    \label{fig:constructs}
\end{figure}

Figure~\ref{fig:constructs} depicts how we position maintainability and productivity in this study. The cloud represents the overall context. In dark gray, we show maintainability as a property of artifact~X, characterized by the ISO~25010 sub-characteristics, which we treat jointly in this work. In light gray, we position productivity as a reflection of a developer's effort when carrying out an adaptation task on the same artifact. The five SPACE dimensions frame our discussion on productivity in this context.

As evident from the SPACE research, the personal perception of productivity plays a central role, i.e., the three dimensions \textbf{S}, \textbf{C}, and \textbf{E} have large subjective components. In this study, we refer to this subset of productivity dimensions as \textit{perceived productivity}. Following previous work on productivity with GitHub Copilot by \citet{ziegler_measuring_2024}, we measure this using Likert scales. These are used both to pre-screen participants' preference for working with AI assistants and to assess their experience with the development tasks.

\subsection{AI-assisted Software Development} \label{sec:ai-gens}

The development of AI assistants is evolving at an astonishing pace. We organize this section into three generations of AI assistants as outlined by~\citet{kim_vibe_2025} 


The \textbf{first generation} of AI assistants refers to LLM-driven \textit{code completion} tools. The most well-known is undoubtedly GitHub Copilot\footnote{\url{https://github.com/features/copilot}}, released in October 2021 -- GitHub recently communicated (May 2025) that it now has 15~million users. Other tools in this category include TabNine\footnote{\url{https://www.tabnine.com/}}, Google Gemini Code Assist\footnote{\url{https://codeassist.google/}}, and IBM's watsonx Code Assistant\footnote{\url{https://www.ibm.com/products/watsonx-code-assistant}}. These tools have in common that they are 1) largely reactive and 2) limited to local context. For example, code completion tools can provide next-token predictions or fill in method signatures based on nearby code. A grounded theory study by \citet{barke_grounded_2023} found that the interaction is bimodal. In \textit{acceleration mode}, developers know what to do and complete it faster with the assistant; in \textit{exploration mode}, they are instead uncertain and use the assistant to quickly explore options. Academic evaluations of this generation primarily focused on Completion Acceptance Rates~\citep{ziegler_productivity_2022}, which are far from the maintainability and productivity constructs central to our study.

The \textbf{second generation} of AI assistants focuses on enabling \textit{chat-based programming}. The interaction follows the prompt-response mode popularized by OpenAI's ChatGPT\footnote{\url{https://chatgpt.com/}}, released in November 2022. Although launched as a general-purpose assistant, ChatGPT quickly demonstrated impressive coding capabilities. While many users still copy-paste from web browsers, specialized IDEs such as Cursor\footnote{\url{https://cursor.com/}} and Windsurf\footnote{\url{https://windsurf.com/}} soon emerged to bring chat-based functionality into the developers' environment. This generation of tools is typically characterized by 1) responding to natural language queries and 2) operating with larger context windows than previous code completion tools. Compared to the first generation, chat-based assistants are often used to generate larger amounts of code, including full functions and boilerplate based on comments. Empirical studies on second-generation tools have typically focused on the generated code rather than developer interactions~\citep{treude_generative_2025} -- our study provides new insights to fill this gap.

The \textbf{third generation} of AI assistants refers to \textit{autonomous coding agents}. This is a rapidly evolving space, with several vendors actively competing at the time of this writing. Notable examples are Anthropic's Claude Code\footnote{\url{https://www.claude.com/product/claude-code}}, Sonargraph's Amp\footnote{\url{https://sourcegraph.com/amp}}, and OpenAI's Codex\footnote{\url{https://openai.com/codex/}} (a name previously referring to an LLM trained on code). These agents accept high-level tasks from developers and follow plan–execute–validate–revise loops until a stopping condition is met. Compared to previous generations, these AI tools are distinguished by 1) persistent memory across chat interactions, 2) better understanding of project-level context, and 3) integration with other development tools such as linters, test runners, and documentation systems. Interoperability is often enabled via command-line interfaces or the Model Context Protocol (MCP)~\citep{anthropic_model_2024}. As reported in Section~\ref{sec:intro}, our study largely predates the third generation tools.

\subsection{Empirical Studies on AI-Assisted Development} \label{sec:rw}
Our current study adds to a growing body of empirical research on AI-assisted development. In this section, we focus on studies where participants 1) complete realistic tasks, and 2) evaluations go beyond Completion Acceptance Rates and simplistic code metrics such as keystrokes and Lines of Code (LoC). We include both controlled experiments and field surveys that capture developers' real-world experiences.

We provide the first RCT that explicitly targets the maintainability of code written by AI-assisted developers. In contrast, the most closely related prior studies have focused on developer productivity, pioneered by GitHub's internal research on GitHub Copilot. The first and most widely cited work is the controlled experiment by \citet{peng_impact_2023}, which recruited freelance programmers (N=95) to implement an HTTP server in JavaScript. Participants were randomly assigned to either a treatment group with GitHub Copilot or a control group without assistance. The authors report that the treatment group completed the task 55.8\% faster on average. Although this study generated substantial attention and was used in marketing, it remains unpublished in a peer-reviewed venue. 

A year later, another GitHub team published a paper combining repository mining of usage data with a questionnaire-based survey (N=2,047)~\citep{ziegler_measuring_2024}. The results included usage telemetry and survey responses structured around the SPACE framework described in Section~\ref{sec:maint}, which we adopt in our current study. By sending the survey to 17,240 users of the GitHub Copilot free technical preview, the authors captured experience from real usage in early 2022. Their findings suggest that the AI assistant had a significant positive impact on perceived productivity across multiple dimensions, e.g., task completion time, quality, cognitive load, enjoyment, and learning. Notably, the reported gains were higher among junior developers.

IBM Research conducted a similar questionnaire-based survey among users of the internal AI assistant watsonx Code Assistant (WCA)~\citep{weisz_examining_2025}. Instead of using the dimensions of SPACE, they designed the instrument to collect attitudinal measures of productivity and surveyed participants (N=669) from a WCA training program. The authors report that the main WCA use case was code comprehension rather than generation. While the respondents generally felt that their work was faster, easier, and of higher quality with WCA, the magnitudes were small. Moreover, the individual differences were substantial -- 42.6\% felt that WCA made them less effective.

Google has also published results from related internal research activities. \citet{paradis_how_2025} conducted an RCT (N=96) with Google developers to assess the impact of three AI features on the completion time of a realistic C++ task: 1) code completion, 2) smart paste, and 3) chat-based programming. Participants were randomly assigned to either the treatment group with all three features enabled, or the control group with all disabled. The authors report a 21\% average speedup with AI support, although this effect was not significant when controlling for developer proficiency and task familiarity. In contrast to previous work, they found that more senior participants were faster with AI than juniors. The task under study in our RCT is larger in terms of both LoC and expected task completion time.

\citet{chatterjee_impact_2024} report on a six-week controlled experiment (N$\approx$100\footnote{Some methodological details are not fully disclosed in that paper.}) at ANZ Bank, in which participants completed algorithmic programming tasks in Python. After two weeks of GitHub Copilot training, the study design included both between-group and within-subject comparisons. The results show that participants using the AI assistant completed tasks 42.3\% faster on average, with developers less proficient with Python benefiting the most. Furthermore, the authors report that the code quality improved, with fewer bugs and code smells. Finally, participants shared that Copilot supported them in understanding, testing, and documenting code.

\citet{weber_significant_2024} conducted a controlled experiment (N=24) using a within-subjects design in which participants completed three Python programming tasks. Each task was paired with one level of AI-assistance: GitHub Copilot, chat-based programming (GPT-3), or traditional web search. The authors report that both types of AI-assistance led to more completed requirements per minute, greater code output, and higher self-reported satisfaction. Compared to our study, we note that the sample was small, the tasks were short, and only nine participants were professional developers, with an average age of 26.8 years, indicating a mostly junior cohort.

\citet{liang_large-scale_2024} conducted a questionnaire-based survey in January 2023 (N=410) with a diverse set of AI-assisted developers to study usability. GitHub Copilot was the most widely used tool (74.6\%), whose users reported that a median of 30.5\% of their code was now generated by the AI assistant. The primary motivations for adopting AI assistants were 1) reducing keystrokes, 2) completing tasks faster, and 3) skipping web searches for code examples. On the other hand, the most common reasons for not using such tools were 1) that the code output often is of subpar quality, and 2) the difficulty of controlling the AI assistants to produce the desired output. Our prescreening survey (N=449, see Section~\ref{sec:presecreening-demo}) is of similar size, but features a notably higher proportion of professional developers (92.2\% vs. 49.5\%).

\citet{butler_dear_2025} conducted a mixed-methods study at a multinational company to examine the impact of introducing GitHub Copilot to new users. The authors combined an RCT (N=51) with surveys, telemetry analysis, and a three-week diary study (N=106). A difference-in-differences analysis of telemetry data found no statistically significant differences, but medium to large effect sizes for most of the metrics. However, the surveys and diaries showed interesting subjective experiences: 84\% reported positive changes in their daily work, e.g., reduced web searching and less manual boilerplate coding. Moreover, 66\% described feeling more positive about work after introducing GitHub Copilot. 

\citet{cui_effects_2025} conducted a massive field experiment combining three RCTs (N=4,867) at Microsoft, Accenture, and an anonymous Fortune 100 company. Participants were randomly assigned to use GitHub Copilot or serve in the control group without access. During a two- to eight-month period, participants using the AI assistant completed 26\% more tasks per week, with larger effects among junior developers. While the sample size is outstanding, no direct measurements of code quality were collected -- only build success rate as a simple proxy. It was significantly lower at Accenture with a medium effect size, but there was no difference in the pooled results.

Finally, \citet{he_speed_2026} recently estimated the causal effect of adopting development with Cursor. In a difference-in-differences study of 806 GitHub projects, they found a large and statistically significant increase in development velocity (measured as commits and added LoC) following adoption. However, these gains were accompanied by a substantial increase in static analysis warnings and code complexity. The authors conclude that initial productivity improvements tend to give way to maintenance burden and technical debt.

In summary, several early studies were conducted by the major tech companies Microsoft, IBM, and Google, leveraging unique access to AI assistants. Across both experiments and survey research, studies consistently report productivity gains from AI assistants -- typically on the order of 20–30\% faster task completion for professionals, and sometimes more for novices or repetitive tasks. However, findings related to code quality and its potential impact on maintainability are less clear. Previous studies did not follow up with a second developer and rarely employed objective maintainability proxies. Our study is designed to address this gap by focusing on downstream evolution tasks and by grounding our analysis in established metrics and frameworks. 

\section{Method} \label{sec:method}
\begin{figure}[h]
    \centering
\includegraphics[width=0.8\linewidth]{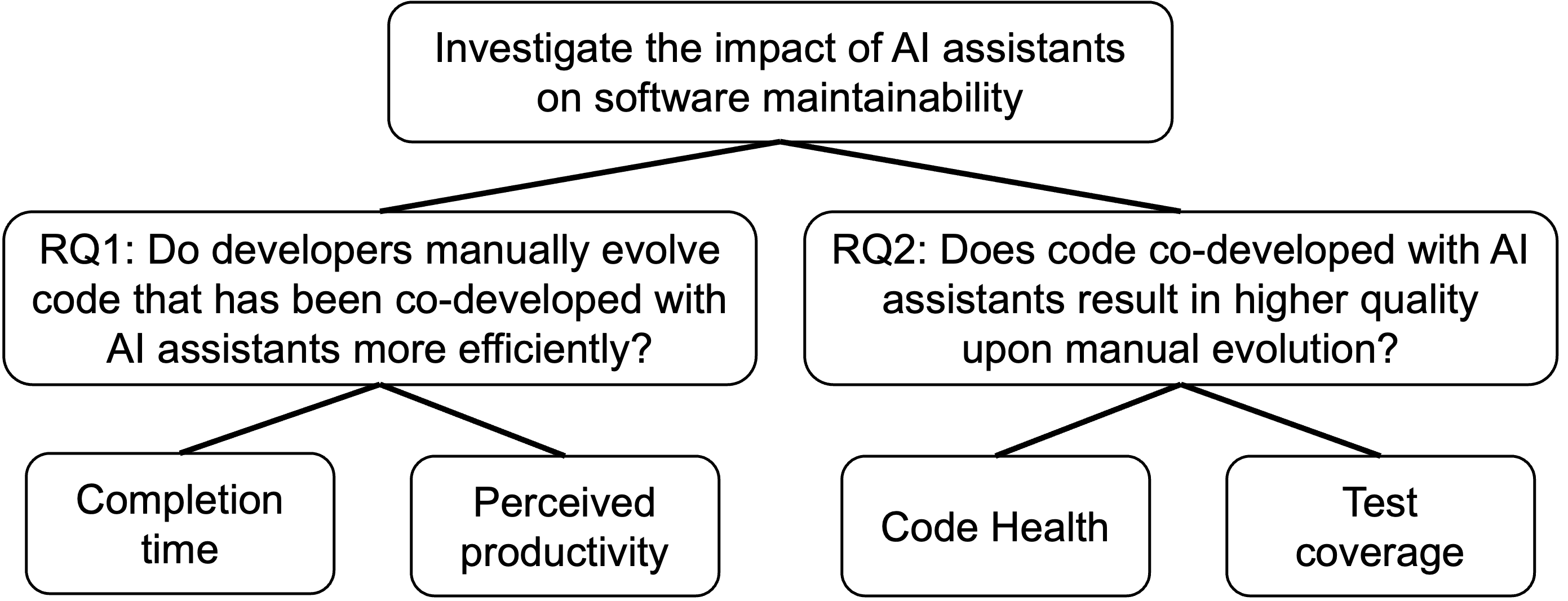}
    \caption{Goal of the study, outlined using the GQM structure.}
    \label{fig:gqm}
\end{figure}

Our goal is to investigate the impact of AI assistants on software maintainability. We break down the goal into two research questions: RQ1) Do developers manually evolve code that has been co-developed with AI assistants more efficiently? and RQ2) Does code co-developed with AI assistants result in higher quality upon manual evolution? We rely on four metrics to answer these RQs: task completion time, perceived productivity according to the SPACE framework, CodeHealth, and test coverage. Fig.~\ref{fig:gqm} summarizes the aim of this study, structured using the GQM model~\citep{basili_goal_1994}. Our preregistered design was guided by the ACM SIGSOFT Empirical Standard for Experiments with Human Participants~\citep{acm_sigsoft_experiments_2024} and we used its essential attributes as a reporting checklist when writing this manuscript.

\subsection{Study Design and Participant Recruitment} \label{sec:design}
\begin{figure}
    \centering
    \includegraphics[width=1\linewidth]{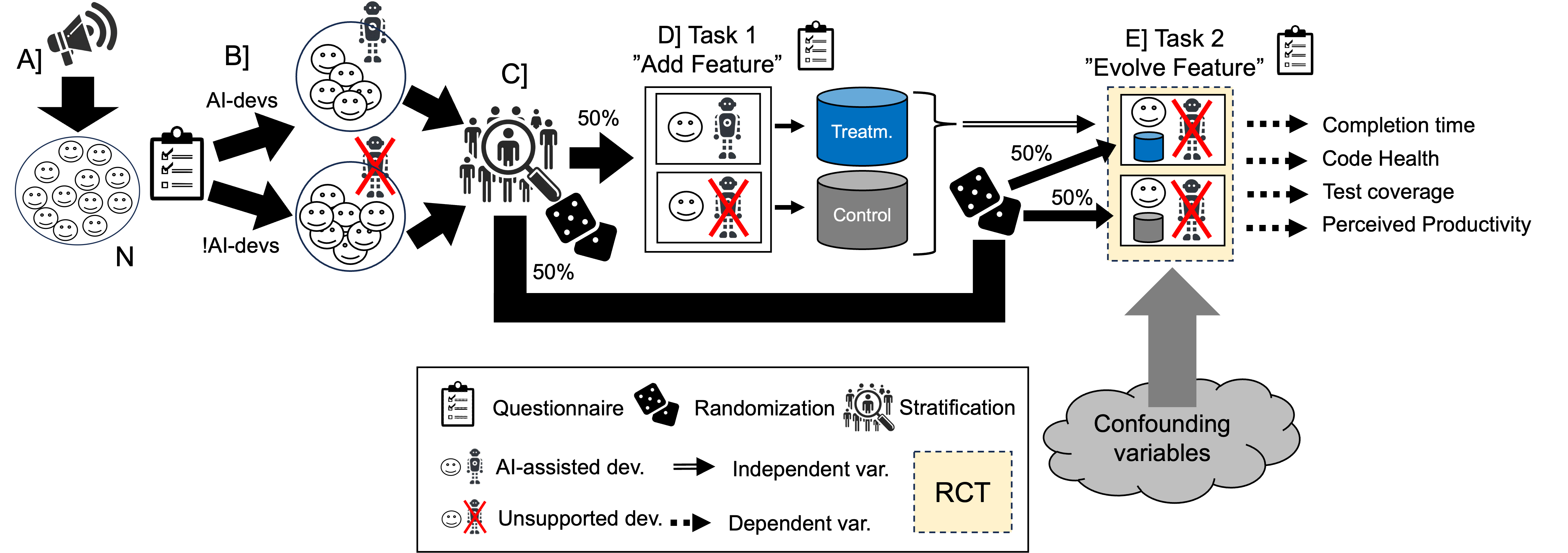}
    \caption{Overview of the study. The part in the yellow box, to which about 50\% of the participants were assigned, constituted the RCT.}
    \label{fig:overview}
\end{figure}

Figure~\ref{fig:overview} shows an overview of our preregistered, two-phase sequential design. In Phase~1, half of the participants prepare artifacts to be used by either the treatment group or the control group. In Phase~2, the remaining participants take part in an RCT, in which all participants receive only one treatment. The rest of this section is organized according to the Steps A] -- E] indicated in the figure, and the experimental variables on the right-hand side are defined in Section~\ref{sec:vars}.

\textbf{Step A]} We called for volunteers to take part in a controlled experiment about ``Software Development with AI Assistants'' (see A] in Fig.~\ref{fig:overview}). All participants volunteered to engage in the research, and agreed to complete assigned tasks remotely in the preferred development environment. We recruited participants through i) social media advertisements on platforms such as YouTube, LinkedIn, and X and ii) using our personal networks. To incentivize participation, all participants were offered a signed copy of the last author's latest book~\citep{farley_modern_2022}, and a chance to win a private online ``Ask Me Anything'' session with him.

\textbf{Step B]} Participants signed up for the study by completing the pre-screening questionnaire outlined in Table~\ref{tab:pre-survey}. In total, 449 participants submitted valid questionnaires. The primary purpose of the pre-screening questionnaire was to facilitate subsequent stratified random sampling into Phase~1 or Phase~2. In Phase~1, we aimed to assign an equal share of participants to either use AI assistants or work without any AI support. We refer to these cohorts as \textbf{AI-devs} and \textbf{!AI-devs}, respectively. In Phase~2, all participants were instructed to work without AI support.

\begin{table*}[h]
\centering
\caption{Outline of the pre-screening questionnaire. A colon indicates a free-text input field. A letter within parentheses shows a mapping to the corresponding SPACE dimension -- item g) in Q1-8 is inverted. As will be explained in Section~\ref{sec:dag}, we refer to participants who strongly agree to Q1-7a as ``habitual AI users.''}
\label{tab:pre-survey}
\resizebox{\textwidth}{!}{%
\begin{tabular}{|p{5cm}|p{1.2cm}|p{11cm}|}
\hline
\textbf{Question} & \textbf{Type} & \textbf{Operationalization} \\ \hline
Q1-1. What is your gender? & Nominal & a) Man b) Woman c) Non-binary d) Prefer not to disclose, e) Prefer to self-describe: \\
\hline
Q1-2. What is your age? & Ordinal & a) 19 or younger b) 20-29, ... g) 70 or older, h) Prefer not to disclose \\
\hline
Q1-3. Where do you live? & Nominal & Closed country list + Prefer not to disclose \\
\hline
Q1-4. Which of the following best describes what you do? & Nominal & a) Student, full-time or part-time, b) Professional programmer, writing code for work, c) Hobbyist programmer, writing code for fun or outside of work, d) Researcher, e) Prefer not to disclose, f) Other: \\
\hline
Q1-5. How proficient are you in software development with Java? & Ordinal & a) Beginner, I can write a correct implementation for a simple function b) Intermediate, I can design and implement whole programs, c) Advanced, I can design and implement a complex system architecture \\
\hline
Q1-6. Do you have experience of working with an AI assistant while programming? & Binary & a) Yes, b) No (ends the questionnaire)\\
\hline
Q1-7. Experience and preferences & 5-point Likert + N/A & Thinking of your experience as a developer and your ways of working, please indicate your level of agreement with the following statements.\\
& & a) I am a habitual user of AI assistants while programming.\\
& & b) I am more productive when using AI assistants. (E)\\
& & c) I complete tasks faster when using AI assistants. (E)\\
& & d) I spend less time searching for information or examples when using AI assistants. (C)\\
& & e) I complete repetitive programming tasks faster when using AI assistants. (E)\\
& & f) Using AI assistants helps me stay in the flow. (E)\\
& & g) Using AI assistants is distracting. (E)\\
& & h) I feel more fulfilled with my job when using AI assistants. (S)\\
& & i) I can focus on more satisfying work when using AI assistants. (S)\\
\hline
\end{tabular}
}
\end{table*}

To facilitate the assignment, the pre-screening collected participant information about 1) experience with AI assistants and 2) whether such assistance was currently the preferred way of working (measured using a Likert scale). To qualify for the \textbf{AI-devs} cohort, participants had to: i) answer yes to Q1-6, ii) agree to statement Q1-7a), and iii) have a positive mean response to the preference questions Q1-7 b)–i) -- taking the inverted question g) into account. 

Allowing participants to adhere to their preferred work methods reduces the likelihood of protocol non-compliance, such as using AI tools when told not to. However, this flexibility might introduce a bias, e.g., less experienced developers could be more inclined to use AI assistants. To compensate for this, we will include Java proficiency as a covariate in our Bayesian model. Note that early research suggests that junior developers might benefit more from AI assistance~\citep{ziegler_measuring_2024}.

A total of 118 participants (26.2\%) who answered confirmatory to both 1) and 2) were split into the \textbf{AI-devs} cohort (see B] in Fig.~\ref{fig:overview} and Section~\ref{sec:datacol} for details). The remaining 331 participants were assigned to the \textbf{!AI-devs} cohort. Notably, the proportion of \textbf{AI-devs} (26.2\%) was closely aligned with our ideal target of one quarter of the participants in this cohort~\citep{borg_does_2024}. As AI-assisted development becomes increasingly prevalent, finding developers without prior exposure to AI will surely become harder -- our study may represent one of the last opportunities to examine developers with limited experience reviewing AI-generated code. 

\textbf{Step C]} We used random stratified sampling to divide the participants into Task 1 or Task 2 of the experiment. Stratification was needed to ensure that we assigned an equal share of \textbf{AI-devs} and \textbf{!AI-devs} to Task 1. The assignment is further described as part of the data collection in Section~\ref{sec:datacol}.

\textbf{Step D]} In Task 1, the \textbf{AI-devs} and \textbf{!AI-devs} cohorts each added a new feature to an existing Java system. Both the system and the task are described in Appendix~\ref{app:system_tasks}. The code submitted by these participants is then handed off to another developer in Task 2 for further evolution. Task 1 was concluded by the exit questionnaire presented in Table~\ref{tab:exit-survey}, and the code was analyzed as part of the submission system described in Section~\ref{sec:datacol}. Only one Task~1 participant did not submit the questionnaire.

\begin{table*}[h]
\centering
\caption{Overview of the exit questionnaire. A colon indicates a free-text input field. A letter within parentheses shows a mapping to the corresponding SPACE dimension -- items d) and j) are inverted.}
\label{tab:exit-survey}
\resizebox{\textwidth}{!}{%
\begin{tabular}{|p{6cm}|p{1.2cm}|p{10cm}|}
\hline
\textbf{Question} & \textbf{Type} & \textbf{Operationalization} \\ \hline
Q2-1. Did you complete the task in one uninterrupted sitting? & Nominal & a) Yes, b) Yes, but with short breaks, c) No\\ \hline
Q2-2. Did you use AI assistants whilst completing the task? & Binary & a) Yes,  c) No\\
\hline
Q2-3. (AI-devs only) Which AI assistant did you work with? & Closed list & a) GitHub Copilot, b) Amazon CodeWhisperer, c) JetBrains AI Assistant, d) Visual Studio IntelliCode, e) TabNine, f) ChatGPT, g) Other: \\\hline
Q2-4. (AI-devs only) How frequently did you interact with the AI assistant? & Ordinal & a) Hardly at all, b) Sometimes, c) Often, d) Almost for every statement I wrote\\
\hline
Q2-5. The task generally resembled development work I have done in the past. & 5-point Likert & a) Strongly disagree, b) Disagree, c) Neutral, d) Agree, e) Strongly agree\\
\hline
Q2-6. Please list any development tools used during the task beyond the standard IDE. Examples include code linting tools, quality analyzers, and vulnerability scanners. & Nominal & a) N/A, b) Used tools: \\
\hline
Q2-7. Perceived productivity & 5-point Likert + N/A & Thinking of your experience with the task, please indicate your level of agreement with the following statements.\\
& & a) I was focused on the task during the programming session. (E)\\
& & b) I was a productive programmer while completing the task. (E)\\
& & c) I felt fulfilled while completing the programming task. (S)\\
& & d) I found myself frustrated while completing the programming task. (S)\\
& & e) I made fast progress despite working with an unfamiliar system. (E)\\
& & f) The code I wrote was of high quality. (S)\\
& & g) I maintained a state of flow during the programming task. (E)\\
& & h) I enjoyed completing this task. (S)\\
& & i) I completed the repetitive programming activities fast during the task. (E)\\
& & j) I spent considerable time searching for information or examples during the task. (C)\\
\hline
Q2-8. Is there anything you would like to add regarding the study or your role in the experiment? & Free-text & : \\
\hline
Q2-9. Please provide your email if you want to receive the report when the study is done. & Free-text & : \\
\hline
\end{tabular}
}
\end{table*}

\textbf{Step E]} In Task 2, which constituted the RCT, new participants were randomly assigned to evolve a valid Task 1 solution. These solutions originated either from the \textbf{AI-dev} cohort (treatment) or the \textbf{!AI-dev} cohort (control). Note that all Task 2 participants worked without AI assistants, and the task itself is described in Appendix~\ref{app:tasks}. Finally, Task 2 was concluded with the same exit questionnaire as Task~1 (see Table~\ref{tab:exit-survey}). All but three Task 2 participants submitted the questionnaire.

\subsection{Experimental Variables and Hypotheses} \label{sec:vars}
We conducted a two-level, single-factor experiment with underlying variations within the two levels. The independent variable was whether participants evolved code that had been co-developed with an AI assistant during Phase~1 (treatment group) or not (control group). As shown in Figure~\ref{fig:overview}, we measured four dependent variables in the Phase~2 RCT. 

\textbf{Completion time} was measured by the submission system as the duration between 1) the moment a participant gained access to their unique GitHub repository and 2) the point at which they irrevocably submitted their solution. In line with the definition of technical debt (see Section~\ref{sec:maint}), we consider the time it takes to implement changes a valid proxy for maintainability. Between 1) and 2), participants were free to push commits to the GitHub repository as frequently as they liked, allowing them to follow their preferred Git workflow. However, many participants reported not completing the task in one uninterrupted session. For this subset, we largely replaced the measurements with self-reported time estimates. This adjustment is further explained in Section~\ref{sec:datacol} and critically discussed in Section~\ref{sec:threats}.

\textbf{CodeHealth (CH)} was measured as a project-level weighted average over all files in the codebase, i.e., not only the files modified by the participant. The weighting is based on LoC calculated using \textit{cloc}\footnote{\url{https://github.com/AlDanial/cloc}}. CH ranges from 1 to 10 on an interval scale (no true zero) and has been shown to align well with expert assessments of maintainability~\citep{borg_ghost_2024}. However, CH has been calibrated over the years for large proprietary systems, where substantial structural problems accumulate over time. To cater to specific needs, CH can be customized by adjusting violation thresholds and the weights of individual rules. In another project, we calibrated CH for smaller development tasks. Together with our partner Codility\footnote{\url{https://www.codility.com/}}, a company developing programming assessment tools used in recruitment and technical screening, we identified a configuration that better detects fine-grained differences in small code samples -- making it a better fit for our study as well. In this custom rule set, two code smells (see Appendix~\ref{app:codescene}) have lower thresholds: \emph{Complex Method} triggers at a cyclomatic complexity of 4 (instead of 9), and \emph{Nested Complexity} at three levels of nesting (instead of 4). The configuration file is available in the replication package. Finally, the average CH was calculated in the continuous integration pipeline using GitHub Actions.

\textbf{Test coverage (TC)} was measured as the line coverage of the final solution's test suite, reported as a percentage on an interval scale. We consider TC an indicator of how much attention participants gave to testing. Participants could freely add test cases, and the resulting coverage was measured using the Java code coverage library JaCoCo, integrated into the continuous integration pipeline via the same GitHub Action. We used JaCoCo’s default metric, which is line coverage rather than statement coverage.

\textbf{Perceived productivity (PP)} is a subjective assessment based on the SPACE framework (see Section~\ref{sec:maint}). We measured PP using a Likert scale composed of ordinal Likert items as part of the exit questionnaire, for which participants received a link upon completing the task. The Likert scale was inspired by the approach described by~\citet{ziegler_measuring_2024} and it allowed us to complement objective measurements with developer perceptions. We adapted the statements to suit both \textbf{AI-devs} and \textbf{!AI-devs}, and avoided references to specific tools. The scale consisted of ten 5-point Likert items (with an optional N/A response), structured according to the SPACE framework; see Q2-7 a)–j) in Table~\ref{tab:exit-survey}.

Several contextual and individual-level variables inevitably influence outcomes in software engineering experiments. While our RCT helps balance both observed and unobserved confounders across treatment and control groups, the realism introduced by remote participation limits experimental control. In addition to the core variables, we preregistered four contextual variables that we expected would particularly influence the results. Specifically, we measured the following in the pre-screening and exit questionnaires to enable post hoc analysis:

\begin{itemize}
\item Whether the participant completed the task in one uninterrupted sitting (Q2-1).
\item Which AI assistant was used during Phase~1 (Q2-3).
\item The extent to which the AI assistant was used during Phase~1 (Q2-4).
\item Whether the participant used additional supportive development tools (Q2-6).
\end{itemize}

\subsubsection{Frequentist Hypotheses} \label{sec:hypo}
In the frequentist analysis, we hypothesize that the use of AI assistants will have a positive impact on the dependent variables in the Phase~2 RCT.
To formally assess these effects, we define four null hypotheses ($H_{0}1$–$H_{0}4$), each corresponding to one of the dependent variables. Each null hypothesis states that there is \textit{no difference} between participants evolving code (in Task~2) co-developed with an AI assistant and those evolving code written without AI assistance (in Task~1).

\vspace{0.2cm}
\begin{itemize}[leftmargin=1cm]
\item[$H_{0}1$] There is no difference in completion time.
\item[$H_{0}2$] There is no difference in CodeHealth (CH).
\item[$H_{0}3$] There is no difference in test coverage (TC).
\item[$H_{0}4$] There is no difference in perceived productivity (PP).
\end{itemize}
\vspace{0.2cm}

\noindent For each null hypothesis, we also define a corresponding non-directional alternative hypothesis ($H_{A}1$–$H_{A}4$), stating that there \textit{is} a difference. 

\subsubsection{Bayesian Causal Analysis} \label{sec:dag}
\begin{figure}
    \centering
    \includegraphics[width=1\linewidth]{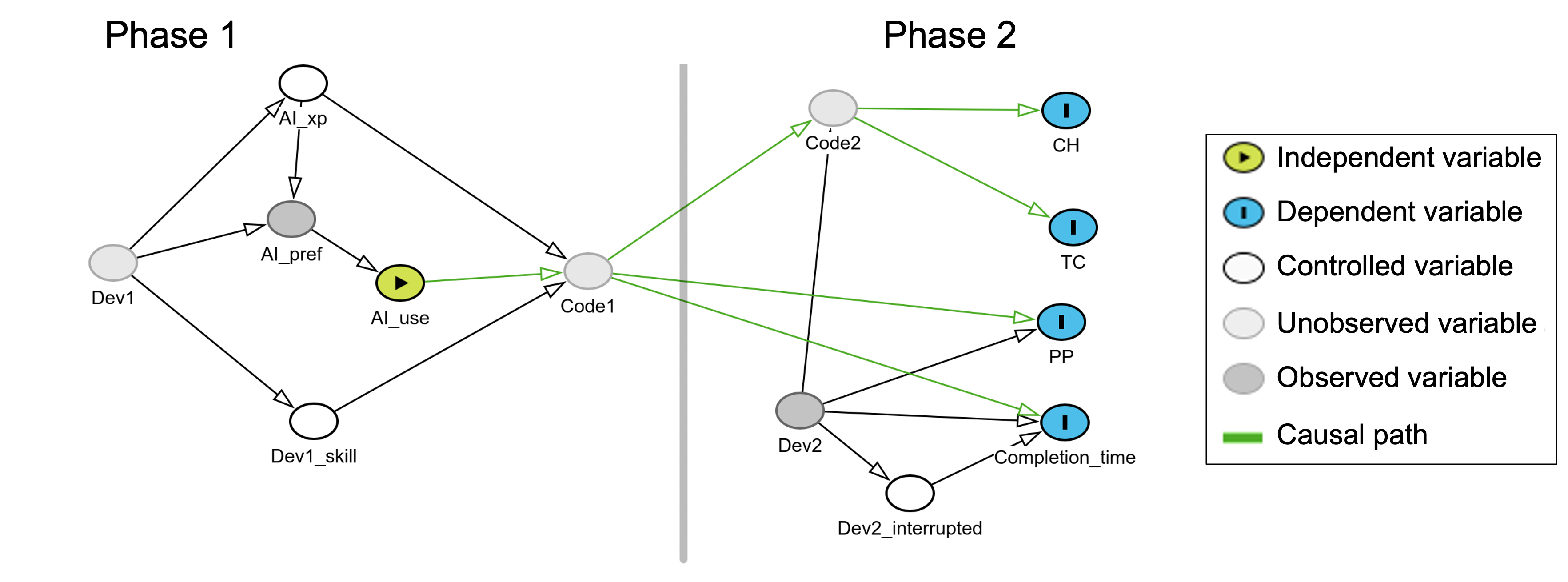}
    \caption{DAGitty causal graph. $Dev1$ and $Dev2$ represent the full complexity of the human participants in Phases 1 and 2, respectively. $Code1$ and $Code2$ are the participants' solutions after Phases 1 and 2, respectively. $AI\_use$ is the independent variable. The other variables are explained in Table~\ref{tab:covariates}.}
    \label{fig:causal}
\end{figure}

To determine which variables to control for, we rely on causal inference. Figure~\ref{fig:causal} shows the causal graph underlying our Bayesian analysis. The directed acyclic graph, created using DAGitty, shows how we connect the variables involved in our two-phased study: each vertex represents a variable, and we trace an arrow between two variables when we consider they have a causal relationship. Based on this graph, we can determine which variables need to be controlled for to estimate causal effects.

As an example, consider the effect of using an AI assistant ($AI\_use$) on TC. Developer 1 ($Dev1$) has a level of Java proficiency level ($Dev1\_skill$), which will influence the quality of their submission ($Code1$). In Phase~2, developer 2 ($Dev2$) continues working on this code, and extends it into a new submission ($Code2$), which will have a test coverage ($TC$). As a result, there is a causal path between $AI\_use$ and $TC$. 

Note, however, that in the causal graph, there are two additional paths between $AI\_use$ and $TC$.
The previously discussed is \emph{causal} ($AI\_use \rightarrow Code1 \rightarrow Code2 \rightarrow TC$), since we follow the arrows. However, there are two other paths that connect the two variables, if we ignore the direction of the arrows: ($AI\_use \leftarrow AI\_pref \leftarrow Dev1 \rightarrow Dev1\_skill \rightarrow Code1 \rightarrow Code2 \rightarrow TC$). These paths are called non-causal or backdoor \citep{mcelreath_statistical_2020}, and indicate the presence of a potential confounder: What if developers who are less proficient at Java prefer to use AI assistants, and more proficient Java developers prefer to avoid them? We would compare two cohorts that are different in \emph{two} ways: More Java-proficient developers who do not want to use AI, and less proficient developers who prefer to use it. As a result, we need to control for the Java proficiency level of developer 1 to isolate the effect of using AI assistants. 

Using the graph, we reason in a similar way to deduce that we should also control for developer 1's habitual use of AI assistants. However, there is an important caveat: when developer 1 does not use an AI assistant, their prior habitual use of such tools becomes irrelevant. We present in Appendix~\ref{app:bayes} how we model this fact. Throughout the remainder of this paper, we refer to participants who answered ``Strongly agree'' to the statement ``I'm a habitual user of AI assistants while programming'' in the prescreening questionnaire (Q1-7a) as ``habitual AI users.'' 

Table~\ref{tab:covariates} summarizes the covariates in the causal graph. The last column indicates whether each variable is directly \textit{controlled} for, to block backdoor paths, or \textit{unobserved} (latent). Further details are available in Appendix~\ref{app:bayes}.

\begin{table*}
\caption{Covariates in the causal analysis. The questionnaire is available in Table~\ref{tab:pre-survey}.}
\label{tab:covariates}
\resizebox{\textwidth}{!}{%
\begin{tabular}{|p{2.7cm}|p{6cm}|p{3.7cm}|p{5.6cm}|}
\hline
\textbf{Variable}                         & \textbf{Description}                                                            & \textbf{Scale and Operationalization}    & \textbf{Role in Analysis}                               \\ \hline
$Dev1$ & Full complexity of the Task 1 participant, including intrinsic motivation, cognitive ability, and psychological state.  & N/A  & Unobserved; assumed balanced by randomization.   \\ \hline

$Dev1\_skill$ & Java programming skill of Dev1.  & Ordinal 3-point scale, Q1-5.  &  Controlled; used for adjustment.  \\ \hline
$AI\_xp$      & Habitual use of AI assistants. & Ordinal 5-point scale, Q1-7a. & Controlled; used for adjustment. \\ \hline
$AI\_pref$      & Stated preference for AI-assisted development. & Binary, derived from Q1-7 items b-i. & Used for preceding group assignment (see Section~\ref{sec:design}), not adjusted. \\ \hline
$Code1$              & Solution produced by Dev1 in Task 1.                             & Only partially observed. & Unobserved; treated as latent mediator.              \\ \hline
$Dev2$              & The full complexity of the human participant in Task 2.                                & N/A  & Unobserved; assumed balanced by randomization.   \\ \hline
$Code2$              & Solution produced by Dev2 in Task 2.                             & Only partially observed. & Unobserved; treated as latent mediator.              \\ \hline
$Dev2\_interrupted$              & To what extent Dev2 completed Task~2 in one uninterrupted session.                             & Ordinal 3-point scale, Q2-1. & Controlled; used for adjustment.              \\ \hline
\end{tabular}
}
\end{table*}

In summary, we state that $AI\_use$ (the independent variable), $AI\_xp$, and $Dev1\_skill$ causally influence $Code1$ in Phase~1. In Phase~2, $Code1$ and $Dev2$ jointly influence the final solution, $Code2$. By adjusting for $Dev1\_skill$ and $AI\_xp$, DAGitty identifies a green causal path that extends through Phase~2 to the four dependent variables. We measure the causal effects on $Code2$ using the two dependent variables CH and TC. The dependent variables PP and Completion time, on the other hand, are causally influenced by both $Code1$ and $Dev2$. When modeling Completion time, we also control for $Dev2\_interrupted$, to improve the precision of our estimates of other effects, as it has a major impact of the Completion time. 

Finally, to validate our controlled variables, we examined how the \textbf{AI-devs} and \textbf{!AI-devs} differ. We report this demographic information in Section~\ref{sec:task1-demo}, but highlight two key findings here. First, \textbf{AI-devs} are more often habitual users of AI assistants, which is expected since we purposefully assigned developers accordingly to encourage compliance. Second, \textbf{!AI-devs} include fewer Java beginners and more advanced Java developers. This aligns with previous findings that experienced developers benefit less from using AI assistants~\citep{ziegler_measuring_2024,cui_effects_2025}. These two findings validated our controlled variables. 



\subsection{Data Collection and Validation}
This section presents how we collected data during the experiments, and how we validated and corrected it afterwards.

\subsubsection{Data Collection} \label{sec:datacol}
The programming tasks were administered using the \textit{snapcode.review}\footnote{\url{https://snapcode.review}} platform, a service used by Equal Experts for coding tests during recruitment processes. \textit{snapcode.review} automates take-home challenges and integrates with GitHub to manage individual repositories per participant. This solution enabled us to trigger acceptance test runs, run CodeScene analyses (to measure CH), execute JaCoCo measurements (for TC), and collect time stamps (for completion time) as part of a continuous integration pipeline. 

We used \textit{Typeform}\footnote{\url{https://www.typeform.com}} to distribute and collect both the prescreening and exit questionnaires. Both questionnaires were piloted with four developers from Equal Experts to ensure clarity and relevance. The feedback led to only minor adjustments.

We opted for a phased rollout of the programming tasks to validate the scalability of our infrastructure. The first batch of participants was recruited through an internal announcement in the Equal Experts developer community on October 10, 2024. Thirty-one developers signed up for the experiment in this round. We assigned eight of them to the Phase~1 \textbf{AI-dev} subgroup based on two prescreening criteria. First, participants had to report whether they were habitual users of AI assistants by answering 4 or 5 to statement Q1-7a. Second, participants had to express a preference for working with AI assistants by providing a median response greater than 3 across items Q1-7b to Q1-7i (taking the inverted Q1-7g into account). The remaining participants were randomly assigned to either the \textbf{!AI-dev} subgroup or Phase~2. In total, only two habitual AI users were assigned to Phase~2 (one to treatment and one to control). Given this small number, we did not conduct a separate subgroup analysis for habitual AI users in Phase~2. After two weeks, we concluded this pilot phase with satisfactory results and opened up the experiment to all volunteers, using the same assignment criteria.

We invited all qualifying \textbf{AI-dev} participants to begin Phase~1 at the end of November, alongside a matching number of randomly selected \textbf{!AI-dev} participants. Every time a Task 1 solution was submitted, three new \textbf{!AI-dev} participants were continuously invited to build on it as part of Task 2 in Phase~2. We selected three to increase the chances of obtaining at least one valid Task~2 solution per submission. 

Both automatic and manual reminder emails were sent to encourage participation. The submission system automatically sent up to two acceptance reminders (after 7 and 12 days) to participants who had not yet accepted their task invitation. Similarly, participants who had accepted, but not yet submitted, received up to two additional reminders. In addition, we manually sent a reminder email to participants approximately one month after they completed the prescreening questionnaire, followed by up to two additional chaser emails. We closed the data collection on January 18, 2025.

\subsubsection{Data Validation and Correction} \label{sec:dataproc}
For both tasks, we excluded participants whose solutions failed to pass the acceptance test suite. We manually inspected all passing solutions and their corresponding exit questionnaire responses. We found no indications of attempts to game or manipulate the system, nor any responses suggesting non-serious engagement. However, we identified three protocol violations:

\begin{itemize}
    \item One \textbf{AI-dev} participant (\textit{anon126}) completed Task 1 twice, both times with AI assistance. Since both these solutions received valid follow-up solutions in Task 2, we chose to keep them. While this means that two Task 1 solutions are not independent, it does not affect the RCT in Phase~2. However, it introduces a minor threat to internal validity when comparing \textbf{AI-dev} and \textbf{!AI-dev} for Task 1, as discussed in Section~\ref{sec:threats}.
    \item One participant assigned to the \textbf{!AI-dev} group self-reported using AI assistance in the exit questionnaire. We resolved this violation by reclassifying the corresponding Task~1 solution as belonging to the \textbf{AI-dev} group -- after we confirmed in the prescreening questionnaire that this participant’s AI experience profile matched the AI-dev group.
    \item About half of the participants (73 out of 151, 48.3\%) self-reported (Q2-1) that they did not complete the task in a single uninterrupted session. As a result, the corresponding submission time stamps recorded by the submission system are highly unreliable. While randomization mitigates the effects in comparisons between groups, we took additional steps to address this issue, as explained next.
\end{itemize}

We reached out via email to the participants\footnote{All but two who explicitly opted out from follow-up questions.} who responded b) or c) to Q2-1 in Table~\ref{tab:exit-survey}. Moreover, we also reached out to participants who responded a) to Q2-1 but had a recorded completion time of more than 8~hours. We asked all of them to provide their best time estimate with 30~min granularity. This way we received 67 time estimates, which we use instead of the recorded time estimates. The longest estimate we received was 15~hours. Thus, we decided to consider all longer time values ($>$ 16~hours) unreasonable for this study and removed 11 values from the analysis (Task~1: 3~\textbf{AI-dev}, 5~\textbf{!AI-dev}; Task~2: 1~treatment, 2~control). Note that for the four participants who did not submit exit surveys, two of them had reasonable completion times of less than 5 hours, which we include in the analysis.

We validated the internal consistency of the Likert scale used to measure PP using Cronbach’s alpha. After inverting the two negatively worded items, the resulting alpha was 0.82, which shows a good internal consistency. Based on this result, we proceeded to compute mean values for PP rather than medians in the frequentist analysis. In the Bayesian analysis, however, we use the individual Likert items as will be explained in Section~\ref{sec:bayes}.

\subsection{Data Analysis}
We follow a mixed-methods approach, using multiple methods \textit{``to collect, analyze, and integrate both qualitative and quantitative data in our analysis''}~\citep{storey_guiding_2025}. Throughout the analysis, we assume independence of observations, i.e., that the participants did not communicate or influence each other. As outlined in the registered report~\citep{borg_does_2024}, we present both frequentist and Bayesian analyses of the quantitative data. The rationale is that Bayesian analysis will explore uncertainties to provide a robust probabilistic understanding, whereas frequentist hypothesis testing will facilitate communication of results to a broader audience. Moreover, we conduct qualitative analyses of free-text responses from the questionnaires and the submitted source code solutions.

\subsubsection{Quantitative Frequentist Analysis} \label{sec:freq}
The purpose of the frequentist analysis is to apply inferential statistics to test the hypotheses in Section~\ref{sec:hypo}. We assessed the normality of the dependent variables using Shapiro-Wilk tests. All results are available in the replication package, a summary of the outcomes follows:

\begin{itemize}
    \item \textbf{Completion time.} Normality was rejected. We proceeded with Wilcoxon rank-sum Test and report Cliff's delta as the effect size.
    \item \textbf{CH.} Normality was not rejected, neither was equal variances. We used Welch's t-test and report Cohen's d as the effect size.
    \item \textbf{TC.} Normality was rejected. We proceeded with Wilcoxon rank-sum Test and report Cliff's delta as the effect size.
    \item \textbf{PP.} Normality was not rejected, neither was equal variances. We used Welch's t-test and report Cohen's d as the effect size.
\end{itemize}

\noindent Given the limited number of hypotheses, we did not apply corrections for multiple testing. To estimate confidence intervals for the dependent variables, we used nonparametric bootstrapping: drawing 1,000,000 samples with replacement from each group.

As part of our pre-registration, we conducted an a priori power analysis using G*Power (v.3.1.9.7) to determine the required sample size for the Phase~2 RCT. Assuming a medium effect size ($d = 0.5$) for the dependent variables, a significance level of $\alpha = 0.05$, and a desired power of 0.80, the analysis indicated a minimum of 128 participants to support frequentist hypothesis testing. As our study design assigns 50\% of participants to Phase~1, this implies a target of 256 completed tasks across both phases. As reported in Section~\ref{sec:design}, 449 participants signed up and 151 participants completed their tasks (33.6\%), i.e., they submitted a solution that passed the corresponding acceptance tests. This means we did not reach our target. 

As will be reported in Section~\ref{sec:res}, we had a total of 75 Phase~2 participants. For this sample size, two-sided tests at $\alpha = 0.05$ achieve 80\% power for standardized mean differences of approximately $d \approx 0.66$. For a medium effect size ($d \approx 0.5$), the power is approximately 0.57, and for a small effect ($d \approx 0.2$) approximately 0.17. Hence, the Phase~2 frequentist tests are well powered to detect large effects of prior AI assistance on Task~2 outcomes, but remain underpowered for small to moderate effects. This further motivates the Bayesian complement.

\subsubsection{Quantitative Bayesian Analysis} \label{sec:bayes}
We conducted a Bayesian analysis to gain a probabilistic understanding of the treatment effects, an approach that remains robust even in the case of smaller sample sizes~\citep{mcelreath_statistical_2020}. 

\paragraph{Statistical Models}

Our choice of statistical models is guided by the type of variables we consider, as dependent and independent variables (also called predictors). Throughout this section, we indicate dependent variables (defined in Section~\ref{sec:vars}) in \textbf{bold} and predictors (defined in Section~\ref{sec:dag}) in \textit{italics}.
\textit{AI\_xp}, \textit{Dev1\_skill}, and \textit{Dev2\_interrupted} are ordinal predictors. \textit{AI\_xp} ranges on a five-point scale based on habitual use of AI assistants (Q1-7a). \textit{Dev1\_skill} refers to the self-reported (Q1-5) Java proficiency using the levels: Beginner, Intermediate, and Advanced. \textit{Dev2\_interrupted} is developer 2's answer to Q2-1, i.e., whether Task~2 was completed in one uninterrupted sitting: Yes, Short breaks, and No.

\textbf{Completion time} is a metric (continuous) variable, strictly positive. \textbf{CH} is a metric variable, on a scale between 1 and 10. \textbf{TC} is a continuous variable on a scale between 0 and 1 (a percentage). \textbf{PP} is measured using a Likert scale in the exit questionnaire. Unlike the frequentist analysis, which uses the mean value, the Bayesian analysis uses the individual Likert items to estimate a latent variable.

Since \textit{AI\_xp}, \textit{Dev1\_skill} and \textit{Dev2\_interrupted} represent ordered categories, we expect, for example, that a developer with an advanced level of Java also has the level of a beginner. Modeling the effects of these variables as nominal or continuous would be incorrect~\citep{burkner_modelling_2020}. We therefore use an appropriate model, described in more detail in Appendix~\ref{app:bayes}. 

For \textbf{CH}, we use a classical linear regression with normally distributed residuals. We choose this model because the CH score for a solution is the weighted average of each file's scores. Because of the central limit theorem, we would expect residuals to be normally distributed. 

For \textbf{TC}, we use a fractional logistic regression model, since the variable is a percentage bounded between 0 and 1. In essence, this model is a classical linear regression on the logit of the percentage \citep{papke_econometric_1996}. We also tested with a classical linear regression, with similar results, since the range of test coverages in our dataset is narrow (65\% to 75\%). 

For \textbf{PP}, we use a more complex model for two reasons. First, each answer is obtained using a set of ten discrete, ordinal Likert items (questions) on a narrow range (1-5). Second, and more importantly, PP is not measured \emph{directly}, but instead approximated via these 10 questions. It is possible that some questions are more relevant than others when estimating a developer's productivity. To model this type of answer, we use an ordinal logistic regression \citep{mcelreath_statistical_2020,gelman_regression_2021}, with a latent variable representing PP. 

\begin{figure}
    \centering
    \includegraphics[width=0.5\linewidth]{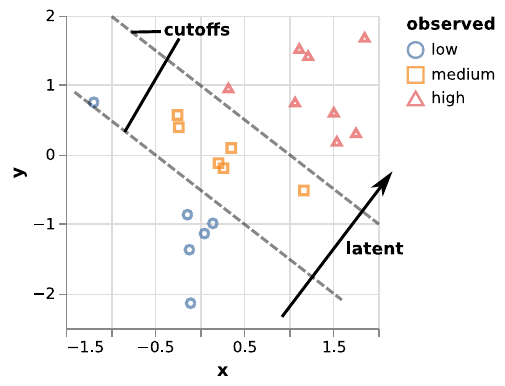}
    \caption{A diagram explaining an ordered logistic regression model. If we observe the two predictors $x$ and $y$, as well as the observed responses (\textit{observed}). The model infers both a latent score (\textit{latent}) for each point (as a linear combination of $x$ and $y$), and cutoffs between the different response levels. In other words, the model finds hyperplanes separating each response level with the next, where all hyperplanes are parallel.}
    \label{fig:ordered-logistic}
\end{figure}

Figure \ref{fig:ordered-logistic} displays a simple example of an ordered logistic regression. In our case, the model represents the productivity of each developer using a latent PP score, predicted from the independent variables. This latent score is then reflected in the answers to each question, where each question has a different set of cutoffs. 

This model has three major advantages: First, it estimates the properties of each question, i.e., the cutoffs between response levels. Second, it uses information from every question in estimating the developer productivity, making our estimates of the developer's productivity more precise. This estimate is then what we use as a response variable when studying the effect of predictors on PP. 

Because we work in a Bayesian inference setting, our estimates of developer productivity are imprecise, which is reflected in the posterior distribution of productivity. This uncertainty is taken into account when we estimate the effect of predictors too. 

\paragraph{Sensitivity Analysis}

In a Bayesian setting, the priors of effects at play can influence the analysis. To study the importance of priors, we constructed three types: i) uninformative priors (neutral), ii) AI-skeptical priors (pessimistic), and iii) AI-enthusiastic priors (optimistic), to reflect different stances on the maintainability impact of AI assistance --- this serves as a sensitivity analysis.

For \textbf{Completion Time}, we used an optimistic prior based on GitHub's research~\citep{peng_impact_2023}, which reported a 55\% improvement. While the GitHub research investigated the direct effects of working with AI, we use the same improvement to inform our optimistic prior for subsequent manual development. For the other response variables, we defined priors that implied an effect of about one standard deviation in the response, where the optimistic prior predicted an improvement, and the pessimistic prior a decline. Table \ref{tab:priors} summarizes our priors for the effect of AI. For full details about our statistical models, we refer to Appendix~\ref{app:bayes}. 

\begin{table}[h!]
\centering
\caption{Summary of priors used in the Bayesian Analysis.}
\label{tab:priors}
\begin{tabular}{|l|l|l|}
\hline
\textbf{Variable} & \textbf{Prior Type} & \textbf{Distribution (Normal)} \\
\hline
\multirow{3}{*}{Completion Time} 
    & Optimistic & $\mathcal{N}(-0.55, 0.3^2)$ \\
    & Neutral    & $\mathcal{N}(0.0, 1.0^2)$ \\
    & Pessimistic& $\mathcal{N}(0.55, 0.3^2)$ \\
\hline
\multirow{3}{*}{Test Coverage} 
    & Optimistic & $\mathcal{N}(0.25, 0.5^2)$ \\
    & Neutral    & $\mathcal{N}(0.0, 1.0^2)$ \\
    & Pessimistic& $\mathcal{N}(-0.25, 0.5^2)$ \\
\hline
\multirow{3}{*}{CodeHealth} 
    & Optimistic & $\mathcal{N}(0.5, 0.5^2)$ \\
    & Neutral    & $\mathcal{N}(0.0, 1.0^2)$ \\
    & Pessimistic& $\mathcal{N}(-0.5, 0.5^2)$ \\
\hline
\multirow{3}{*}{Perceived Productivity} 
    & Optimistic & $\mathcal{N}(0.25, 0.5^2)$ \\
    & Neutral    & $\mathcal{N}(0.0, 1.0^2)$ \\
    & Pessimistic& $\mathcal{N}(-0.25, 0.5^2)$ \\
\hline
\end{tabular}
\end{table}

\paragraph{Reported Statistics}

For each model, we report the influence of a Task~1 participant using AI, compared to another Task~1 participant of the same skill. Now, the effect of using AI presents an \emph{interaction} with AI experience: If the participant has more experience with AI ($AI\_xp=5$), i.e., a habitual AI user as defined in Section~\ref{sec:dag}, the effect of using AI can be expected to be more important. We focus our reporting on the effects of evolving habitual AI users' code, but this does not mean that we excluded non-habitual AI users from the analysis. They are part of the data, but our Bayesian model assumes that AI experience has a monotonic effect, so for developers with lower experience, the effect of using AI is expected to be closer to zero, in comparison (see Appendix \ref{app:bayes} for the full model specification).

Besides the effect of AI, we also report the effect of higher skill on the outcome. We define ``higher skill'' as the difference between outcomes for developers with $Dev1\_skill=3$ vs developers with $Dev1\_skill=1$, keeping all other factors equal.

For each estimate, we report the probability of direction ($P(\Delta > 0)$ or $P(\Delta < 0)$), i.e., the posterior probability that the parameter is strictly above (or below) zero. When the probability of direction exceeds 95\%, we consider the effect to be significant. This probability is somewhat similar to p-values in frequentist statistics. We also provide 95\% Credible Intervals (CrI), which are the Bayesian equivalent of confidence intervals. 
When relevant, we plot the posterior distribution, or perform posterior predictive simulations to compare the Task~2 results one might expect when $Dev1$ uses an AI assistant versus when $Dev1$ does not.

\subsubsection{Qualitative Analysis}
This section describes how we conducted qualitative analyses of Free-text Responses and Source Code, respectively.

\paragraph{Free-text Responses}

We collected qualitative feedback from participants through free-text responses in the exit survey. In total, we received 59 responses from Phase~1 and 45 from Phase~2. Additionally, more than a third of the participants shared insights via (often rich) follow-up emails, typically in response to clarification questions we sent after task completion. Despite conducting the experiment remotely, we believe we gained a solid understanding of participants' perceptions.

The first and sixth author started by independently conducting inductive coding of the free-text responses. Already at this stage, we observed strong agreement on high-level themes, supported by the participants' focused and relevant feedback. Notably, there were no junk responses. The two authors discussed discrepancies and collaboratively refined the initial coding scheme into a two-level structure. The first author then re-applied the coding, and the sixth author validated the outcome. The final coding scheme is presented in Table~\ref{tab:coding}, organized around three high-level topics 1) Development without AI, 2) AI-assisted development, and 3) Task reflections. The number of times the codes were applied is also reported, which will be discussed in Section~\ref{sec:res-freetext}.

\begin{table}[ht]
\centering
\caption{Coding scheme for free-text responses. Numbers in parentheses show the number of occurrences.}
\label{tab:coding}
\begin{tabular}{|p{3.8cm}|p{8.2cm}|}
\hline
\multicolumn{2}{|c|}{\textbf{1) Development without AI (30)}} \\
\hline
Refactoring (9) & Manual restructuring of code to improve readability, design, or maintainability. \\
Testing and debugging (5) & Manually creating or running tests, identifying and reproducing defects, and fixing them. \\
Learning and onboarding (7) & Understanding of the tech stack or onboarding into the specific project without AI support. \\
Misc. (9) & Other development activities. \\
\hline
\multicolumn{2}{|c|}{\textbf{2) AI-assisted development (46)}} \\
\hline
Refactoring (4) & Using AI to suggest or implement code improvements or restructuring. \\
Testing and Debugging (4) & Leveraging AI to create tests, identify and reproduce defects, and assist in fixing them. \\
Learning and onboarding (8) & Using AI to explore unfamiliar technologies, frameworks, or project-specific structures. \\
Productivity boost (5) & Reports of increased speed or efficiency as a result of AI assistance. \\
AI limitations (10) & Situations where AI failed, gave incorrect suggestions, or hindered progress. \\
Interaction mode (10) & Descriptions of how participants engaged with AI. \\
Misc. (5) & Other development activities. \\
\hline
\multicolumn{2}{|c|}{\textbf{3) Task reflections (101)}} \\
\hline
Instruction uncertainty (9) & Confusion or ambiguity related to task requirements or scope. \\
Tech stack (28) & Comments on familiarity or challenges with the programming tools or frameworks. \\
Setup friction (25) & Difficulties encountered during environment setup, configuration, or initial project use. \\
Quality expectations (21) & Uncertainty or reflections about how much to improve, test, or refactor the code. \\
Simplistic task (18) & Perceptions that the assignment was too easy, trivial, or lacked meaningful complexity. \\
\hline
\end{tabular}
\end{table}

\paragraph{Source Code}
We also spent substantial effort in qualitatively analyzing the submitted source code. For Task 1, we applied a systematic approach to investigate the impact of AI assistants. We began by sampling five random \textbf{AI-dev} and five random \textbf{!AI-dev} solutions. The first, third, fourth, and sixth author independently analyzed them to identify interesting variation points. We then met to discuss our findings and compiled a list of 15 aspects that appeared to characterize the solutions. We organized them into four categories: 1) refactoring, 2) solution style, 3) testing, and 4) misc. Furthermore, we found it useful to describe a) the overall solution type, and b) specific changes to the search method presented in Figure~\ref{lst:code}. 

Following this joint meeting, we independently re-coded the same sample of Task 1 solutions and found satisfactory inter-rater agreement. To scale the analysis, we designed regular expressions to automatically identify keywords indicative of specific attributes of Task~1 development. Running these expressions over the \texttt{git diff} output between each modified and original repository enabled us to efficiently identify the addition of unit tests (e.g., assertions) or comments, as well as code patterns suggesting a shift to the functional paradigm (e.g., map, collect, and reduce).

To complement this manual analysis, we also ran detailed static analysis of the Task~1 solutions using CodeScene (v6.19.15) and PMD (v7.13.0 with default settings) to identify patterns by comparing with the base \textit{RecipeFinder} (see Appendix~\ref{app:recipefinder}). Moreover, we ran RefactoringMiner 2.0 by~\citet{tsantalis_refactoringminer_2022} on all Task~1 \texttt{git diffs} to investigate what types of refactoring operations the participants undertook.

\section{Results} \label{sec:res}
We organize this section as follows. First, we present participant demographics and descriptive statistics related to Tasks~1 and~2. Second, we report the results from the preregistered RCT, organized per RQ. Third, we present observational findings on the use of AI assistants in Task~1, which provide context for interpreting our findings. Finally, we summarize insights from the qualitative analysis of free-text answers to add additional nuance to the quantitative results.

Throughout this section, we complement the quantitative findings with quotes from participants. The quotes add qualitative nuance and constitute either representative examples of a theme or contrasting perspectives. All mentioned participants are professional developers unless otherwise stated. Shorter quotes are inlined, longer quotes are separated and indicated with a blue vertical line. When presenting longer quotes, we report the following metadata (the four dependent variables listed in parentheses):

\begin{itemize}
    \item Participant ID, Task \{1, 2\}, \{\textbf{AI-dev}/\textbf{!AI-dev}\} [HABITUAL] (if Q1-7a=5) or \{Treatment/Control\} , Role (Q1-4)+Java proficiency (Q1-5), AI assistants used (if applicable), resemblance agreement (R1-5, Q2-5), added LoC, (\textit{time}, \textit{CH}, \textit{TC}, and \textit{PP}).
\end{itemize}

\subsection{Descriptive Statistics} \label{sec:presecreening-demo}
In total, 449 participants signed up for the experiment by completing a valid prescreening survey. While controlled experiments in software engineering tend to largely rely on student participants~\citep{feldt_four_2018}, this study stands out: 92.2\% of our participants were professional developers. The median age group was 40-49 years, underscoring the experience level of our volunteer base.

\begin{figure}
    \centering
    \includegraphics[width=1\linewidth]{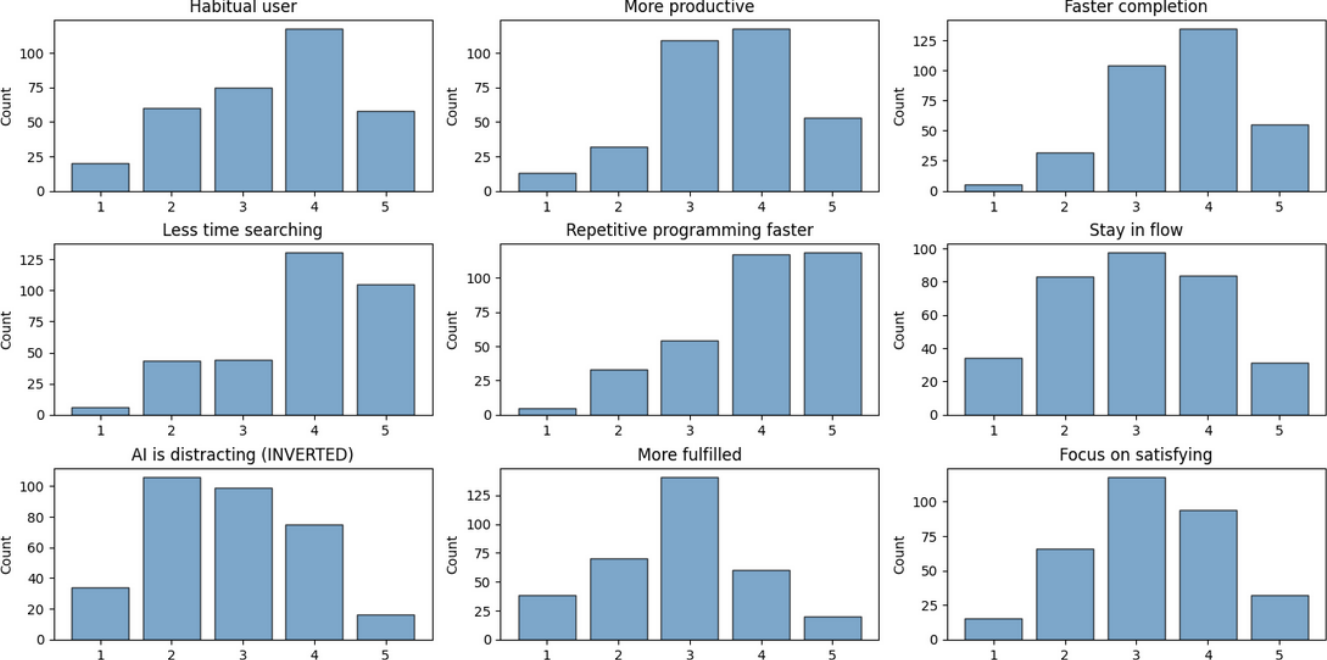}
    \caption{Respondents' experience and preference with AI assistants (Q1-7 in Table~\ref{tab:pre-survey}), 1=Strongly disagree, 5=Strongly agree.}
    \label{fig:ai-likert}
\end{figure}

The prescreening revealed that 73.6\% of the volunteers had prior experience with AI assistants (Q1-6 in Table~\ref{tab:pre-survey}). We asked this subset to self-report their experience using the Likert scale in Q1-7, with results presented in Figure~\ref{fig:ai-likert}. These responses provide a timely snapshot of AI-related development experience and preferences at the end of 2024 -- a very fast-moving area of practice.

We assigned the 118 participants who expressed a preference for AI-assisted development to the \textbf{AI-dev} group in Task 1. Among the remaining participants, 106 were randomly selected for the \textbf{!AI} group in Task 1. The remaining 225 participants were reserved for the RCT in Task 2. The remainder of this section presents demographics and task-level summary statistics.

\subsubsection{Task 1 Demographics} \label{sec:task1-demo}
Task 1 was designed to establish a baseline for the RCT in Task 2. The task concluded with 39 valid \textbf{AI-dev} solutions and 37 \textbf{!AI-dev} solutions. We consider this a solid starting point for the RCT. More specifically, the Task 1 completion rates were as follows:
\begin{itemize}
    \item \textbf{AI-devs} submitted 38 valid solutions (38 of 118, 32.2\%) and a corresponding exit questionnaire. Sixteen of them (42.1\%) were habitual AI users. For the remaining 80 participants 
    \begin{itemize}
        \item 51 participants never started the task,
        \item 20 participants never submitted a valid solution, and
        \item 9 participants actively dropped out.
    \end{itemize}
    \item \textbf{!AI-devs} submitted 38 valid solutions (38 of 106, 35.8\%) -- three of them never submitted the exit questionnaire, but are still included. As reported under protocol violations in Section~\ref{sec:dataproc}, one participant reported using an AI assistant and was thus moved to \textbf{AI-dev}, which explains the 39/37 split. Note that this person was not a habitual AI user. For the remaining 67 participants
    \begin{itemize}
        \item 36 participants never started the task,
        \item 23 participants never submitted a valid solution, and
        \item 8 participants actively dropped out.
    \end{itemize}
\end{itemize}

\begin{figure}[ht]
    \centering
    \begin{subfigure}[b]{0.48\textwidth}
        \centering
        \includegraphics[width=\linewidth]{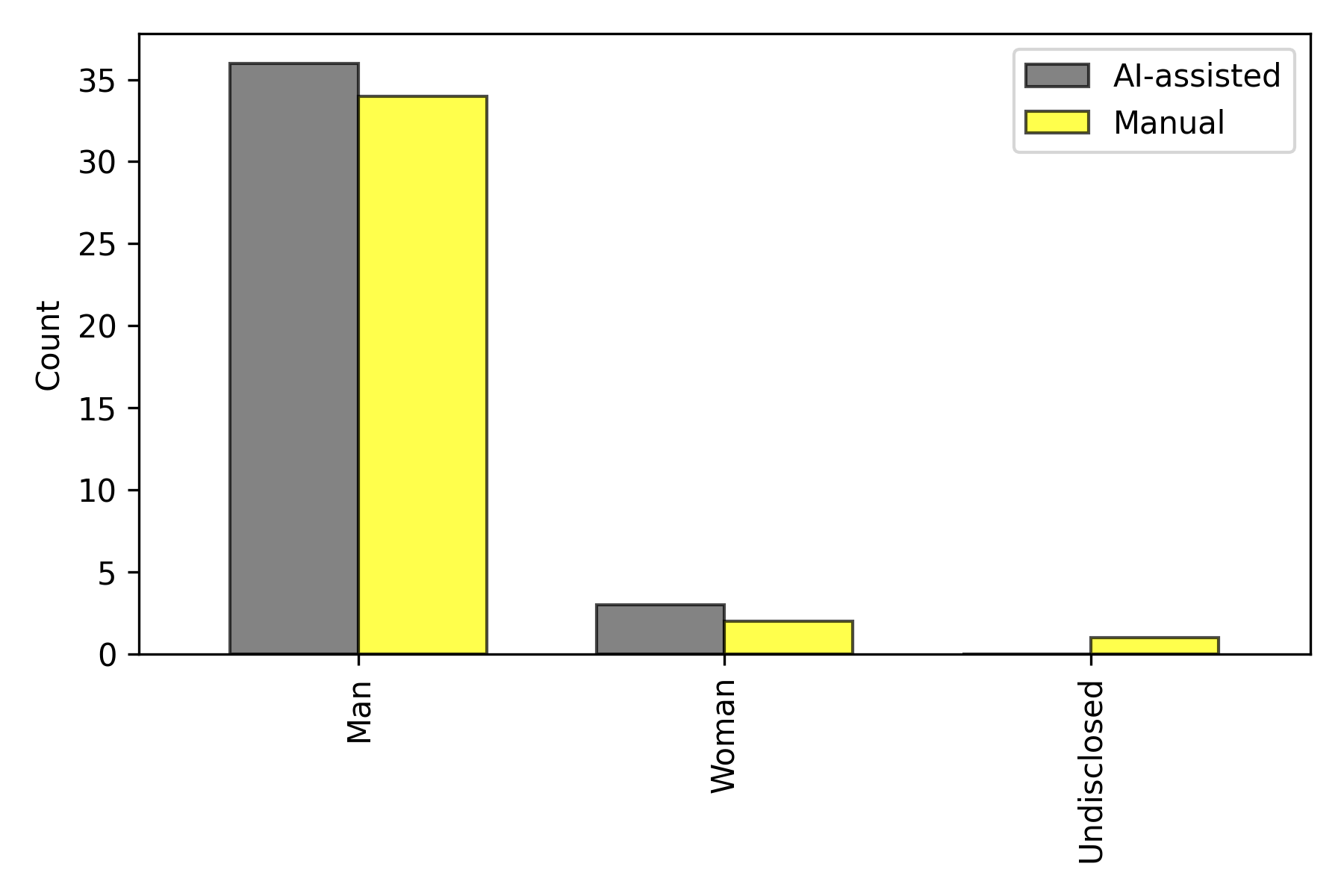}
        \caption{Gender}
        \label{fig:task1-gender}
    \end{subfigure}
    \hfill
    \begin{subfigure}[b]{0.48\textwidth}
        \centering
        \includegraphics[width=\linewidth]{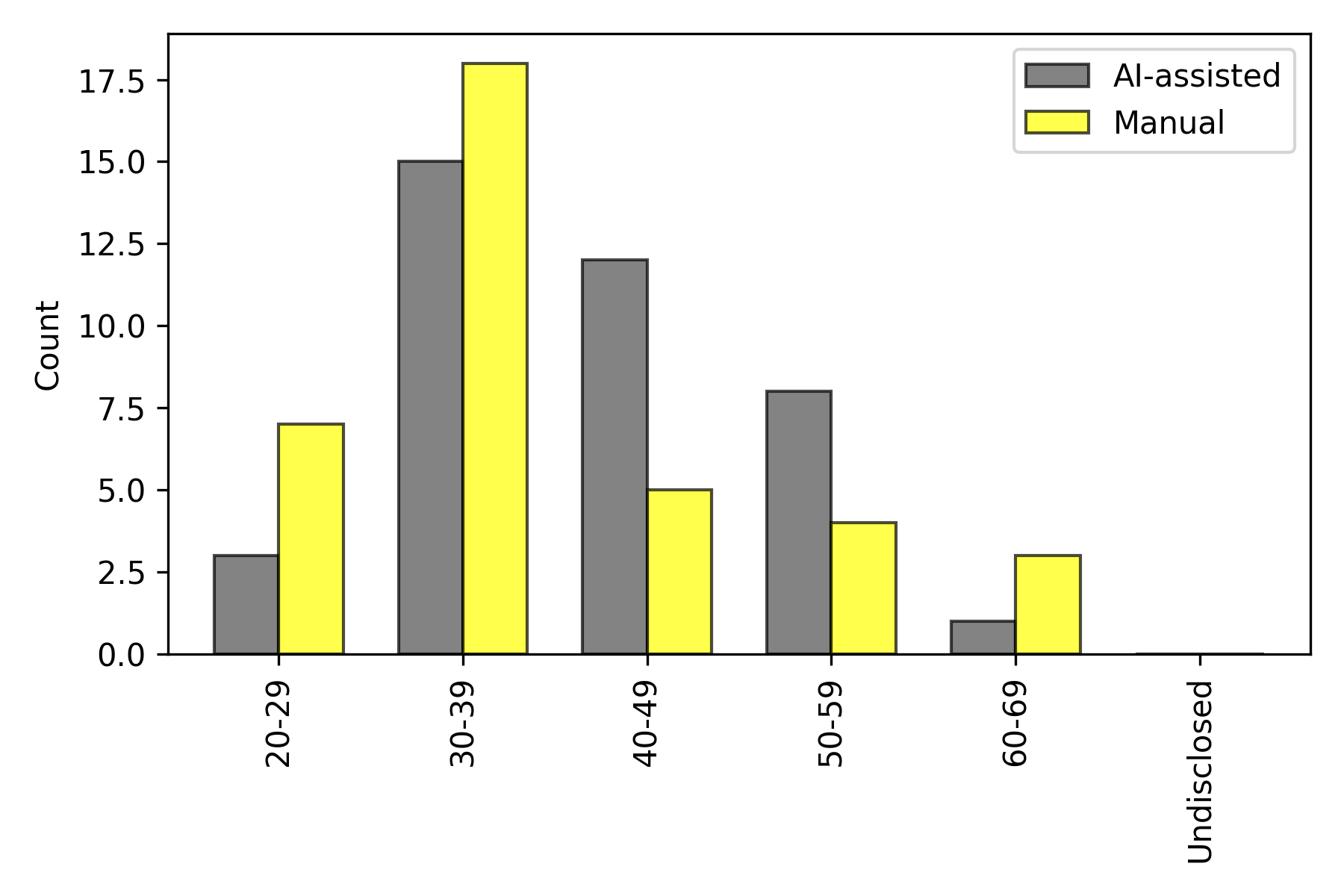}
        \caption{Age}
        \label{fig:task1-age}
    \end{subfigure}
    \begin{subfigure}[b]{0.48\textwidth}
        \centering
        \includegraphics[width=\linewidth]{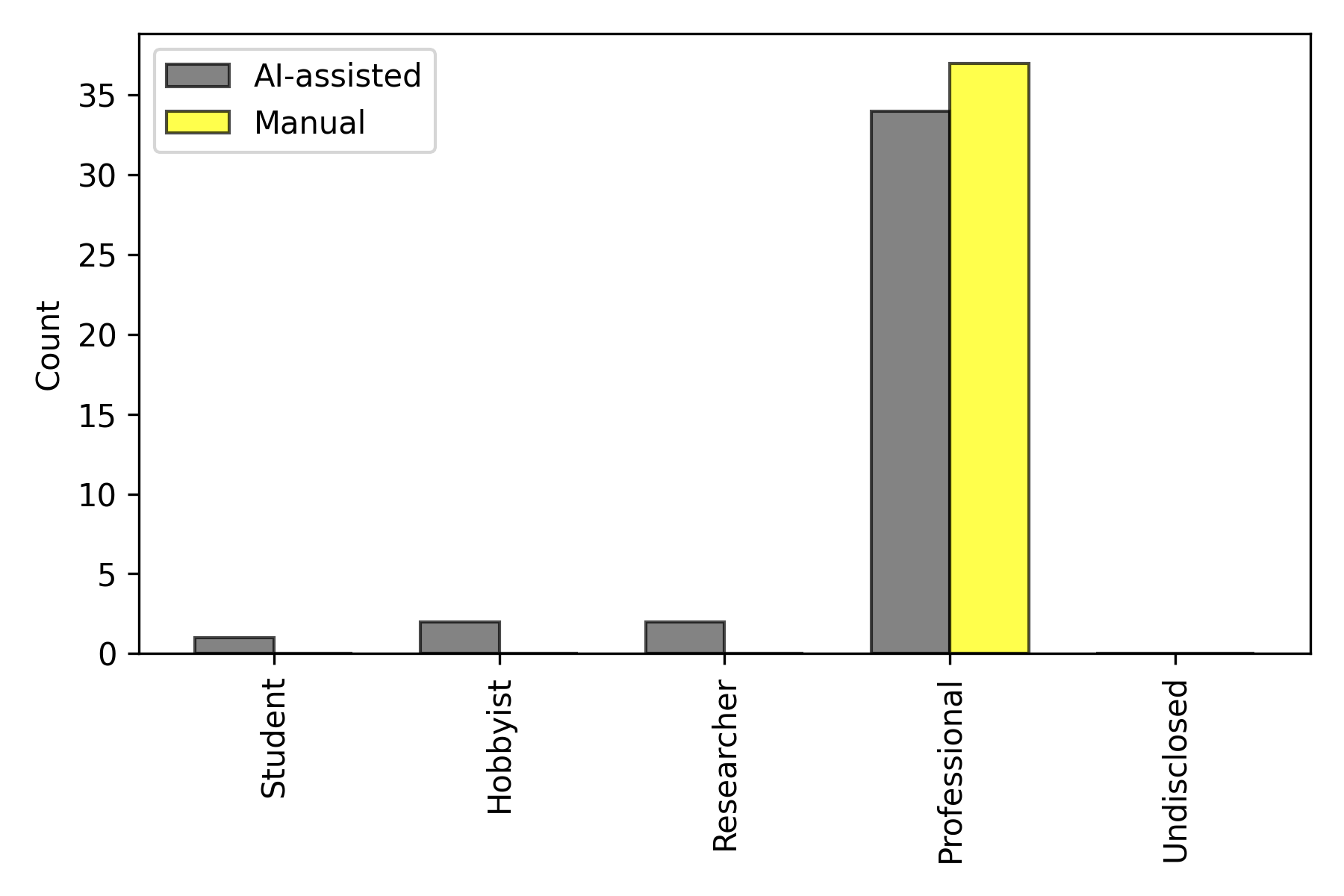}
        \caption{Role}
        \label{fig:task1-role}
    \end{subfigure}
    \hfill
    \begin{subfigure}[b]{0.48\textwidth}
        \centering
        \includegraphics[width=\linewidth]{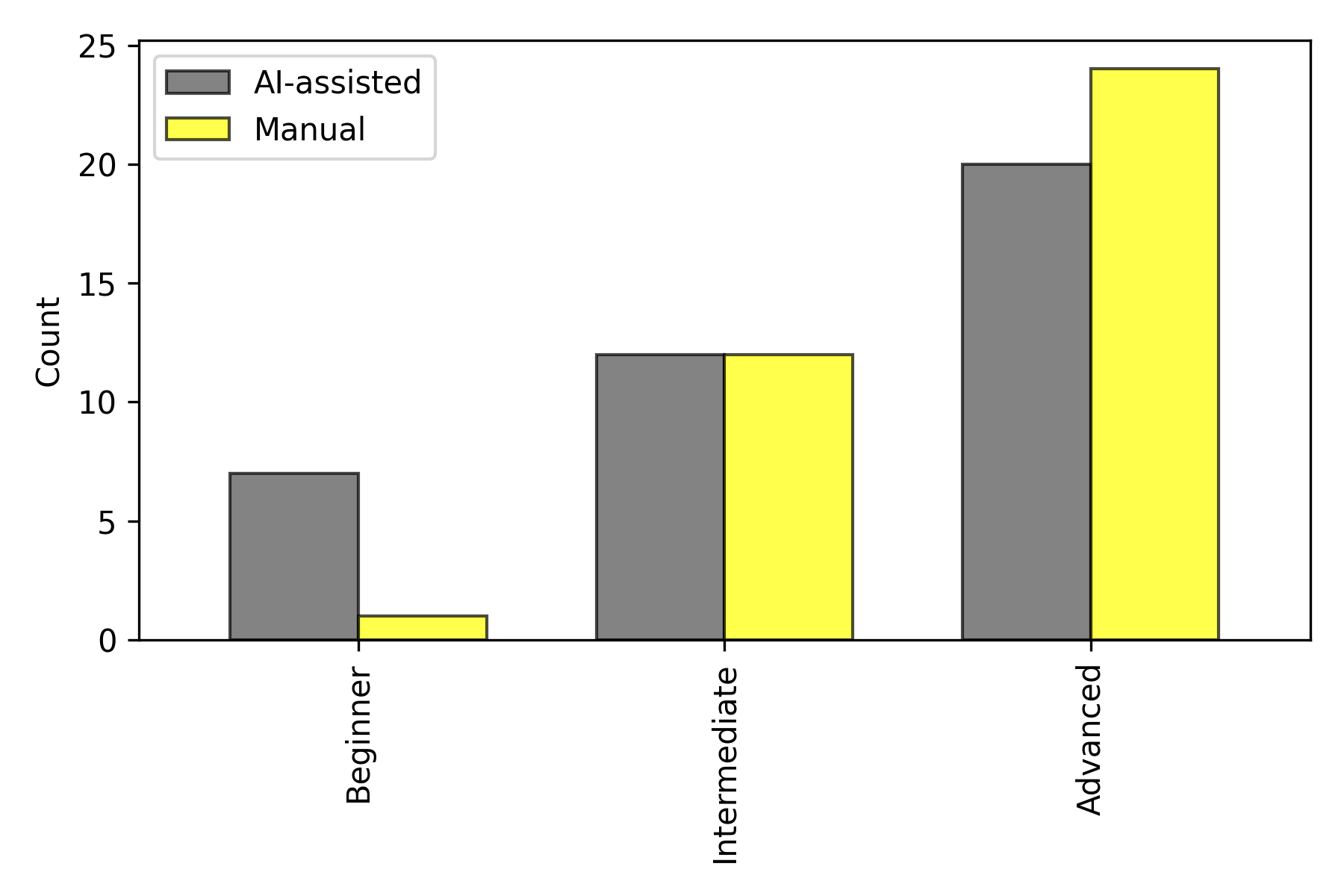}
        \caption{Java Skill Level}
        \label{fig:task1-java}
    \end{subfigure}
    \caption{Demographic distribution of Task 1 participants.}
    \label{fig:task1-demo}
\end{figure}

Figure~\ref{fig:task1-demo} shows demographics for the Task~1 participants. We found that most participants were men and between 30-49 years old. Most participants were professional developers, but there were slightly more Java beginners in the \textbf{AI-dev} group. This aspect will be further analyzed later. Overall, we do not observe any major systematic biases that threaten the validity of the RCT.

\begin{figure}
    \centering
    \includegraphics[width=0.99\linewidth]{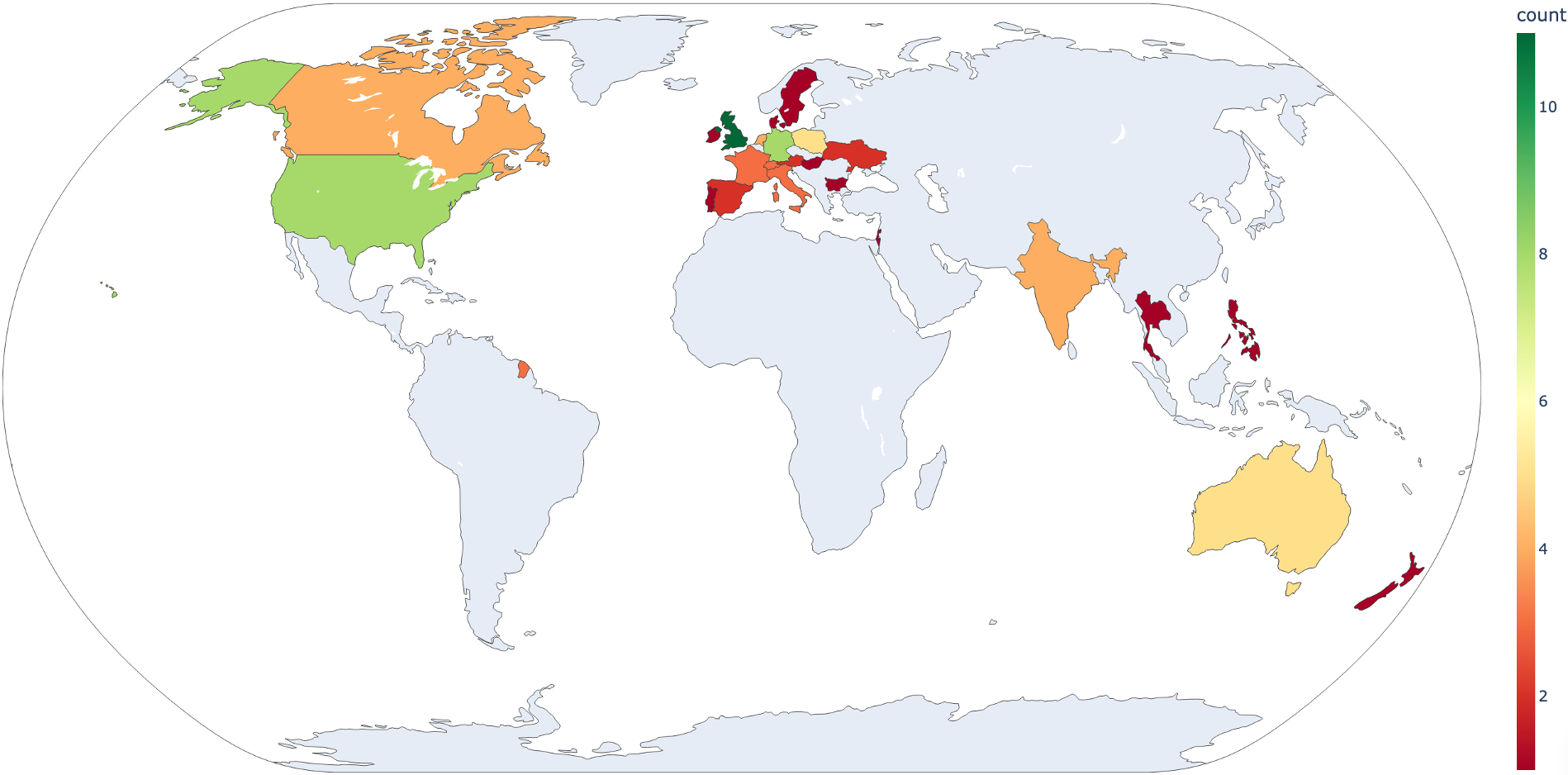}
    \caption{Location of Task 1 participants.}
    \label{fig:task1-country}
\end{figure}

Figure~\ref{fig:task1-country} shows the geographic distribution of Task~1 participants. We notice participation from many Western countries. However, we find that neither South America nor Africa is represented. Notably, several major countries with large developer populations, including China, Japan, and South Korea, are also absent.

Table~\ref{tab:ai-tools} shows the frequencies of AI tool usage reported by \textbf{AI-devs} in Task~1 (Q2-3), both for all participants and habitual AI users only. The bolded alternatives were explicitly listed in the questionnaire, whereas all others were entered manually by participants. We note that GitHub Copilot was clearly the most used AI assistant, followed by ChatGPT and Cursor. We observed no instances of fully autonomous agents being tasked with the entire problem in Task~1, although recent versions of Cline and Claude can be used in such workflows. Still, some participants relied heavily on autonomous code generation for parts of the task -- shifting into product managers and treating the AI assistant as a junior developer. Regarding the frequency of AI usage (Q2-4), we obtained: Every statement 17, Often 10, and Sometimes 12.

\begin{table}[ht]
\centering
\caption{Frequencies of AI tools reported in Task 1 (Q2-3).}
\label{tab:ai-tools}
\begin{tabular}{c|*{16}{c}}
\toprule
\textbf{AI tool} &  
\rotatebox{90}{\textbf{GitHub Copilot}} &
\rotatebox{90}{\textbf{ChatGPT}} &
\rotatebox{90}{Cursor} &
\rotatebox{90}{\textbf{JetBrains AI}} &
\rotatebox{90}{Claude} &
\rotatebox{90}{Tailwind AI} &
\rotatebox{90}{\textbf{VS IntelliCode}} &
\rotatebox{90}{Mixtral} &
\rotatebox{90}{Microsoft Copilot} &
\rotatebox{90}{Windsurf} &
\rotatebox{90}{Cline} &
\rotatebox{90}{Grok} &
\rotatebox{90}{Gemini} &
\rotatebox{90}{Supermaven} &
\rotatebox{90}{\textbf{CodeWhisperer}} &
\rotatebox{90}{\textbf{TabNine}} \\
\midrule
\textbf{All participants} &
21 & 13 & 9 & 5 & 4 & 2 & 2 & 1 & 1 & 1 & 1 & 1 & 1 & 1 & 0 & 0 \\
\textbf{Habitual AI users} &
8 & 7 & 7 & 1 & 0 & 2 & 0 & 0 & 0 & 1 & 0 & 0 & 1 & 0 & 0 & 0 \\
\bottomrule
\end{tabular}
\end{table}

Some Task 1 participants reported using (non-AI) development tools beyond a standard IDE. In \textbf{AI-dev}, three used SonarQube and one used Docker. In \textbf{!AI-dev}, the following tools were each reported once: SonarQube, GitHub Codespaces, Firefox, and Chrome. The two web browsers' debug windows were used to check the web application. The four participants using SonarQube submitted solutions with CH 8.35, 8.34, 8.23, 8.16, respectively. These scores do not stand out (see Figure~\ref{fig:task1-ch}), and we consider the influence of additional development tools to be negligible and thus controlled in the rest of the analysis.

\subsubsection{Task 2 Demographics}
\begin{figure}
    \centering
    \includegraphics[width=1\linewidth]{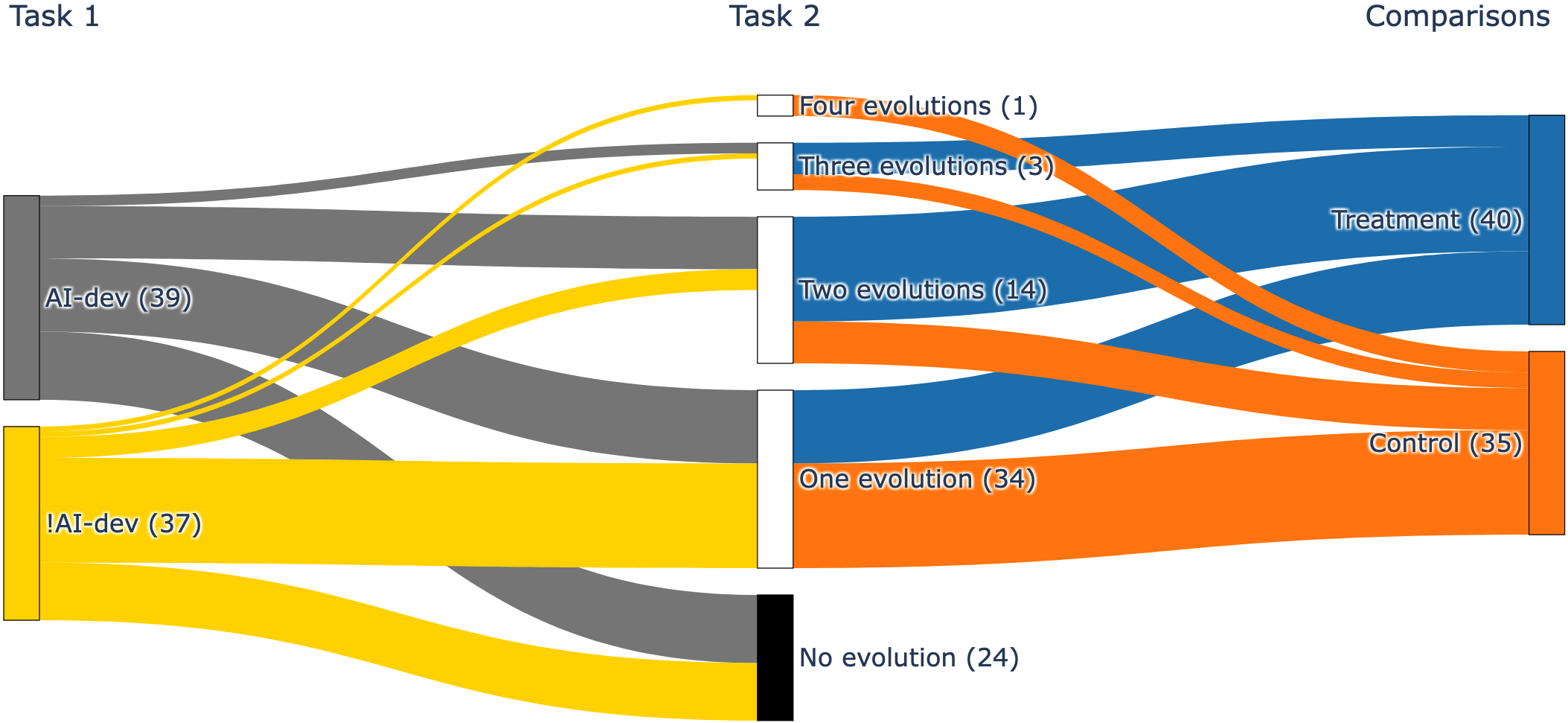}
    \caption{Flow of Task 1 solutions into Task 2 evolutions by experimental group. The treatment group evolves solutions originally completed with AI assistance (AI-dev), whereas the control group evolves solutions completed without AI assistance (!AI-dev).}
    \label{fig:sankey-tasks}
\end{figure}

Task 2 involved extending the feature developed by an unknown Task~1 developer. In the end, 75 out of 225 (33.3\%) assigned participants submitted valid Task~2 solutions. The completion rate matches what we observed for both \textbf{AI-dev} and \textbf{!AI-dev} groups in Task~1. Figure~\ref{fig:sankey-tasks} shows an overview of how the Task 1 solutions were evolved by Task 2 participants. The number of Task~2 submissions building on Task~1 solutions varied between 0 and 4\footnote{On a few occasions, we assigned more than three participants to a single Task 1 solution -- essentially due to rounding in the assignment logic.}. Of the 75 Task 1 solutions, 24 were not evolved further (see the black box in Figure~\ref{fig:sankey-tasks}). Note that these 24 solutions had CH scores comparable to the 51 that were evolved (mean 8.32 vs. 8.35; median 8.26 vs. 8.34). The most common outcome was that a Task~1 solution was evolved by one Task~2 participant. In some cases, individual Task 1 solutions were evolved by multiple Task~2 participants. More specifically, 14~Task~1 solutions were each used as the starting point for two different Task~2 participants, 3 solutions for three different participants, and 1 solution for four different participants.


\begin{figure}[ht]
    \centering
    \begin{subfigure}[b]{0.48\textwidth}
        \centering
        \includegraphics[width=\linewidth]{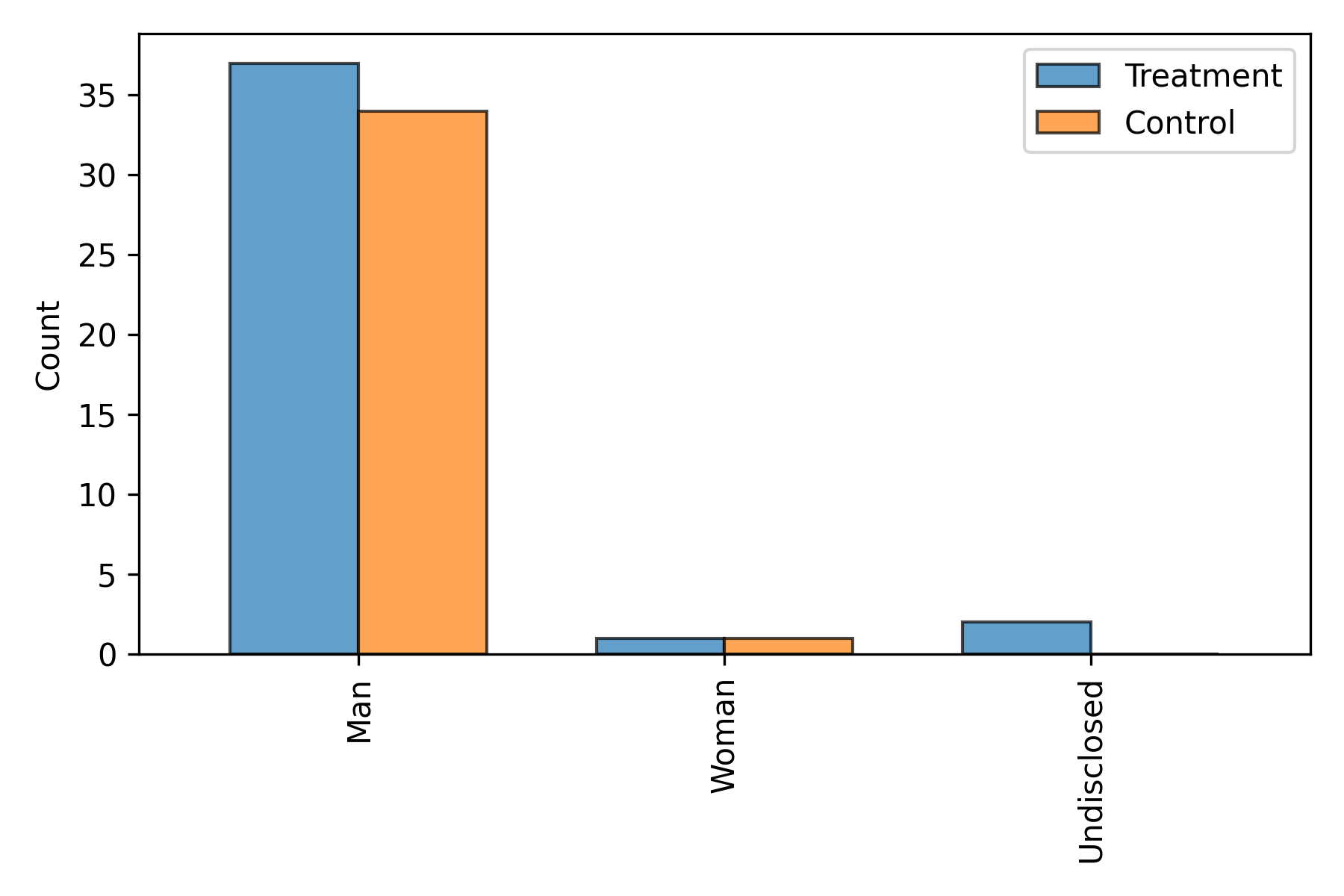}
        \caption{Gender}
        \label{fig:task2-gender}
    \end{subfigure}
    \hfill
    \begin{subfigure}[b]{0.48\textwidth}
        \centering
        \includegraphics[width=\linewidth]{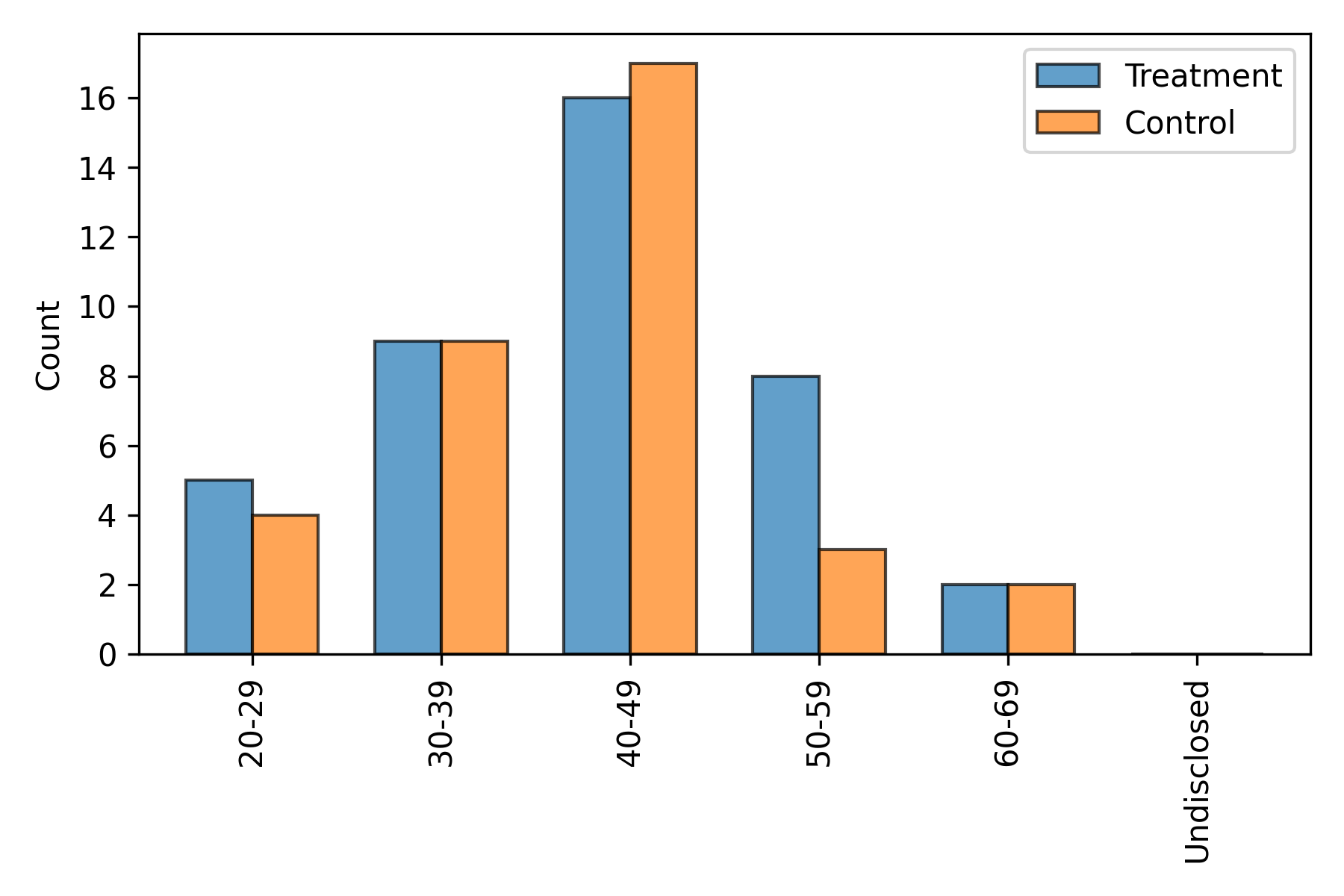}
        \caption{Age}
        \label{fig:task2-age}
    \end{subfigure}
    \begin{subfigure}[b]{0.48\textwidth}
        \centering
        \includegraphics[width=\linewidth]{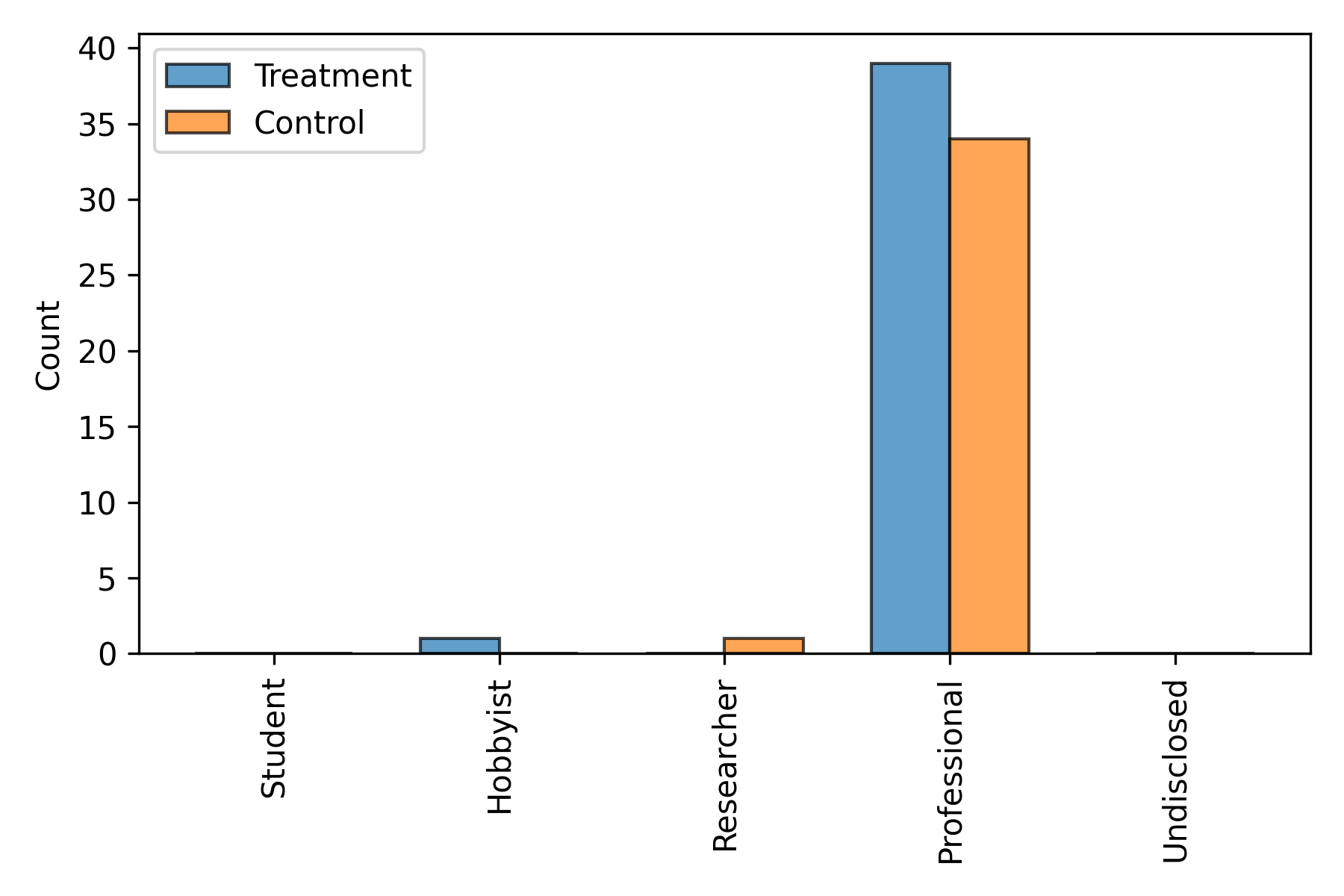}
        \caption{Role}
        \label{fig:task2-role}
    \end{subfigure}
    \hfill
    \begin{subfigure}[b]{0.48\textwidth}
        \centering
        \includegraphics[width=\linewidth]{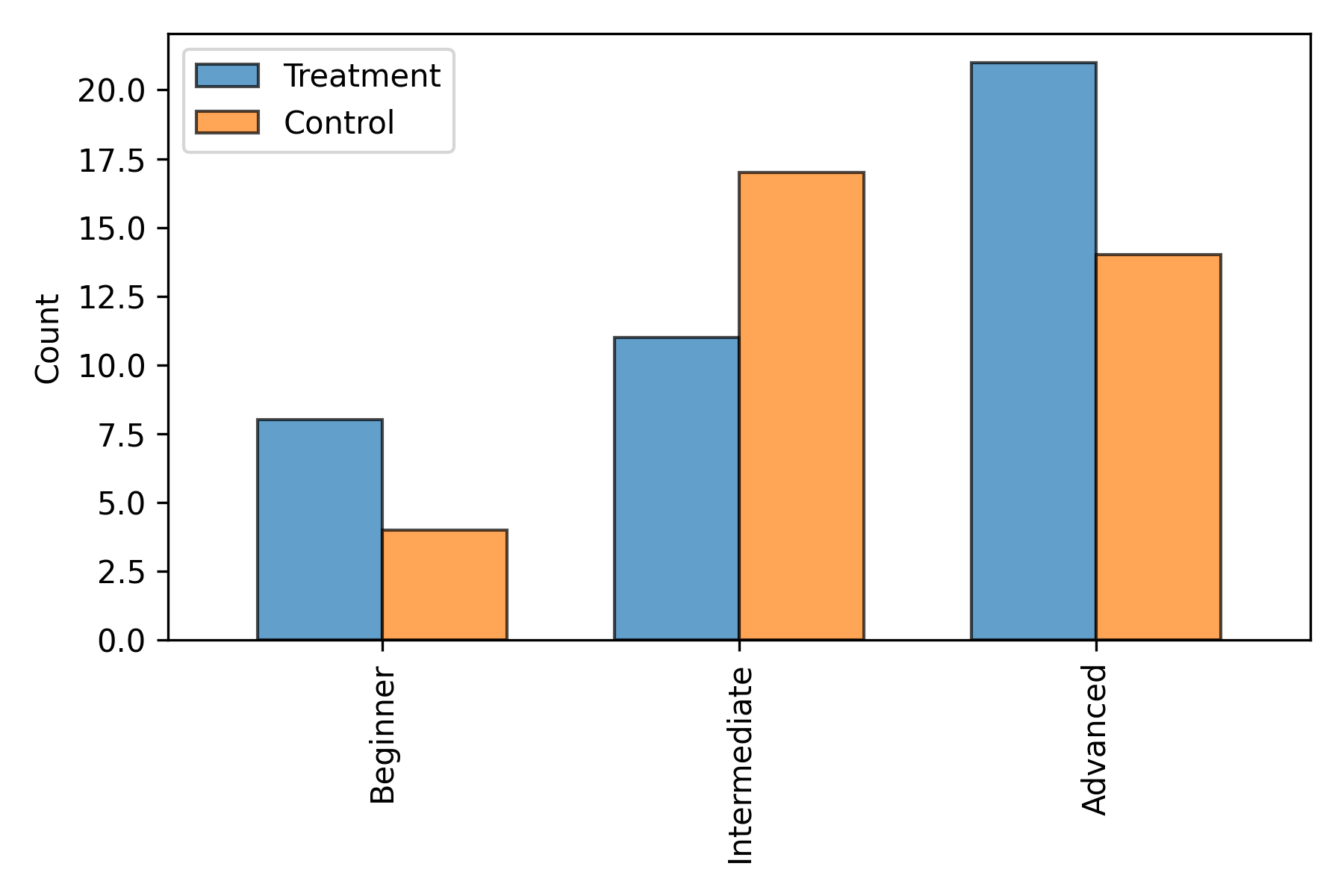}
        \caption{Java Skill Level}
        \label{fig:task2-java}
    \end{subfigure}
    \caption{Demographic distribution of Task 2 participants.}
    \label{fig:task2-demo}
\end{figure}

Figure~\ref{fig:task2-demo} shows the demographics of the 75 participants who completed Task~2. As in Task~1, most participants were men, and the median age in both groups was 40-49 years. Similarly, the Task~2 participants are predominantly professional developers. The reported levels of Java mastery differed slightly between the treatment and control groups, which we account for later. 

\begin{figure}
    \centering
    \includegraphics[width=0.99\linewidth]{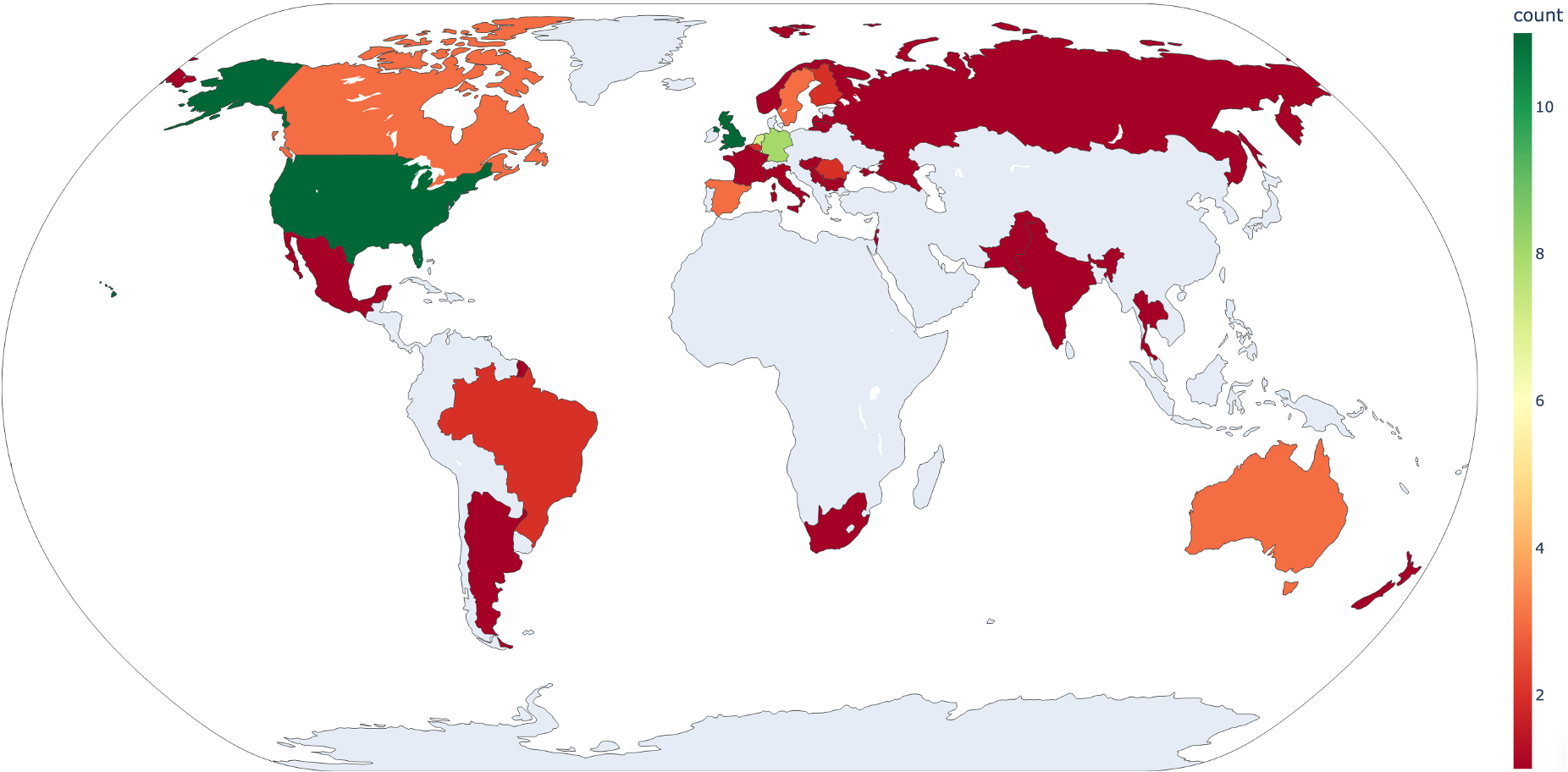}
    \caption{Location of Task 2 participants.}
    \label{fig:task2-country}
\end{figure}

Figure~\ref{fig:task2-country} shows the geographic distribution of Task~2 participants. The distribution resembles what we found for Task~1, but now both South America and Africa are represented. However, there are still no participants from China, Japan, or South Korea.

Eight Task 2 participants reported using development tools beyond a standard IDE. In the treatment group, three used SonarQube and one used GitHub Codespaces. In the control group, three used SonarQube and one used Postman. The six participants who used SonarQube submitted solutions with CH 8.84, 8.55, 8.54, 8.32, 8.25, and 8.24, respectively. The highest score (8.84) was obtained by the ambitious \textit{anon050}, who spent 5.5 hours on the task and also achieved the top TC score (90\%). The two lowest scores belong to 1) a Java beginner who stated \textit{``I found it fun working in an unfamiliar language. My main language is C\# and I work normally in Visual Studio and with .NET.''} and 2) an advanced developer who complained about the lack of unit tests: \textit{``I'm very used to TDD so finding no to very few tests, specially unit tests, is somewhat frustrating.''} The remaining three CH scores align well with the overall distribution (see Figure~\ref{fig:task2-ch}), and we find no indications of systematic differences caused by the tool usage.

\subsubsection{Task-Level Summary Statistics} \label{sec:desc-stats}
Table~\ref{tab:git-stats} shows summary statistics of git activity in Task 1 and Task 2. In Task 1, we notice that four \textbf{AI-devs} skew the distribution by adding more than 2,000 lines of code -- well above the maximum number of 660 observed in the \textbf{!AI-dev} group. Still, the median number of added lines for \textbf{!AI-dev} is lower than for \textbf{AI-dev}. The four most prolific \textbf{AI-devs} were:

\begin{itemize}
    \item[anon126] Java beginner. Used Cursor and Tailwind AI. Also the participant who completed Task~1 twice (see Section~\ref{sec:dataproc}). In the first submission, \textit{anon126} explained pushing AI to the maximum, resulting in 10,788 added lines. Despite the volume, the CH was the second highest in Task 1: 8.88. Note that the added lines did not include extra features beyond the task; roughly 90\% consisted of Javadoc comments, test code, and a large Tailwind library.
    \item[anon054] Advanced Java developer. Used Cursor, ChatGPT, and Cline. Added 3,094 lines, and achieved the highest recorded CH: 9.12 with 98\% TC. Explained:
    \interviewquote{My approach was to learn how well it works letting the AI do most of the work on its own and me being the mentor like with a new colleague at work steering where necessary. [...] Cline did most of the work autonomously based on the existing code and description.}{\textit{anon054}, Task 1, Pro+Advanced, \textbf{AI-dev} [HABITUAL], ChatGPT+Cline+Cursor, R4, 3094 LoC, (\textit{5h}, \textit{CH=9.12}, \textit{TC=98\%}, \textit{PP=4.4})}
    \item[anon139] Java beginner. Used GitHub Copilot, ChatGPT, VS IntelliCode, and Claude. Added 2,032 lines. Explained \emph{``going a bit deep in making it more maintainable''}, but the CH remained at a modest 8.48. 
    \item[anon073] Advanced Java developer. Used Cursor. Added 2,635 lines and reached CH 8.38. Provided a rich comment: 
    \interviewquote{I haven't programmed in Java for 15 years, and never touched Spring Boot. It would have taken me a lot longer without the help of an AI. The interaction with the AI felt like pair programming with someone with a vast experience but lazy who sometimes gives hacky answers. I had to constantly remind the AI to use the most recent and best practice information to get better code. I do have 30 years experience in development and multiple frameworks. I felt that that experience was important with identifying subpar code or dead leads that the AI suggested. Just pointing out those to the AI was enough for it to come up with better code. I hardly had to tamper with the code manually.}{\textit{anon073}, Task 1, AI-dev [HABITUAL], Pro+Advanced, Cursor, R4, 2635 LoC, (4.5h, CH=8.38, TC=67\%, PP=4.2)}
\end{itemize}

\noindent We note that the two Task~1 participants who added the most code using AI also obtained the highest CH. This suggests that heavy use of AI assistants can result in code that is, on average, easily understandable by humans. However, maintainability effort also increase with the overall size of the codebase. We revisit this dilemma in Task~2. 

\begin{table}[ht]
\centering
\caption{Descriptive statistics of git activity.}
\label{tab:git-stats}
\begin{tabular}{lcccc}
\toprule
 & \multicolumn{2}{c}{\textbf{Task 1}} & \multicolumn{2}{c}{\textbf{Task 2}} \\
 & AI-dev (n=39) & !AI-dev (n=37) & Treatment (n=40) & Control (n=35) \\
\midrule
\textbf{commits}      &  &   &  &   \\
mean / std                     & 6.85 / 13.66 & 5.16 / 7.48 & 5.62 / 12.16 & 4.40 / 4.72\\
min / median / max                   & 1 / 3 / 82 & 1 / 3 / 38 & 1 / 2 / 74 & 1 / 3 / 21 \\
\midrule
\textbf{added lines}    &   &   &  &   \\
mean / std                    & 642.74 / 1801.63 & 216.41 / 187.41 & 158.62 / 202.54 & 154.23 / 172.97 \\
min / median / max                   & 18 / 123 / 10,788 & 22 / 166 / 660 & 6 / 56.5 / 703 & 7 / 80 / 703 \\
\midrule
\textbf{deleted lines}    &  &   &   &  \\
mean / std                     & 128.54 / 141.45 & 102.32 / 111.79 & 72.53 / 129.67 & 103.26 / 196.53\\
min / median / max                   & 8 / 74 / 642 & 4 / 57 / 512 & 0 / 10 / 562 & 0 / 17 / 854 \\
\midrule
\textbf{changed files} &   &  &  &  \\
mean / std                   & 10.92 / 14.92 & 9.65 / 6.71 & 5.92 / 5.90 & 6.60 / 6.04 \\
min / median / max                   & 2 / 7 / 88 & 2 / 8 / 29 & 1 / 4 / 29 & 2 / 4 / 27 \\
\bottomrule
\end{tabular}
\end{table}

The Task 2 git activity of the treatment and control groups (see Table~\ref{tab:git-stats}) appears similar. The median number of added and deleted lines is slightly higher in the control group, but the difference is modest. We have no other patterns to report.

\begin{figure}[ht]
    \centering
    \begin{subfigure}[b]{0.42\textwidth}
        \centering
        \includegraphics[width=\linewidth]{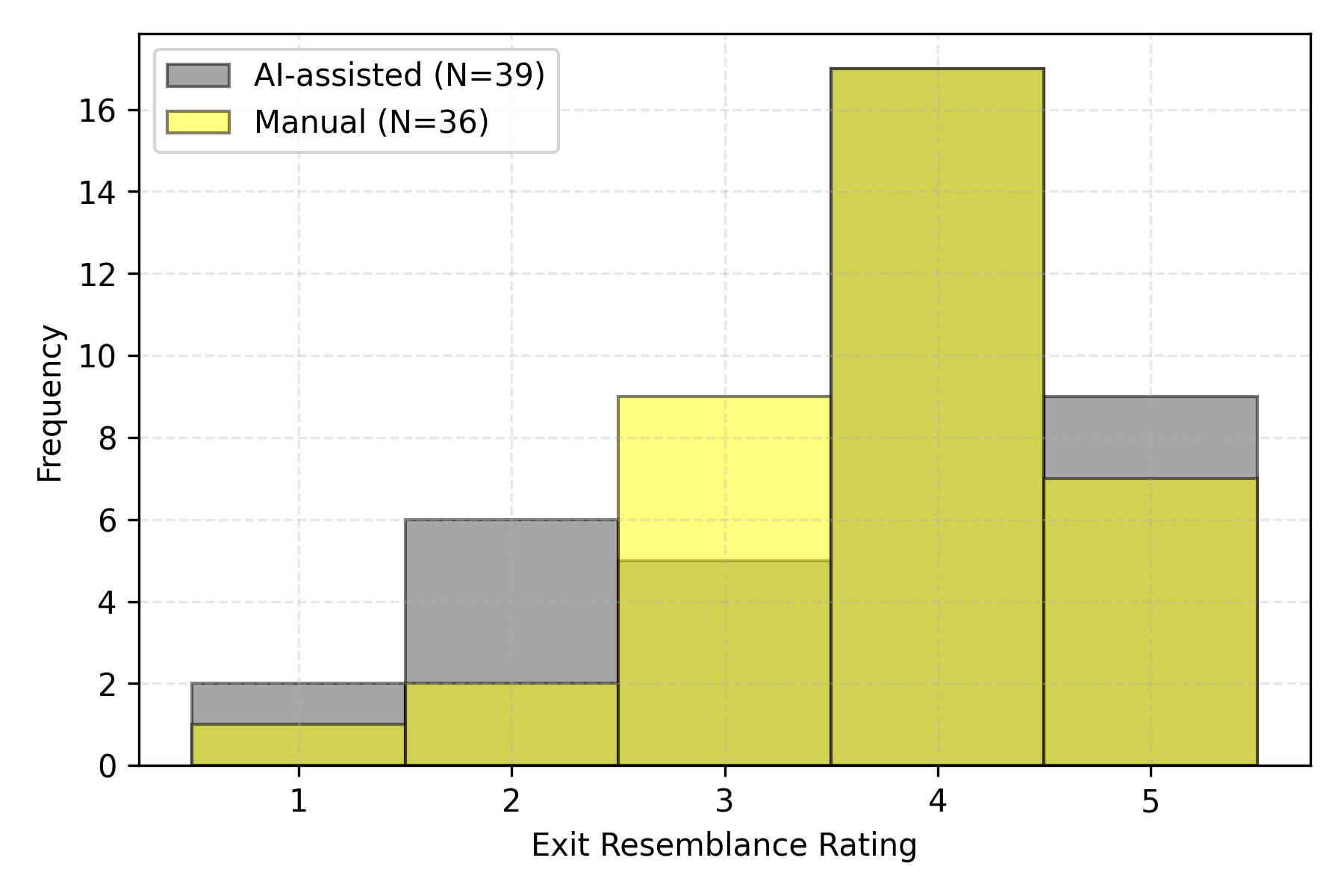}
        \caption{Task 1.}
        \label{fig:task1-resemblance}
    \end{subfigure}
    \hfill
    \begin{subfigure}[b]{0.42\textwidth}
        \centering
        \includegraphics[width=\linewidth]{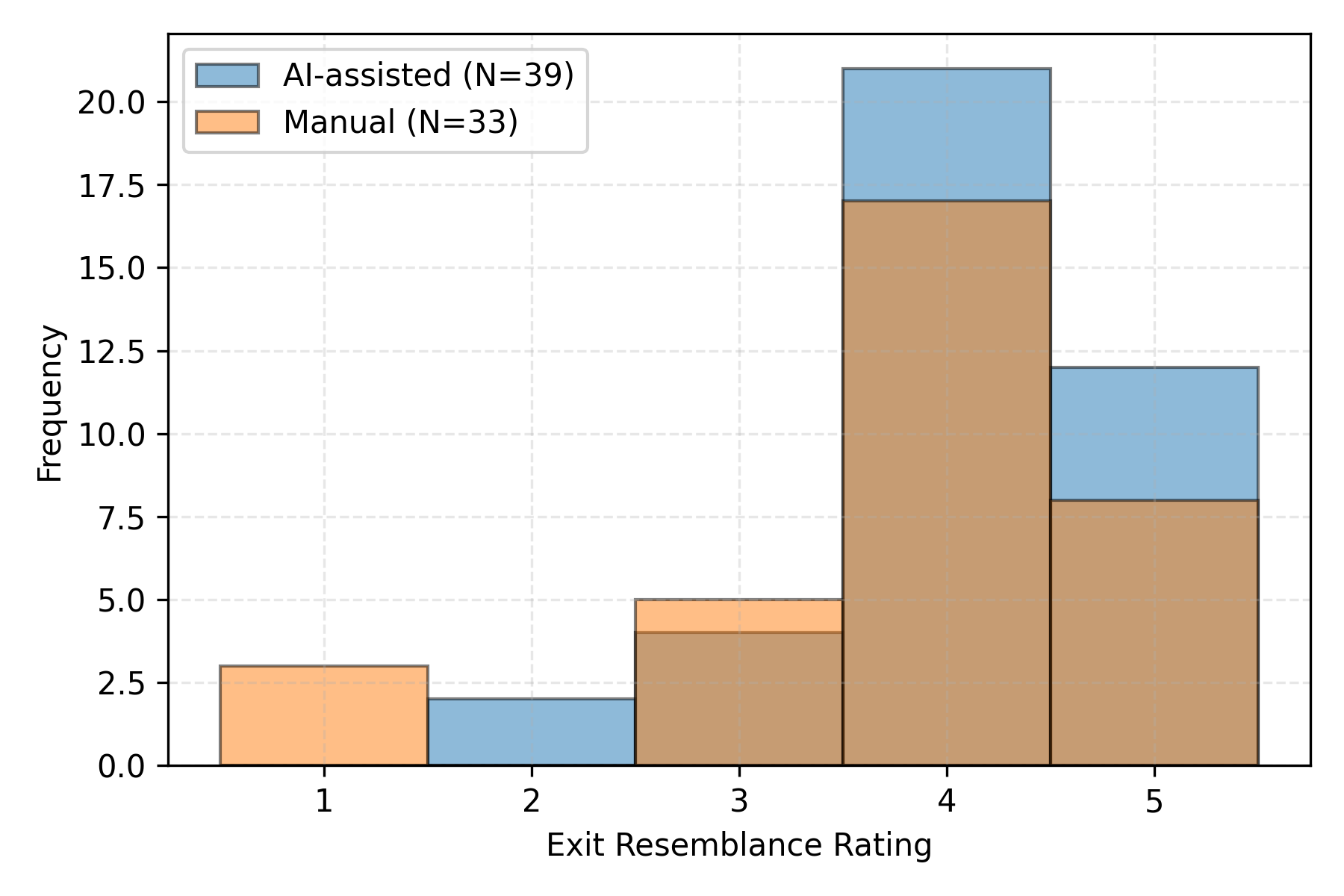}
        \caption{Task 2.}
        \label{fig:task2-resemblance}
    \end{subfigure}
    \caption{Perceived resemblance to prior development work (Q2-5: 1=Strongly disagree, 5=strongly agree).}
    \label{fig:task-resemblance}
\end{figure}

Figure~\ref{fig:task-resemblance} shows answers to Q2-5, i.e., the level of agreement to the statement ``The task generally resembled development work I have done in the past.'' The results show that both Tasks 1 and 2 were perceived as realistic development tasks, which supports the validity of our study.
 
\subsection{RQ1: More Efficient Manual Evolution?} \label{sec:res-rq1}
This section reports results related to the two dependent variables Completion time and Perceived Productivity (PP) (see Figure~\ref{fig:gqm}). For each variable, we discuss the distribution followed by frequentist inferential statistics and a Bayesian analysis. 
We conclude by summarizing the effect of the independent variable, expressed as treatment minus control, in a textbox with the following structure:
\begin{itemize}
    \item Frequentist analysis: mean/median difference; 95\% confidence interval for the difference; $p$-value; effect size.
    \item Bayesian analysis: Probability of direction (positive/negative); posterior mean percentage change; 95\% credible interval for the change.
    \item A brief textual synthesis of these statistics.
\end{itemize}

\subsubsection{Completion Time} \label{sec:rq1-time}
\begin{figure}[ht]
    \centering
    \begin{subfigure}[b]{0.58\textwidth}
        \centering
        \includegraphics[width=\linewidth]{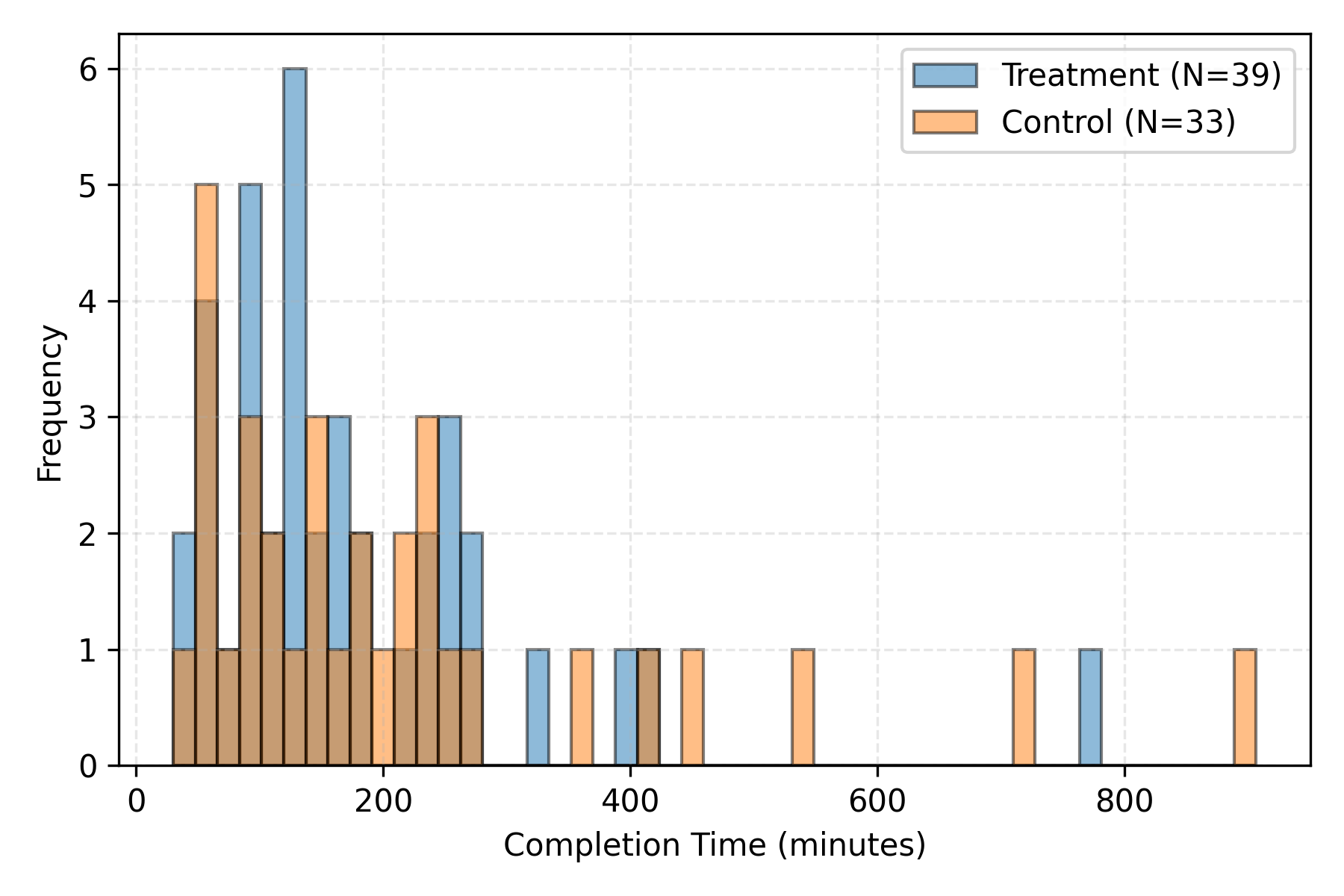}
        \caption{Distribution of completion times.}
        \label{fig:task2-time-dist}
    \end{subfigure}
    \hfill
    \begin{subfigure}[b]{0.34\textwidth}
        \centering
        \includegraphics[width=\linewidth]{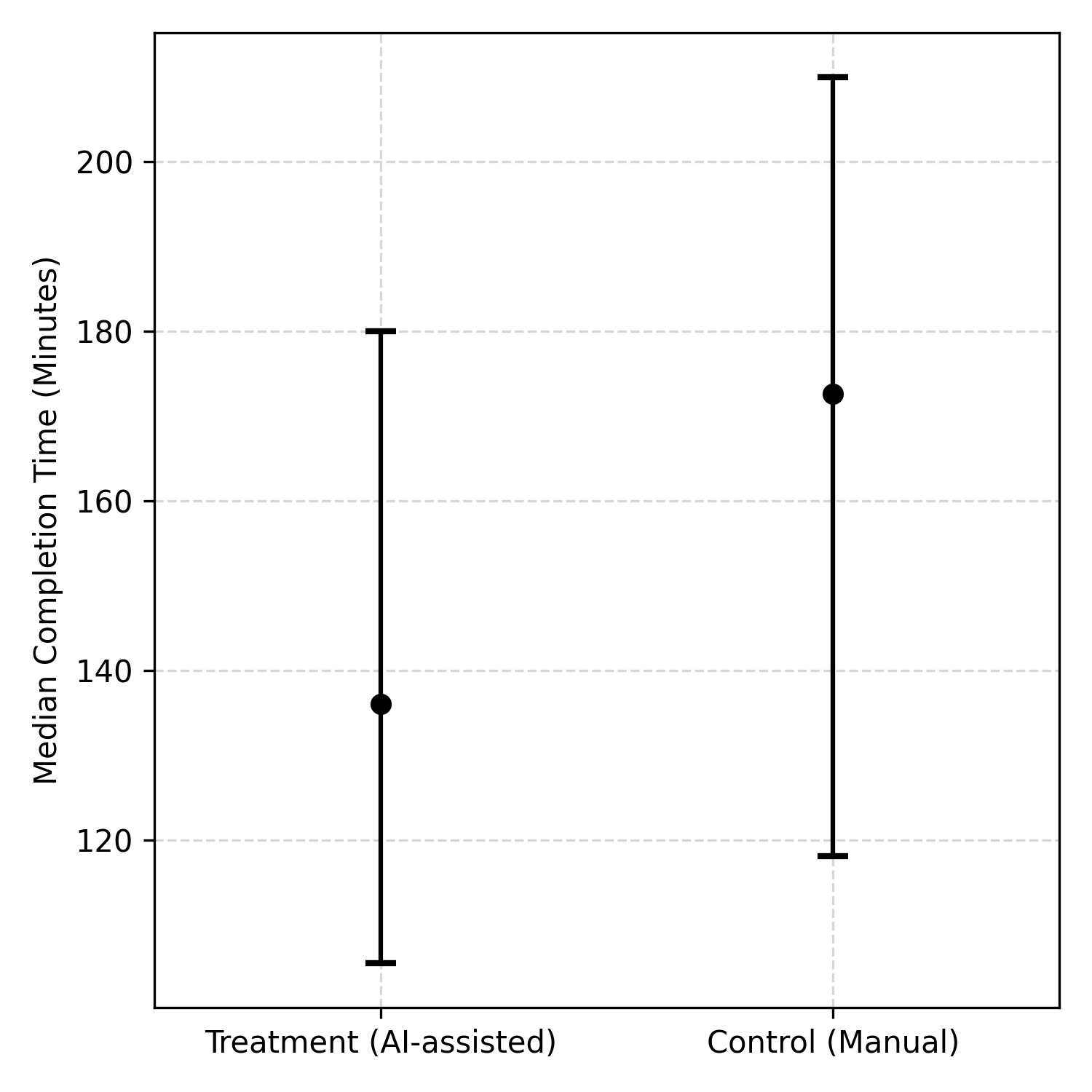}
        \caption{95\% confidence intervals.}
        \label{fig:task2-time-ci}
    \end{subfigure}
    \caption{Task~2 completion time.}
    \label{fig:task2-time}
\end{figure}

Figure~\ref{fig:task2-time-dist} shows the distribution of 72 completion times for Task~2. Most participants completed the task within 300 minutes, and as described in Section~\ref{sec:freq}, the data is not normally distributed. The longest time was 13 hours, based on an estimated breakdown from one particularly ambitious participant, an advanced Java developer, who explained their process as follows:

\interviewquote{I'd say I've spent around 12-16 hours studying the stack -- last time I worked with Java it was Java 7, and a lot has changed. In that time I did, of course, spend some time experimenting with different solutions which did not make it into the project. After settling with the JPA Specification API, it took me 2-4 hours to have the acceptance test green [...] Sticking to the role I was playing as a consultant, I then invested some time in refactoring, which then took another 8-12 hours, I'd say. I wasn't sure if this was expected for the task, but its description gave me the impression that the assessment would be reviewed as if I were hired to do a job.}{anon123, Task 2, Treatment, Pro+Advanced, R5, 645 LoC, (13h, CH=8.89, TC=71\%, PP=2.7)}

\noindent We excluded the initial 12-16 hours of exploratory learning and summed the remaining estimates to 13 hours. This is potentially an underestimate, and it shows that some participants were highly motivated to deliver high-quality solutions.

\paragraph{Frequentist Analysis}
Figure~\ref{fig:task2-ch-ci} shows 95\% confidence intervals for the median completion times. The treatment group had a median of 136 minutes (95\% CI: 105.5--180.0), while the control group had a median of 173 minutes (95\% CI: 118.1--210.0). Cliff’s $\delta$ was -0.079, indicating a \textit{negligible effect size}. The difference was \textit{not statistically significant} (Wilcoxon rank-sum test, $p$=0.56).

\begin{figure}
    \centering
    \includegraphics[width=0.75\linewidth]{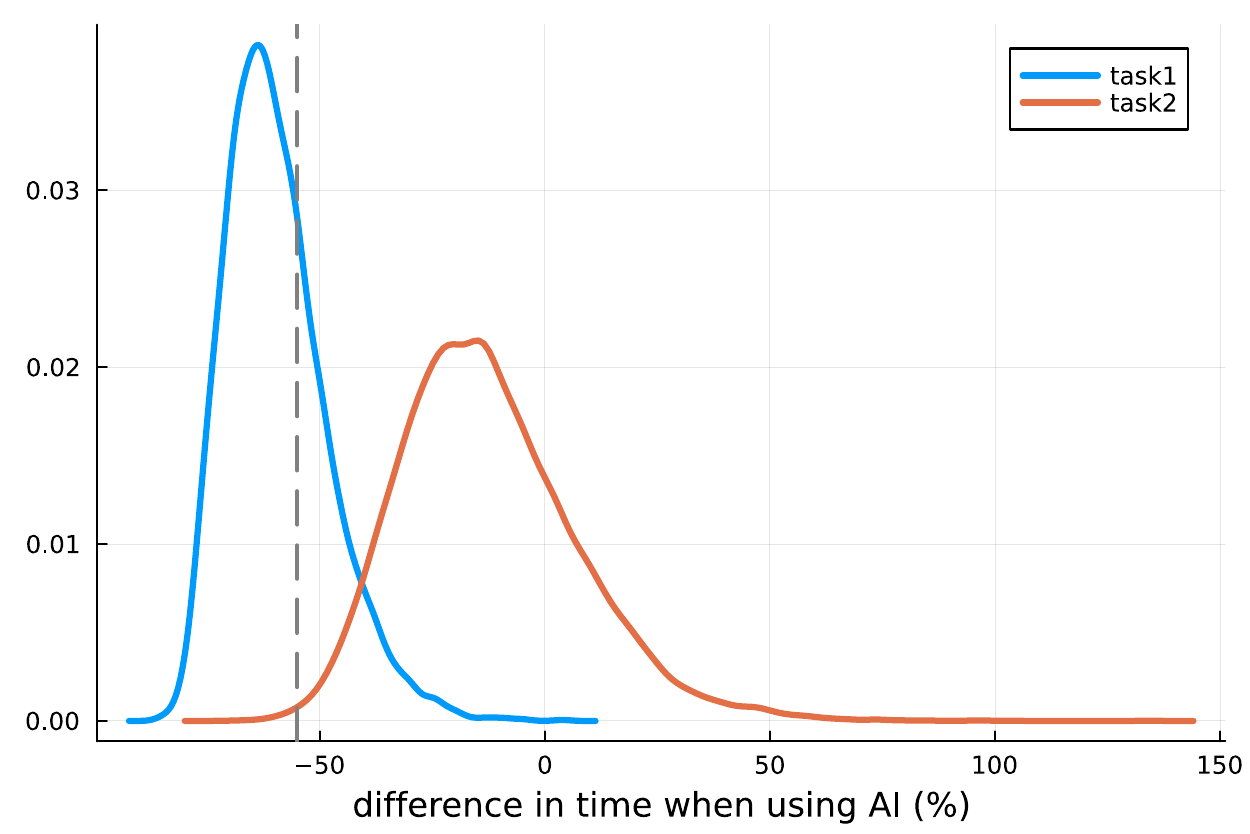}
    \caption{Posterior effect on completion time when a habitual AI user applies AI assistants in Task~1. The orange curve shows the effect on manual evolution in Task~2, i.e., the focus of our study.}
    \label{fig:bayes-time}
\end{figure}

\paragraph{Bayesian Analysis}
For Task 1, our Bayesian model shows that using an AI assistant reliably reduced completion time ($P(\Delta < 0) > 99\%$). Figure~\ref{fig:bayes-time} shows the posterior distribution of the treatment effect on completion time when the $Dev1$ was a habitual AI user. AI-assisted babitual AI users finished Task~1 approximately 60\% faster than \textbf{!AI-devs} on average (CrI: $[-77.11\%, -30.64\%]$). Our results are consistent with those of \citet{peng_impact_2023}, indicated by the dashed line in Figure \ref{fig:bayes-time}.

As a point of comparison, changing $Dev1$'s skill from ``Beginner'' to ``Advanced'' had a non-significant effect ($P(\Delta<0) = 83\%$), with a mean effect of $-21.4\%$ (CrI: $[-60.46\%, +34.28\%]$).

For Task 2, however, the positive effect of prior AI assistance observed for habitual AI users largely vanishes. We find no significant effect on $Dev2$'s completion time ($P(\Delta < 0) = 76\%$). 
The posterior mean effect of using AI is $-12.59\%$, but the posterior distribution is wide (CrI: $[-45.77\%, +32.68\%]$), encompassing both substantial speedups and slowdowns. The effect of higher $Dev1$'s skill is also not significant: the mean effect is $-27.89\%$, (CrI: $[-57.98\%, +12.72\%]$).

Moreover, the posterior probability that prior use of AI made $Dev2$ faster is 76\%, but this estimate is sensitive to the choice of priors. When using a pessimistic prior, the posterior probability that prior AI use decreases Task~2 completion time fell to about 38-39\%, which shows that we do not have conclusive results yet.

In contrast, a major influence on Task~2 completion time is whether $Dev2$ completed the task in one uninterrupted session or not (see $Dev2\_interrupted$ in Figure~\ref{fig:causal}). When $Dev2$ took breaks (interruption levels 2 or 3), they finished $173.66\%$ later than when they did not (CrI: $[+37.69\%, +416.88\%]$).


\begin{tcolorbox}[
  colback=lightblue,
  colframe=blue!75!black,
  coltitle=white,
  title=Quantitative Results – Task 2 Completion Time (Trt-Ctrl),
  fonttitle=\bfseries,
  enhanced,
  breakable,
  boxrule=0.5pt,
  leftrule=1pt,
  rightrule=1pt,
  toprule=1pt,
  bottomrule=1pt
]
\begin{itemize}
  \item \textbf{Frequentist.} Median difference $=-37$ minutes; 95\% CI $[-90,\,+38]$; Wilcoxon $p=0.56$; Cliff’s $\delta=-0.079$.
  \item \textbf{Bayesian.} When Task~1 was completed by a habitual AI user: P($\Delta <0)$ = $76\%$; Posterior mean change $=-12.6\%$; 95\% CrI $[-45\%,\,+33\%]$.
  \item Both analyses are consistent with a null effect on Task~2 completion time.
\end{itemize}
\end{tcolorbox}

\subsubsection{Perceived Productivity} \label{sec:res-pp}
\begin{figure}[ht]
    \centering
    \begin{subfigure}[b]{0.58\textwidth}
        \centering
        \includegraphics[width=\linewidth]{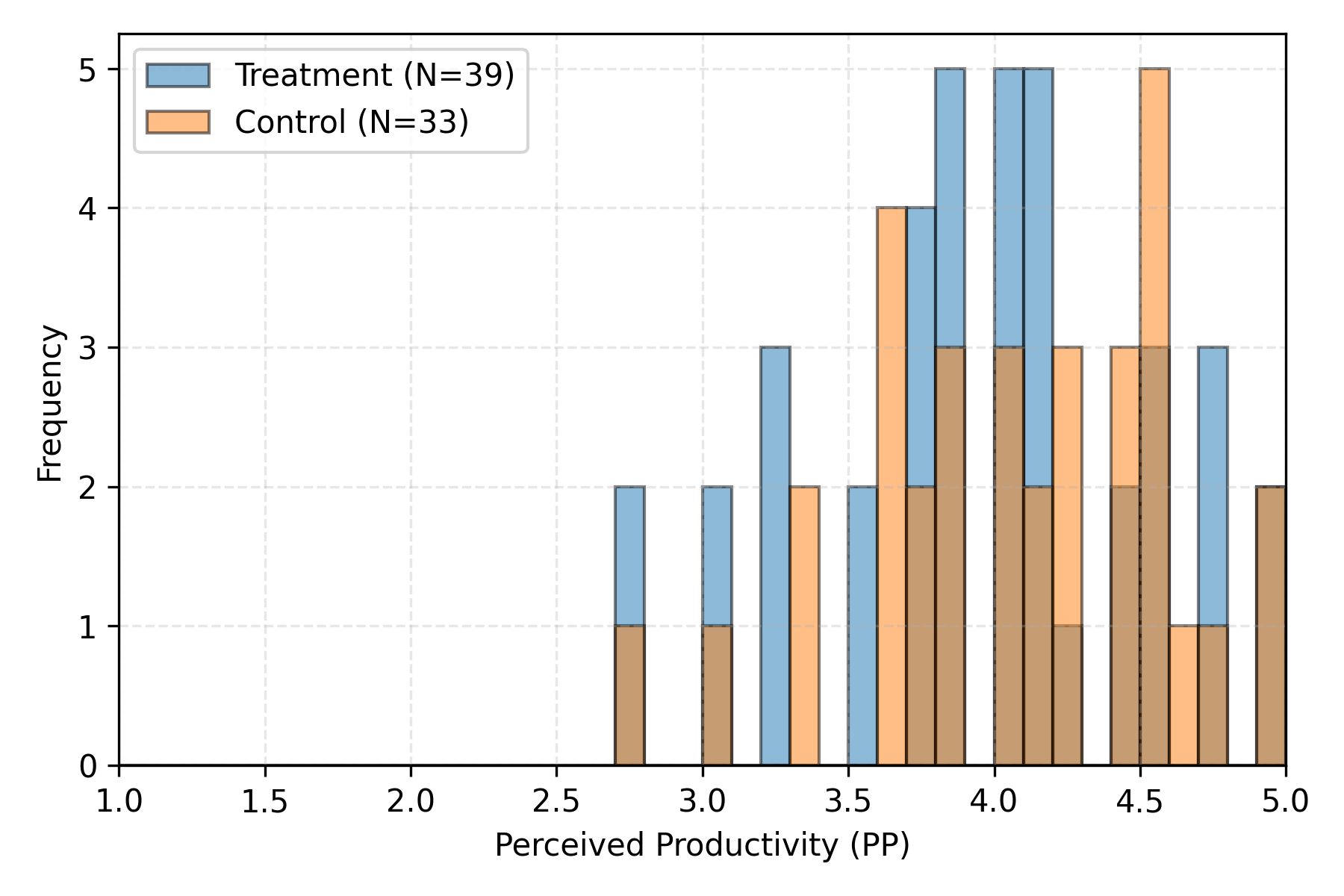}
        \caption{Distribution of Perceived Productivity (PP).}
        \label{fig:task2-pp-dist}
    \end{subfigure}
    \hfill
    \begin{subfigure}[b]{0.34\textwidth}
        \centering
        \includegraphics[width=\linewidth]{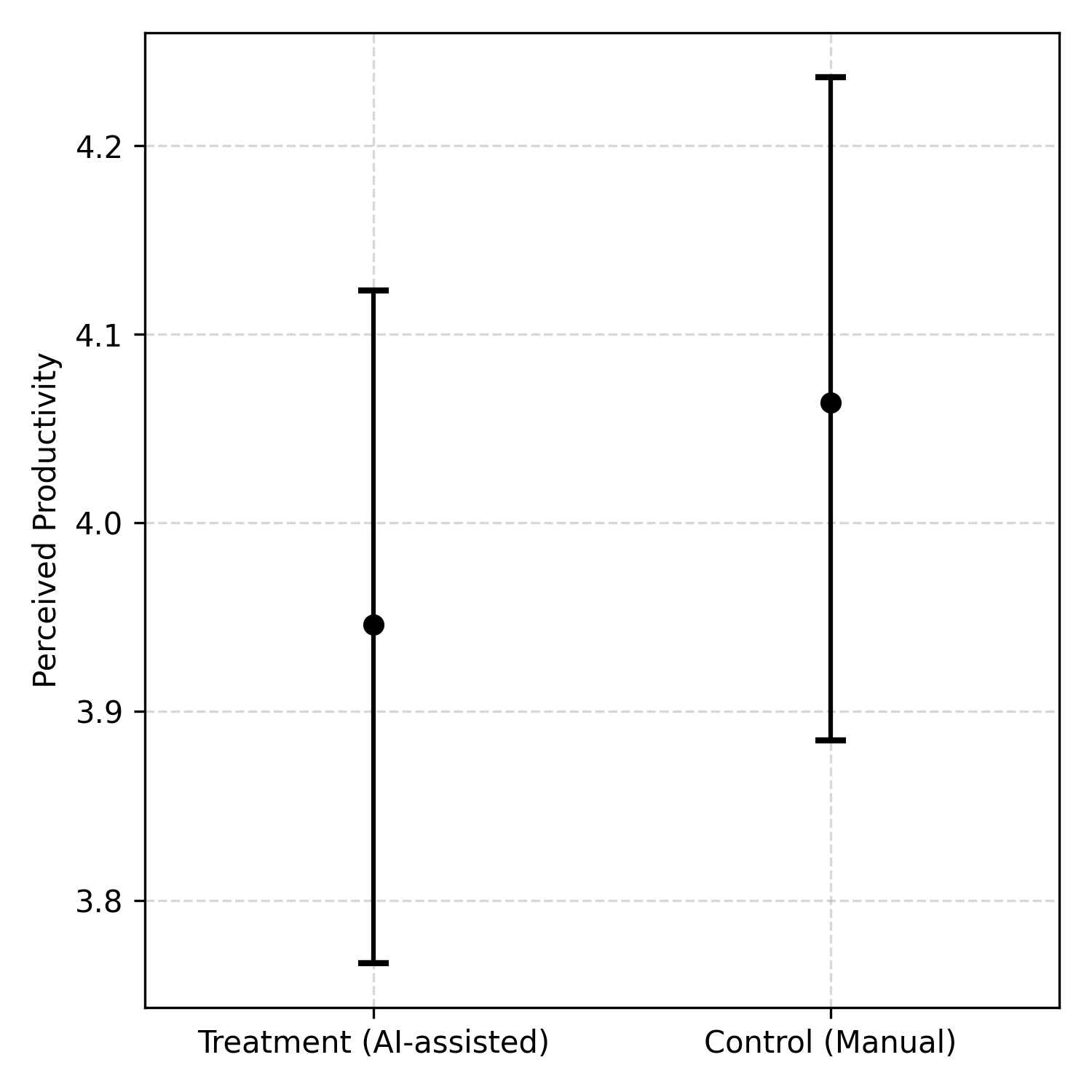}
        \caption{95\% confidence intervals.}
        \label{fig:task2-pp-ci}
    \end{subfigure}
    \caption{Task~2 Perceived Productivity.}
    \label{fig:task2-pp}
\end{figure}

Figure~\ref{fig:task2-pp-dist} shows the distribution of PP for 72 Task~2 solutions. As described in Section~\ref{sec:freq}, we consider the data approximately normally distributed. We continue by describing the seven participants who stand out in Figure~\ref{fig:task2-pp-dist}, namely four with the highest PP scores and three with the lowest. All of them agreed or strongly agreed that the task resembled previous development work (Q2-5). Three of them reported low PP (2.7):

\begin{itemize}
    \item[\textit{anon123}] This participant, discussed in Section~\ref{sec:rq1-time} as the one who spent the most time, evolved \textbf{AI-dev} and made substantial changes, i.e., 14 files in 16 commits. The low PP is explained in Q2-8: 
    \interviewquote{It’s been years since I’ve touched any Java code, even longer since using Spring [Boot]. The broadness of the task also left it open to which optimizations should be done, but I feel much more satisfied and know the system is easier to work with after working on it}{\textit{anon123}, Task 2, Treatment, Pro+Advanced, R5, 645 LoC, (13h, CH=8.89, TC=71\%, PP=2.7)}
    \item[\textit{anon014}] A beginner Java developer who spent 12 hours on the task, evolving \textbf{!AI-dev} code. Despite the long time, only four files were changed across three commits. The low PP is attributed to the participant's inexperience with Java web apps:
    \interviewquote{I also spent about 6 additional hours on non-essential environment setup tasks (i.e. Containerizing the app so I didn't have to run a JDK locally). In case this is relevant, I am not primarily a Java developer, and this was my first time working with a Java web application.}{\textit{anon014}, Task 2, Control, Pro+Beginner, R4, 33 LoC, (12h, CH=8.33, TC=70\%, PP=2.7)}
    \item[\textit{anon119}] An advanced Java developer who spent 134 minutes on the task and evolved \textbf{AI-dev} code. Surprisingly, the participant made many changes in a short time, i.e., 7 files in 20 commits, but still felt unproductive.
\end{itemize}

\noindent Four participants, all advanced Java developers, reported a PP of either 5 or 4.9. 

\begin{itemize}
    \item[\textit{anon100}] Evolved \textbf{AI-dev} code. Worked for 4.5 hours, changing 6 files in 21 commits. 
     \interviewquote{I adhered to the Pomodoro technique. For around every 24 minutes, I'd take a 6-minute break. I'd estimate I spent 3.5 hours on task typing.}{\textit{anon100}, Task 2, Treatment, Pro+Advanced, R5, 290 LoC, (4.5h, CH=8.61, TC=69\%, PP=5.0)}
    \item[\textit{anon109}] Evolved \textbf{AI-dev} code. Solved the task in a single commit with minimal churn. Reported that the task was a bit simple and that they finished it in under 30 minutes even though not knowing the code. 
    \item[\textit{anon146}] Evolved \textbf{!AI-dev} and finished the task in 88 minutes. Changed 4 files in a single commit.
    \item[\textit{anon144}] Evolved \textbf{!AI-dev} and finished the task in 3 hours. Changed 12 files in 21 commits.
    \interviewquote{In total I think I spent about 2.5 hours on the actual task [...] I had some initial set up issues the previous day and couldn’t get the project to build correctly. I probably spent 45 mins in total messing mainly with IntelliJ config.}{\textit{anon144}, Task 2, Treatment, Pro+Advanced, R5, 290 LoC, (4.5h, CH=8.61, TC=69\%, PP=5.0)}
\end{itemize}

\paragraph{Frequentist Analysis}
Figure~\ref{fig:task2-pp-ci} shows 95\% confidence intervals for the mean PP. The treatment group had a mean of 3.95 (95\% CI: 3.77--4.12), while the control group had a mean of 4.06 (95\% CI: 3.88--4.24). Cohen's $d$ was -0.21, indicating a \textit{small negative effect size}. The difference was \textit{not statistically significant} (Welch's t-test, $p$=0.37). Given our limited statistical power (see Section~\ref{sec:freq}), our study cannot reliably detect differences corresponding to such small effect sizes.

\paragraph{Bayesian Analysis}

\begin{figure}
    \centering
    \includegraphics[width=0.8\linewidth]{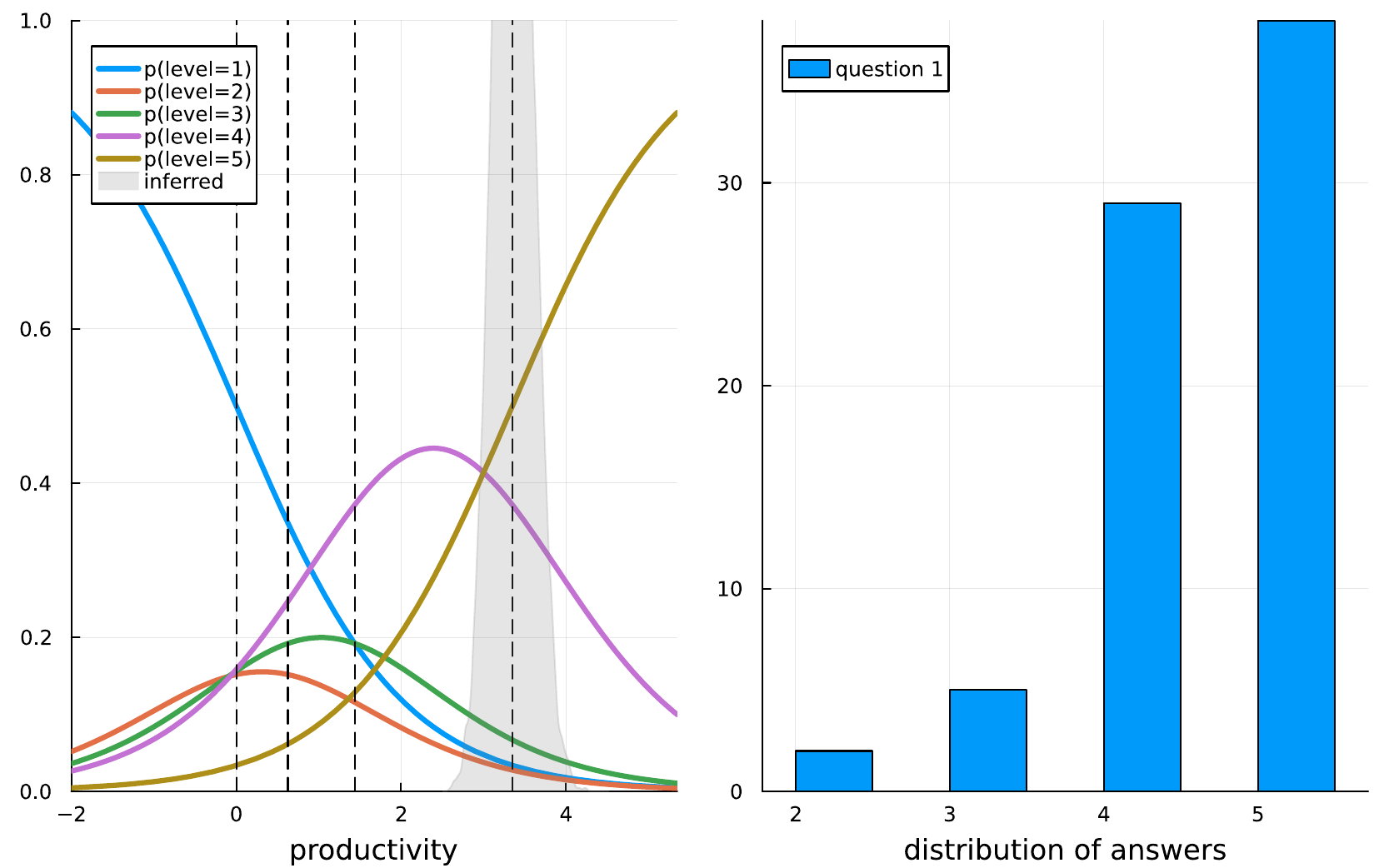}
    \caption{Relationship between inferred productivity and responses to question 1.
    The left-hand plot shows the probability of each response level being observed based on the productivity on the x-axis. 
    The shaded area shows an inferred productivity approximately between 3-4 points.
    The dashed lines indicate the inferred cutoffs between response levels.
    Looking at the curves, we can see that for a productivity in that range, the most likely answers for question 1 are level 5 (highest curve) or 4, and perhaps a few 3.
    This is consistent with the observed responses, showed on the right-hand plot.}
    \label{fig:bayes-productivity-explain}
\end{figure}

As explained in section \ref{sec:bayes}, our ordered logistic model infers a latent productivity score on a continuous scale for each developer. Figure~\ref{fig:bayes-productivity-explain} illustrates the relationship between inferred productivity and responses for question 1 (Q2-7a in Table~\ref{tab:exit-survey}), other questions follow the same model (taking the two inverted questions into account). To make our estimates a bit more intuitive, we report them on a scale of standard deviations of latent productivity.

For Task~1, using an AI assistant has a clear positive effect on latent productivity for habitual AI users, as $P(\Delta>0) > 99\%$, with a mean posterior effect of +1.71 standard deviations in latent productivity (CrI: $[+1.44, +1.97]$).
In other words, AI-assisted users felt more productive when working with their preferred tools. However, this positive effect on habitual AI users does not carry over to the manual evolution in Task~2 ($P(\Delta<0) = 84\%$, with a wide posterior distribution. In Bayesian terms, this pattern suggests that negative effects are more likely than positive ones, but the evidence is too weak for firm conclusions. The posterior mean effect is -1.26 standard deviations (95\% CrI: $[-2.46,+2.31]$) when manually evolving solutions by AI-assisted habitual AI users. Our sensitivity analysis shows similar results with optimistic, pessimistic, and neutral priors.

When looking at developer skill, we observed that developers with higher skill clearly reported feeling less productive in Task 1 ($P(\Delta<0) = 98\%$). The mean effect was $-1.24$ standard deviations (CrI: $[-1.58, -0.87]$). As one might expect, how highly-skilled $Dev1$s perceived their own productivity did not have a significant effect on $Dev2$s' perceptions ($P(\Delta<0) = 66\%$). The mean effect was $-0.11$ standard deviations (CrI: $[-2.82, +2.58]$).

\begin{figure}
    \centering
    \includegraphics[width=0.6\linewidth]{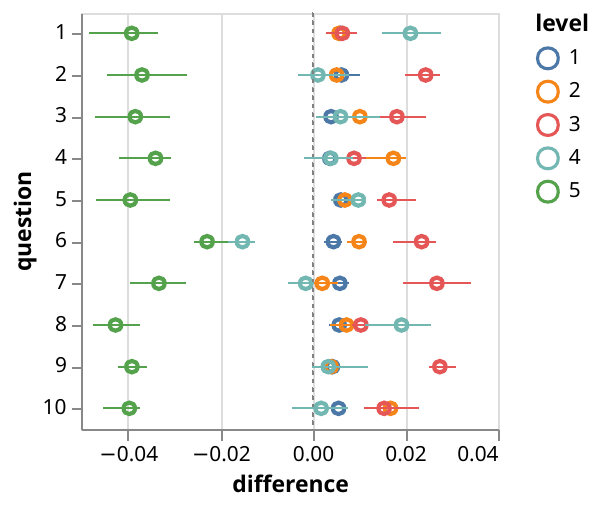}
    \caption{Difference in predicted responses for a habitual AI user with an advanced level of Java, comparing evolution of AI-assisted vs. non-assisted solutions. Each row represents a survey question and the x-axis shows changes in response probabilities. Developers evolving AI-assisted code are about 4 percentage points less likely to answer 5 (“fully agree”) and slightly more likely to answer 3 (“neutral”).}
    \label{fig:bayes-productivity-outcome}
\end{figure}

To clarify how this influences the responses to productivity questions, we compare two simulations for habitual AI users with high Java proficiency and compare outcomes for an \textbf{AI-dev} and \textbf{!AI-dev}.
Figure \ref{fig:bayes-productivity-outcome} presents a comparison of predicted responses in both cases.
When evolving an AI-assisted solution, the model predicts that developers are 3 percentage points less likely to answer 5 (``fully agree'') to each of the questions. Instead, the developer is slightly more likely to answer 3 (``neutral'') or 4 (``agree''). For question~1, the probability of answering 5 shifts from 51\% to 47\%. For question 9, the probability of answering 3 shifts from 28\% to 31\%. The effects are small.



\begin{tcolorbox}[
  colback=lightblue,
  colframe=blue!75!black,
  coltitle=white,
  title=Quantitative Results – Task 2 Perceived Productivity (Trt-Ctrl),
  fonttitle=\bfseries,
  enhanced,
  breakable,
  boxrule=0.5pt,
  leftrule=1pt,
  rightrule=1pt,
  toprule=1pt,
  bottomrule=1pt
]
\begin{itemize}
  \item \textbf{Frequentist.} Mean difference $=-0.10$; 95\% CI $[-0.40,\,+0.10]$; \\Welch's $t$-test $p=0.37$; Cohen’s $d=-0.21$.
  \item \textbf{Bayesian.} For habitual AI users in Task~1: $P(\Delta<0)=84\%$; \\Posterior mean change $=-1.26$ SD; 95\% CrI: $[-2.46,\,+2.31]$.
  \item Both analyses are consistent with a null effect on PP in Task~2.
\end{itemize}
\end{tcolorbox}

\subsection{RQ2: Higher Quality Upon Evolution?}
This section reports results related to the two dependent variables CodeHealth and Test Coverage (see Figure~\ref{fig:gqm}). We follow the same structure as for RQ1, including the use of textbox summaries described in Section~\ref{sec:res-rq1}.

\subsubsection{CodeHealth}
\begin{figure}[ht]
    \centering
    \begin{subfigure}[b]{0.58\textwidth}
        \centering
        \includegraphics[width=\linewidth]{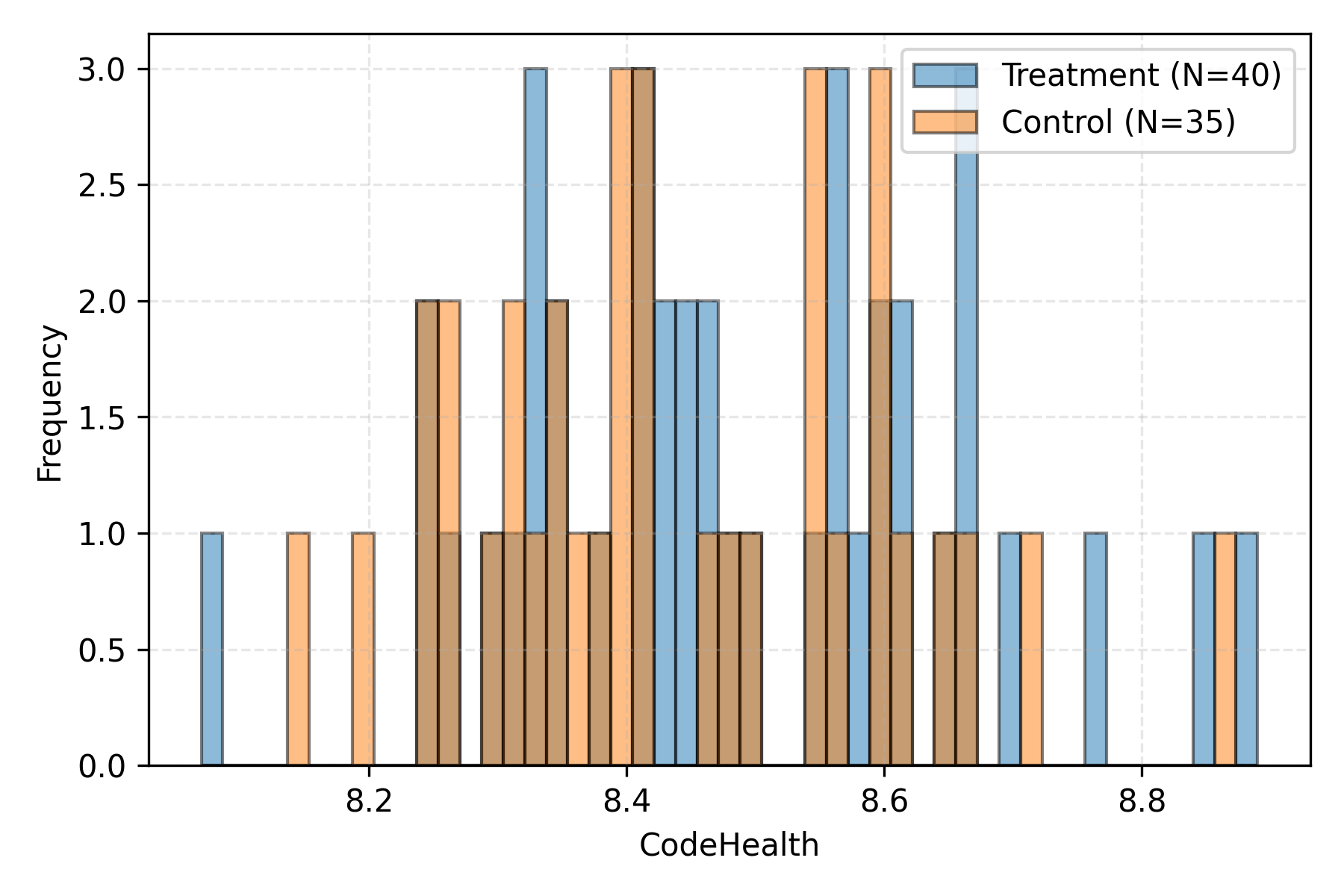}
        \caption{Distribution of CodeHealth (CH).}
        \label{fig:task2-ch-dist}
    \end{subfigure}
    \hfill
    \begin{subfigure}[b]{0.34\textwidth}
        \centering
        \includegraphics[width=\linewidth]{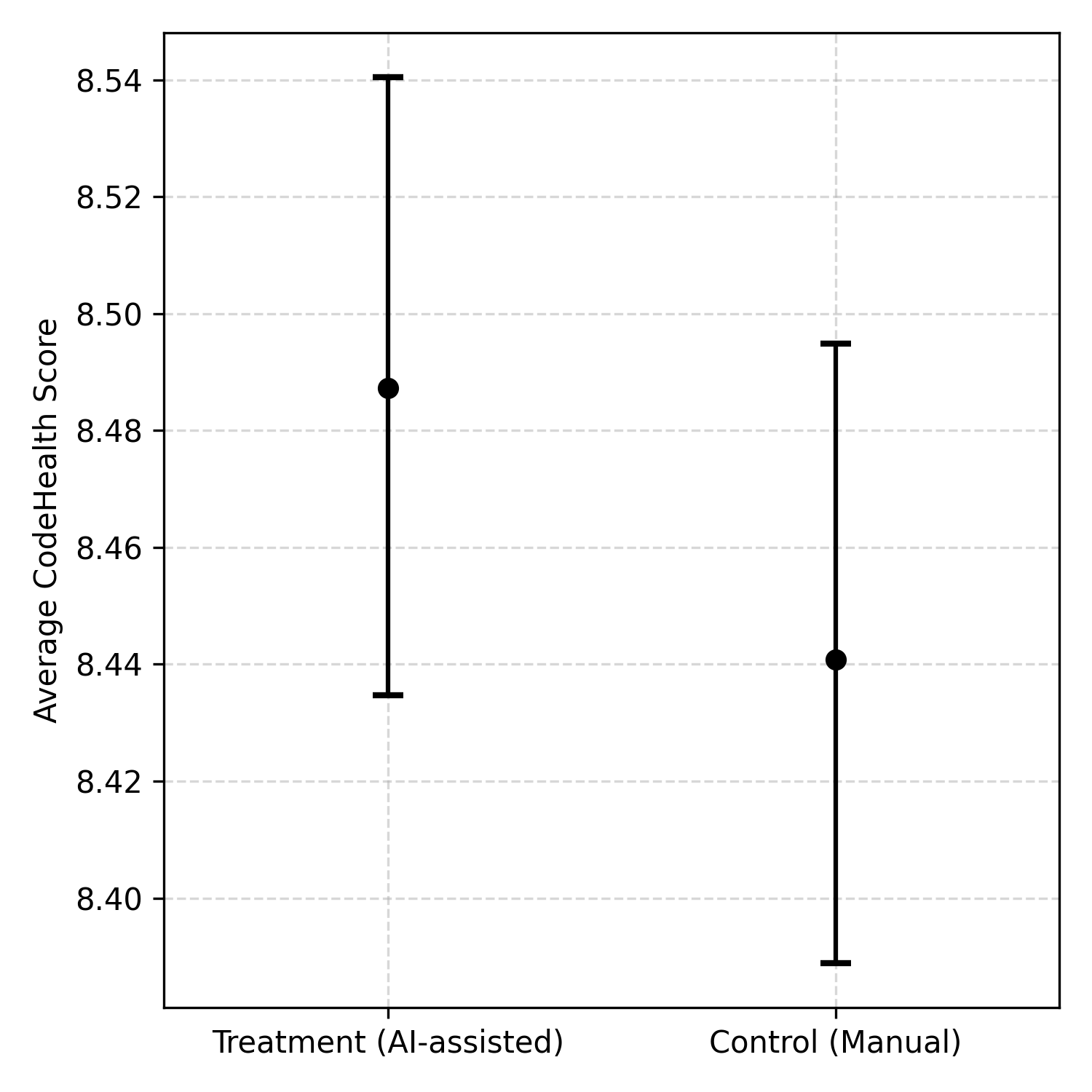}
        \caption{95\% confidence intervals.}
        \label{fig:task2-ch-ci}
    \end{subfigure}
    \caption{Task~2 CodeHealth.}
    \label{fig:task2-ch}
\end{figure}

Figure~\ref{fig:task2-ch-dist} shows the distribution of CH for 75 Task~2 solutions. As described in Section~\ref{sec:freq}, we consider the data approximately normally distributed. The lowest observed CH (8.07) was submitted by a participant who evolved AI-assisted code. At the other end of the spectrum, three participants submitted solutions with CH scores greater than 8.8. 

Certain patterns emerge in how CH evolves for participants in the control and treatment groups, as shown in Table~\ref{tab:task2-delta-combined}. The treatment group built on codebases developed by \textbf{AI-devs}, while the control group worked on code from \textbf{!AI-devs}. The CH characteristics of those Task~1 solutions are detailed in Section~\ref{sec:code-comparison} (see Tables~\ref{tab:task1-code-smells} and~\ref{tab:smell-delta-combined}).

We observe that the treatment group degraded the \textit{Complex Method} code smell significantly more often (55\%, 22 solutions) compared to the control group (31\%, 11 solutions). This suggests that \textbf{AI-devs} tended to include methods with multiple logical paths, and that evolving these solutions for Task 2's requirements often led to the addition of yet another logical path. Manual inspection confirms that this additional logical path is generally related to the handling of the new \texttt{cost} parameter in the \texttt{search} method (see the Task~2 description in Appendix~\ref{app:tasks}). At the same time, the treatment group also removes more \textit{Complex Method} code smells (20\%, 8 solutions) than the control group (9\%, 3 solutions), indicating that more opportunities for simplification existed in the \textbf{AI-devs} code.

The \textit{Complex Conditional} code smell was significantly introduced more frequently in the treatment group (18\%, 7 solutions) compared to the control group (3\%, 1 solution). This pattern typically emerged when participants extended existing null-checks (for \texttt{query} from the base system and \texttt{time} from Task~1) with a third condition for the Task~2 \texttt{cost} parameter, resulting in a long compound conditional expression. Alternative solutions used cleaner branching strategies, separating the \texttt{cost} and \texttt{cost} query logic.

Finally, the \textit{Primitive Obsession} code smell appeared significantly more frequently in the control group (26\%, 9 solutions) than in the treatment group (8\%, 3 solutions). Manual inspection suggests that \textbf{!AI-devs} tended to extract smaller methods, leading to more parameter passing -- typically simple types such as \textit{int}, \textit{long}, and \textit{String} for the search time, cost, and query arguments. In contrast, \textbf{AI-devs} more often used a functional style with fewer method boundaries, reducing the number of primitives exposed between methods.

\begin{table}[ht]
\centering
\caption{Overview of Task~2 changes in code smells grouped by delta types and code smells, showing \textit{mean / std / median [min-max]}.}
\label{tab:task2-delta-combined}
\begin{tabular}{ll|cc}
\toprule
\textbf{Delta Type} & \textbf{Code Smell} & \textbf{Treatment (n=40)} & \textbf{Control (n=35)} \\
\midrule

\multirow{5}{*}{Removed}
  & Bumpy Road Ahead & 0.10 / 0.30 / 0 [0-1] & 0.06 / 0.24 / 0 [0-1] \\
  & Code Duplication & 0.05 / 0.32 / 0 [0-2] & 0.03 / 0.17 / 0 [0-1] \\
  & Complex Conditional & 0.03 / 0.16 / 0 [0-1] & 0.03 / 0.17 / 0 [0-1] \\
  & Complex Method & 0.23 / 0.48 / 0 [0-2] & 0.09 / 0.28 / 0 [0-1] \\
  & Primitive Obsession & 0.03 / 0.16 / 0 [0-1] & 0.00 / 0.00 / 0 [0-0] \\
\midrule

\multirow{1}{*}{Improved}
  & Complex Method & 0.07 / 0.27 / 0 [0-1] & 0.09 / 0.28 / 0 [0-1] \\
\midrule

\multirow{5}{*}{Unchanged}
  & Bumpy Road Ahead & 0.05 / 0.22 / 0 [0-1] & 0.09 / 0.28 / 0 [0-1] \\
  & Code Duplication & 0.10 / 0.30 / 0 [0-1] & 0.17 / 0.45 / 0 [0-2] \\
  & Complex Method & 0.03 / 0.16 / 0 [0-1] & 0.09 / 0.28 / 0 [0-1] \\
  & Primitive Obsession & 0.05 / 0.22 / 0 [0-1] & 0.06 / 0.24 / 0 [0-1] \\
\midrule

\multirow{3}{*}{Degraded}
  & Bumpy Road Ahead & 0.05 / 0.22 / 0 [0-1] & 0.06 / 0.24 / 0 [0-1] \\
  & Complex Conditional & 0.03 / 0.16 / 0 [0-1] & 0.00 / 0.00 / 0 [0-0] \\
  & Complex Method & 0.60 / 0.59 / 1 [0-2] & 0.31 / 0.47 / 0 [0-1] \\
\midrule

\multirow{4}{*}{Introduced}
  & Bumpy Road Ahead & 0.07 / 0.27 / 0 [0-1] & 0.00 / 0.00 / 0 [0-0] \\
  & Code Duplication & 0.10 / 0.30 / 0 [0-1] & 0.11 / 0.32 / 0 [0-1] \\
  & Complex Conditional & 0.17 / 0.38 / 0 [0-1] & 0.03 / 0.17 / 0 [0-1] \\
  & Complex Method & 0.15 / 0.36 / 0 [0-1] & 0.37 / 0.77 / 0 [0-3] \\
  & Primitive Obsession & 0.10 / 0.38 / 0 [0-2] & 0.26 / 0.44 / 0 [0-1] \\

\bottomrule
\end{tabular}
\end{table}

\paragraph{Frequentist Analysis}
Figure~\ref{fig:task2-ch-ci} shows 95\% confidence intervals for the mean CH. The treatment group had a mean of 8.49 (95\% CI: 8.43--8.54), while the control group had a mean of 8.44 (95\% CI: 8.39--8.49). Cohen's $d$ was 0.28, indicating a \textit{small positive effect size}. The difference was \textit{not statistically significant} (Welch's t-test, $p$=0.24). As reported for PP (see Section~\ref{sec:res-pp}), our limited statistical power means we cannot reliably detect differences corresponding to so small effect sizes.

\paragraph{Bayesian Analysis}

For Task 1, among habitual AI users, the posterior mean difference in CH between \textbf{AI-dev} and \textbf{!AI-dev} solutions was $+0.13$ points ($P(\Delta>0)=98\%$, CrI: $[+0.02, +0.24]$). This effect did carry over to Task 2 submissions, since submissions which were initially co-written by AI by a habitual AI user had a slightly higher CH score ($P(\Delta>0) = 98\%)$, with a mean effect of $+0.1$ (CrI: $[+0.02, +0.19]$). These estimates are small but stable under the alternative priors considered in our sensitivity analysis.

As a point of comparison, when comparing $Dev1$s with high Java proficiency versus beginners ($Dev1\_skill$ of 1 vs. 3) in Task 1, we observed that CH did not increase significantly ($P(\Delta>0)=69\%$). The mean effect was $0.03$ points (CrI: $[-0.1, +0.13]$). An increased skill for $Dev1$ did seem to have a small positive effect on CH in Task 2 ($P(\Delta>0)>99\%$). The mean effect was $+0.16$ points (CrI: $[+0.07,+0.26]$).

\begin{figure}
    \centering
\includegraphics[width=0.75\linewidth]{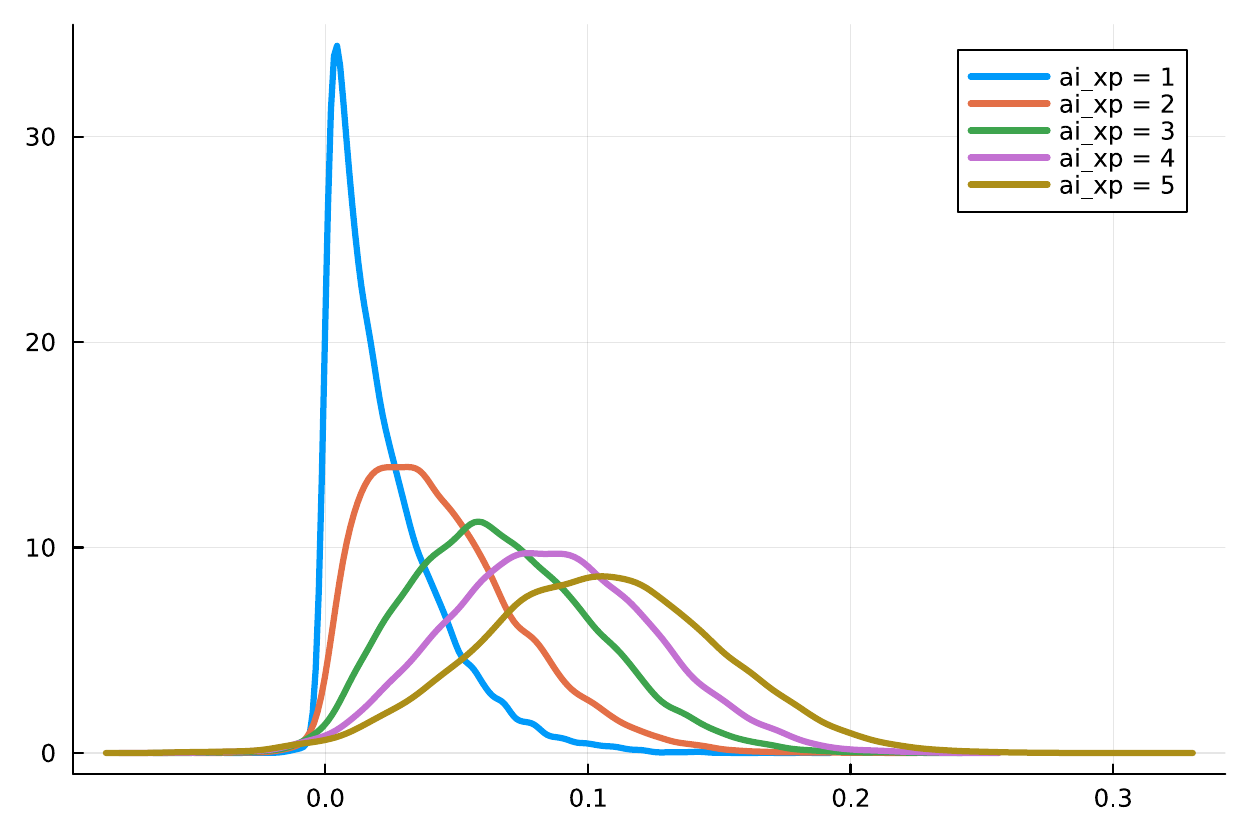}
    \caption{Posterior effect of AI use in Task~1 on CH after manual evolution in Task~2. The effect is negligible when Task~1 developers had minimal AI experience but increases with more habitual use.}
    \label{fig:bayes-codehealth}
\end{figure}

Figure~\ref{fig:bayes-codehealth} shows the posterior distributions for the effects of using AI assistants in Task~1 on the CH of the manually evolved Task~2 solution. The five curves correspond to different levels of AI habituality ($AI\_xp$) reported in the prescreening questionnaire. Each estimate is computed while holding Java proficiency constant, allowing a fair comparison across AI experience levels.


\begin{tcolorbox}[
  colback=lightblue,
  colframe=blue!75!black,
  coltitle=white,
  title=Quantitative Results – Task 2 CodeHealth (Trt-Ctrl),
  fonttitle=\bfseries,
  enhanced,
  breakable,
  boxrule=0.5pt,
  leftrule=1pt,
  rightrule=1pt,
  toprule=1pt,
  bottomrule=1pt
]
\begin{itemize}
  \item \textbf{Frequentist.} Mean difference $=+0.05$; 95\% CI $[-0.03,+0.12]$; \\Welch's $t$-test $p=\,$0.24; Cohen’s $d=\,$0.28.
  \item \textbf{Bayesian.} When Task~1 was completed by a habitual AI user: \\
  $P(\Delta>0)=0.98$; Posterior mean effect $=+0.10$; 95\% CrI $[+0.02,+0.19]$.
  \item Bayesian analysis suggests a small CH advantage for the treatment group.
\end{itemize}
\end{tcolorbox}

\subsubsection{Test Coverage}

Figure~\ref{fig:task2-ch-dist} shows the distribution of TC for 75 Task~2 solutions. Most submissions achieved around 70\% test coverage. As noted in Section~\ref{sec:freq}, the data is not normally distributed. The highest observed TC was 0.90, submitted by Participant \textit{anon050}, who evolved a Task~1 solution with TC=0.91 -- the second highest score in Task~1. Note that we received no Task~2 solution building on the Task 1 solution with the highest TC (0.98). At the lower end, three participants submitted solutions with TC=0.59:

\begin{itemize}
    \item[\textit{anon042}] Decreased the TC of a \textbf{!AI-dev} solution by 7 percentage points. They worked on the task for 9 hours, modifying 27 files across 10 commits (703 added lines, 749 deleted lines). Although new test cases were added, the participant did not maintain TC. The average CH, however, increased by 0.09.
    \item[\textit{anon145}] Decreased the TC of an \textbf{AI-dev} solution by 11 percentage points. They worked for 7 hours, editing 29 files in 74 commits (753 added lines, 562 deleted lines). The average CH increased by a substantial 0.60. The participant actively modified test code: \textit{``I was refactoring test code as well, for me that is also production code, refactoring only the main service would be significantly faster.''}
    \item[\textit{anon080}] Increased the TC of a \textbf{!AI-dev} solution by 3 percentage points and the average CH by 0.21. They worked for 2.5 hours, and changing 21 files in 12 commits (399 added lines, 180 deleted lines).     
\end{itemize}

\noindent We note that \textit{anon042} and \textit{anon145} were among the participants who decreased TC the most. Both were also among the participants who added the most new LoC, which might explain the phenomenon. 

\begin{figure}[ht]
    \centering
    \begin{subfigure}[b]{0.58\textwidth}
        \centering
        \includegraphics[width=\linewidth]{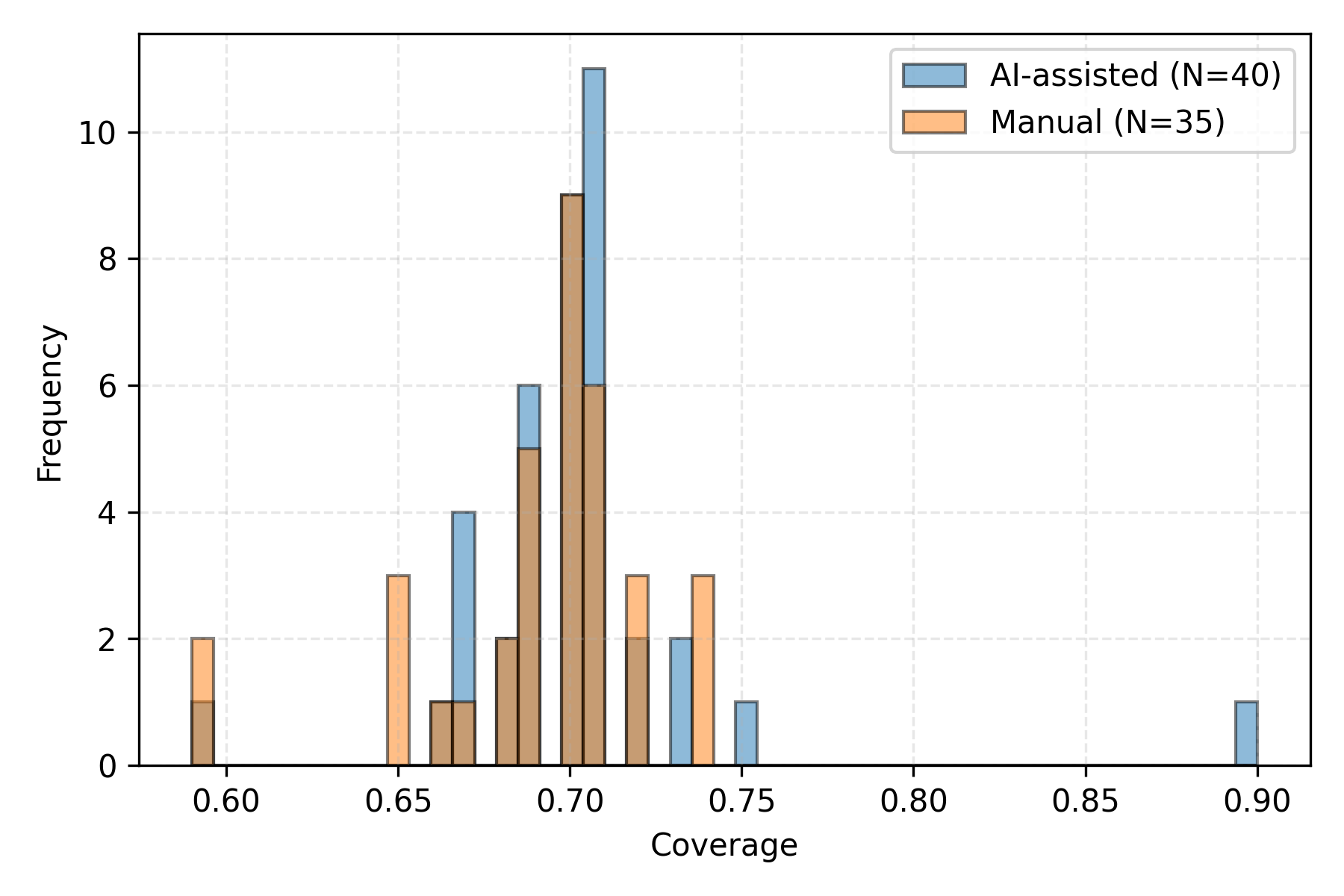}
        \caption{Distribution of Test Coverage (TC).}
        \label{fig:task2-tc-dist}
    \end{subfigure}
    \hfill
    \begin{subfigure}[b]{0.34\textwidth}
        \centering
        \includegraphics[width=\linewidth]{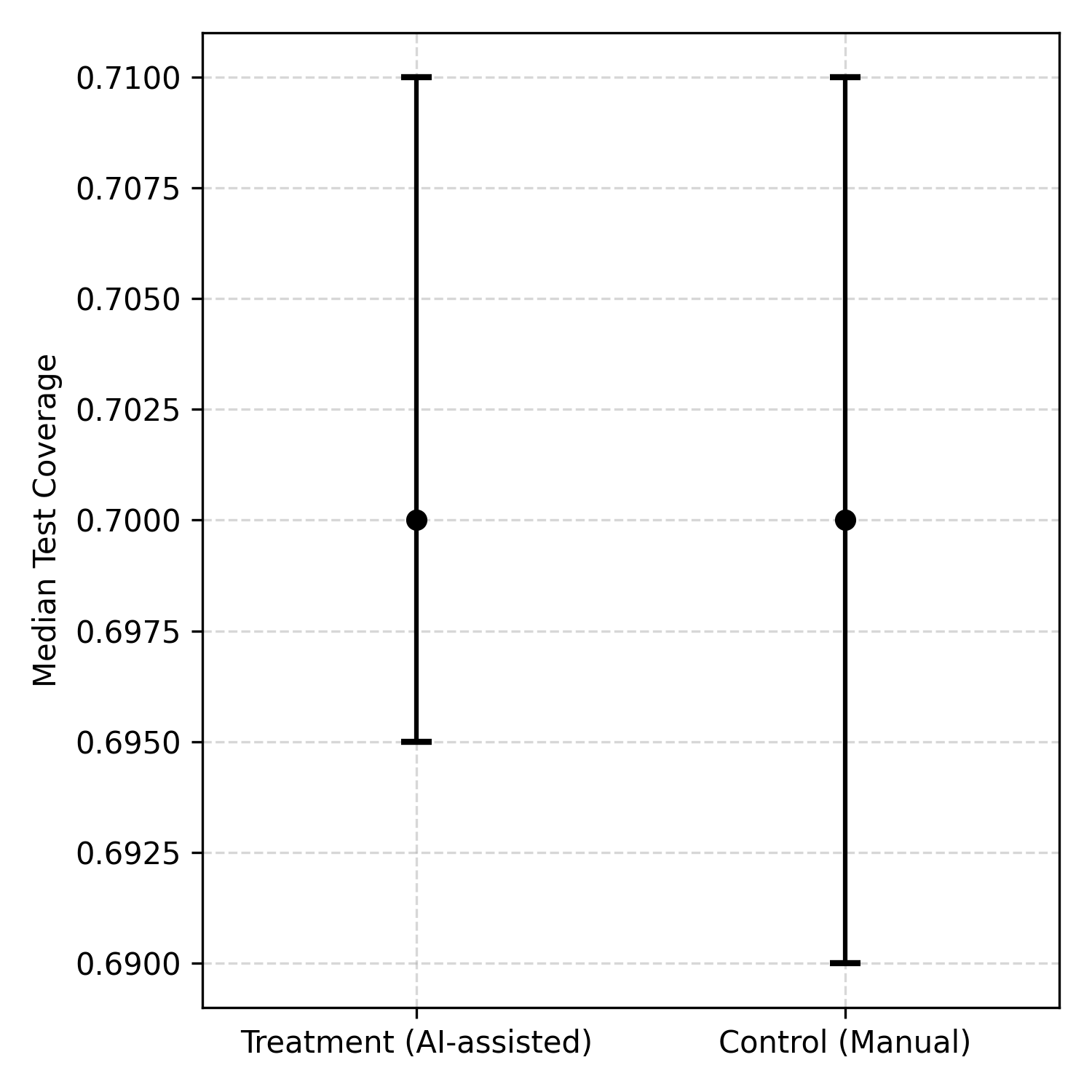}
        \caption{95\% confidence intervals.}
        \label{fig:task2-tc-ci}
    \end{subfigure}
    \caption{Task~2 Test Coverage.}
    \label{fig:task2-tc}
\end{figure}

\paragraph{Frequentist Analysis}
Figure~\ref{fig:task2-tc-ci} shows 95\% confidence intervals for the median TC. The treatment group had a median of 0.70 (95\% CI: 0.695--0.71) which was matched by the control group with a median of 0.70 (95\% CI: 0.69--0.74). Cliff's $\delta$ was 0.09, indicating a \textit{negligible effect size}. The difference was \textit{not statistically significant} (Wilcoxon rank-sum test, $p$=0.49).

\paragraph{Bayesian Analysis}
For Task 1, using AI did not have a significant effect on TC ($P(\Delta>0)=92\%$). The posterior mean effect on the logit scale is $+0.13$ (CrI: $[-0.05, +0.32]$). Because we use a logit transform, these coefficients are hard to interpret directly \citep{gelman_regression_2021}. Instead, we may look at predictions from the model: for a habitual AI user, the model predicts that the mean TC increases by 1 percentage point (from 69\% to 70\%) when using an AI assistant. The CrIs for predicted TC are $[66\%, 71\%]$ for \textbf{!AI-devs}, and $[68\%, 72\%]$ for \textbf{AI-devs}. Note that these estimates account for uncertainty in parameters only and do not consider residual noise.

When looking at Task~2, we also see that prior AI use has no significant effect on TC ($P(\Delta > 0) = 91\%$). The posterior mean effect is $+0.10$ (CrI: $[-0.04, + 0.26]$). If we look at predictions from the model for $Dev1$s with high Java proficiency who are habitual AI users, the model predicts that the TC will go from 68\% (CrI: $[67\%, 70\%]$) to 69\% (CrI: $[68\%, 71\%]$), when using an AI assistant. We see the same results when using optimistic, neutral, and pessimistic priors. 

As a point of comparison, higher Java programming skill ($Dev1\_skill$) has a significant negative effect on TC in Task~1 ($P(\Delta<0) > 99\%$). $Dev1$s with higher skill have on average 10 percentage points less TC (CrI: $[-12pp, -5pp]$).
In Task~2, this effect is slightly attenuated, while still significant ($P(\Delta<0) = 98\%$), as the mean effect was about $-4$ percentage points (CrI: $[-7pp, 0pp]$).

\begin{tcolorbox}[
  colback=lightblue,
  colframe=blue!75!black,
  coltitle=white,
  title=Quantitative Results – Task 2 Test Coverage (Trt - Ctrl),
  fonttitle=\bfseries,
  enhanced,
  breakable,
  boxrule=0.5pt,
  leftrule=1pt,
  rightrule=1pt,
  toprule=1pt,
  bottomrule=1pt
]
\begin{itemize}
  \item \textbf{Frequentist.} Median difference $=0$; 95\% CI $[-0.01,0.01]$; \\Wilcoxon $p=0.49$; Cliff’s $\delta=0.09$.
  \item \textbf{Bayesian.} For habitual AI users in Task~1, with high Java proficiency: $P(\Delta>0)=91\%$; Model-predicted mean change $=+1$ pp.
  \item Both analyses are consistent with a null effect on TC in Task~2.
\end{itemize}
\end{tcolorbox}

\subsection{Observational Findings on AI Assistant Use in Task 1}
While Task~1 is not the main target of this study, analyzing potential differences between \textbf{AI-dev} and \textbf{!AI-dev} solutions is important for interpreting the results. Furthermore, our findings can be discussed in light of the related work introduced in Section~\ref{sec:rw}. That said, we remind the reader that the analyses in this section were not preregistered and Phase~1 was not randomized with respect to AI use, i.e., the findings are observational and do not support causal claims.

\subsubsection{Task 1: Experimental Variables}
\begin{figure}[ht]
    \centering
    \begin{subfigure}[b]{0.48\textwidth}
        \centering
        \includegraphics[width=\linewidth]{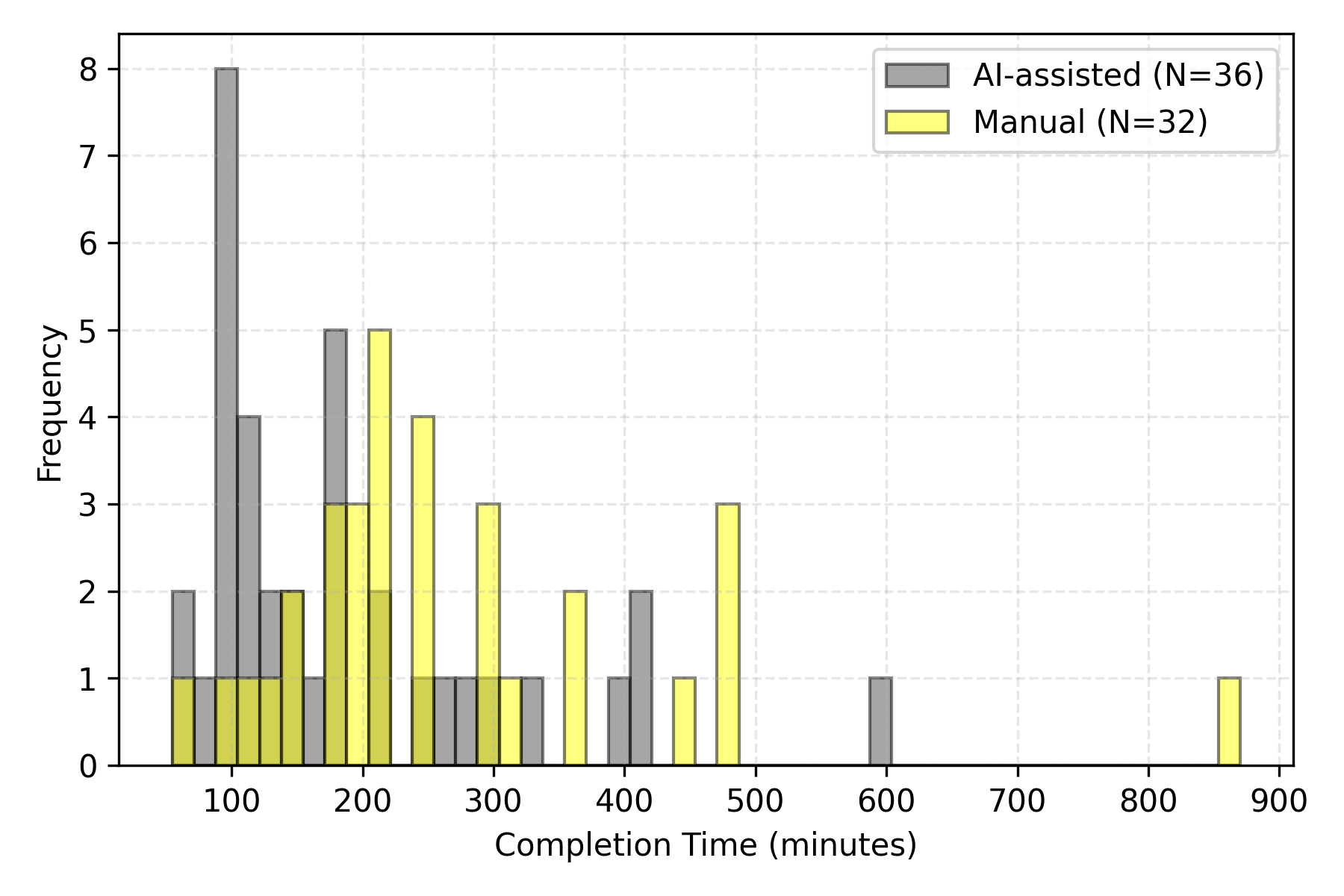}
        \caption{Completion time}
        \label{fig:task1-time}
    \end{subfigure}
    \hfill
    \begin{subfigure}[b]{0.48\textwidth}
        \centering
        \includegraphics[width=\linewidth]{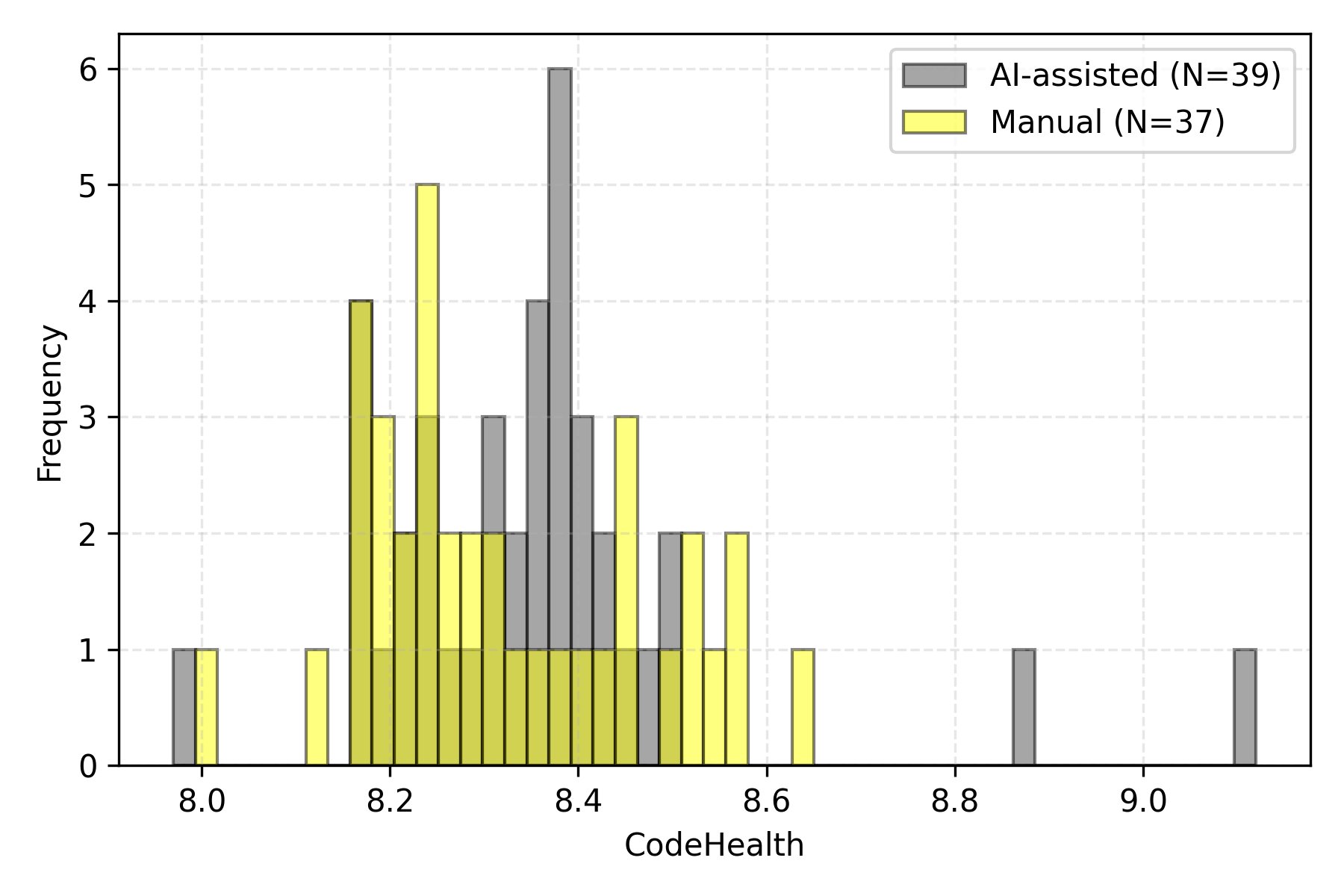}
        \caption{CodeHealth (CH)}
        \label{fig:task1-ch}
    \end{subfigure}
    \begin{subfigure}[b]{0.48\textwidth}
        \centering
        \includegraphics[width=\linewidth]{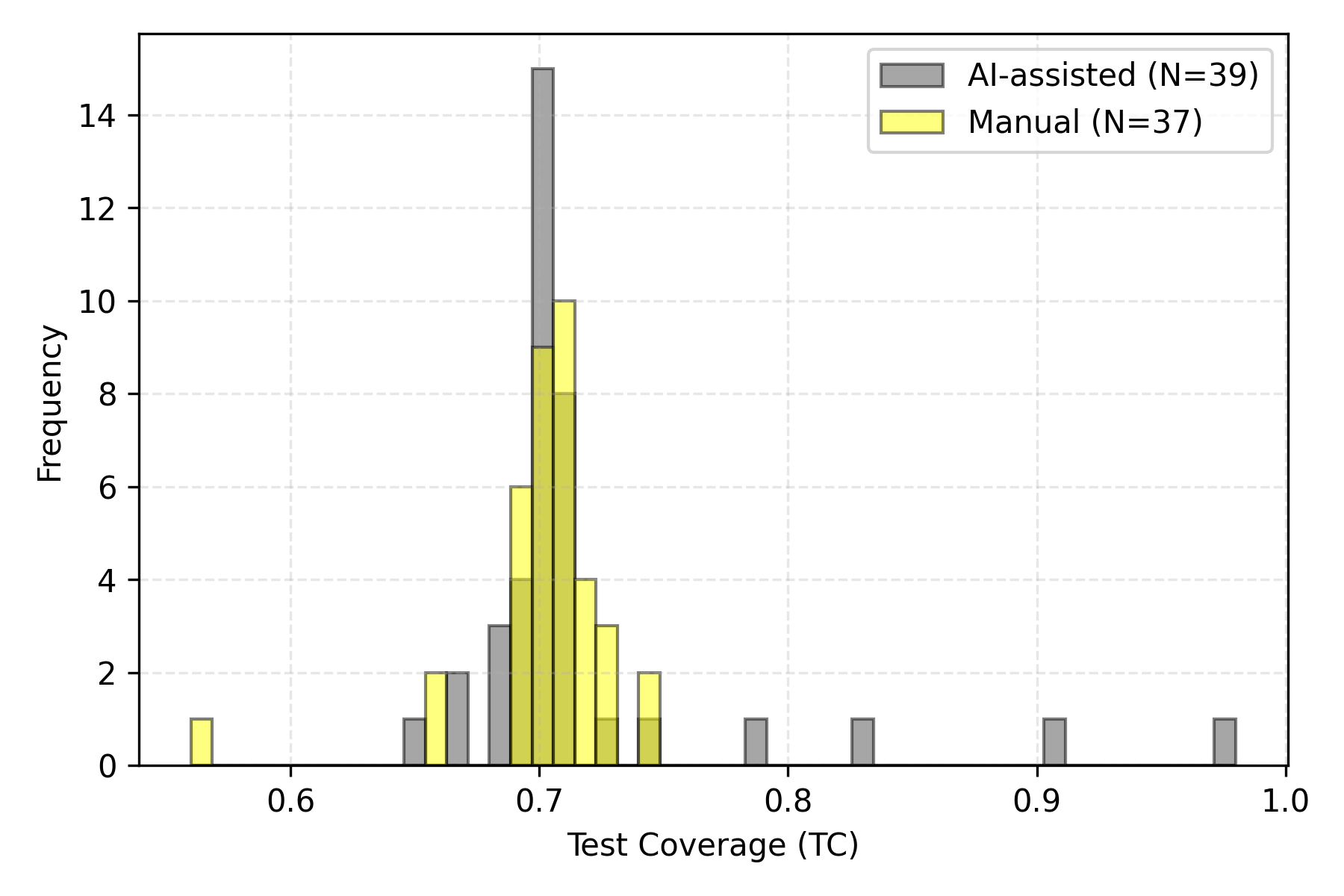}
        \caption{Test coverage (TC)}
        \label{fig:task1-tc}
    \end{subfigure}
    \hfill
    \begin{subfigure}[b]{0.48\textwidth}
        \centering
        \includegraphics[width=\linewidth]{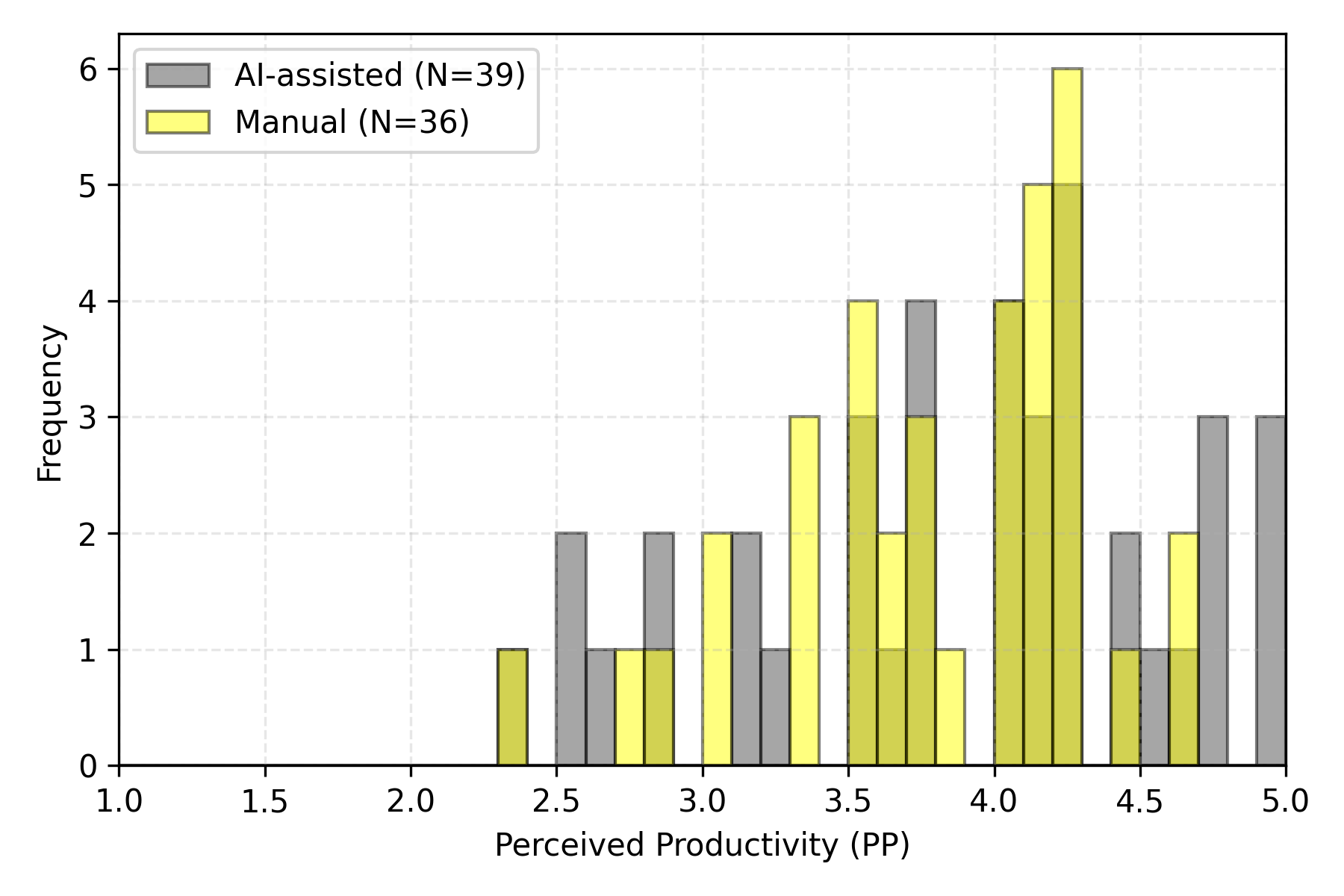}
        \caption{Perceived productivity (PP)}
        \label{fig:task1-pp}
    \end{subfigure}
    \caption{Descriptive statistics of Task 1 solutions.}
    \label{fig:task1-vars}
\end{figure}

Figure~\ref{fig:task1-vars} shows the distribution of the four variables: completion time, CH, TC, and PP. Completion times (see Figure~\ref{fig:task1-time}) are not normally distributed and the Wilcoxon Rank-Sum Test shows that there are significant differences ($p=0.0029$). There is a medium effect size of using an AI-assistant (Cliff's $\delta$=-0.42) and we observed that \textbf{AI-devs} had a 30.7\% shorter median completion time. The \textbf{!AI-dev} anomaly with a very long completion time, \textit{anon081}, explained: \textit{``Not fluent with Java, so a lot of reading up of the basics was required.''} The Bayesian analysis in Section~\ref{sec:rq1-time} will later confirm the positive effect of using AI assistants on Task~1 completion time.

CH for \textbf{!AI-devs} is not normally distributed, but we cannot reject this hypothesis for \textbf{AI-devs}. Wilcoxon Rank-Sum Test shows that there are no significant differences between the groups ($p=0.52$) and the effect size is negligible (Cliff's $\delta$=0.09). However, we notice that the spread of CH is lower for \textbf{AI-devs}, which explains why the normality assumption holds. The two \textbf{AI-devs} (\textit{anon054} and \textit{anon126}) who reached the highest CH (9.12 and 8.88, see Figure~\ref{fig:task1-ch}) both heavily used capable AI-assistants for more than three hours (ChatGPT+Cline+Cursor and Cursor+Tailwind AI, respectively), iterated a lot (54 and 287 test runs, respectively), and added substantial amounts of code (3,094 and 10,788 added LoC, respectively). We conclude that the AI-generated code integrated by these two participants was of high quality and raised the average CH substantially -- note that the subpar 2,500 LoC of the base system is modest in comparison.

TC consistently clustered around 70\% in Task 1. We note that the four highest TC scores were all obtained by \textbf{AI-devs}, while the lowest outlier corresponds to an \textbf{!AI-dev} (see Figure~\ref{fig:task1-tc}). The highest two TC scores (98\% and 91\%, respectively) belong to \textit{anon054} and \textit{anon126} just mentioned -- both users of AI assistants that go beyond code completion.

PP appears similar across the \textbf{AI-dev} and \textbf{!AI-dev} groups. However, Figure~\ref{fig:task1-pp} shows that the six participants who reported the highest scores all worked with AI assistants. While it is reasonable that such tools increase PP -- that is aligned with their purpose -- we acknowledge the potential for confirmation bias. \textbf{AI-devs} may have been more inclined to evaluate their productivity positively due to their belief in the value of AI assistance.

\subsubsection{Task 1: Detailed Code-Level Differences} \label{sec:code-comparison}
This section describes results from our detailed analysis of how \textbf{AI-devs} and \textbf{!AI-devs} changed the code in Task~1. We remind the reader that a handful of AI users generated large volumes of code, thus skewing mean values.

\paragraph{Regular Expressions} New \textit{unit tests} were added by 46\% of \textbf{AI-devs}, compared to 31\% of \textbf{!AI-devs}. Thus, there was no clear tendency for participants to use AI assistants to supplement the existing test suite. 

\textbf{AI-devs} added \textit{comments} more frequently than \textbf{!AI-devs} -- 79\% versus 54\%. AI-generated code is known for generating noisy surface-level comments. Such comments can aid the LLMs in subsequent code evolution, but offer limited value to human readers. In fact, noisy comments risk bloating the codebase and introducing discrepancies between documentation and logic if they are not kept aligned. 

Finally, we observe that both \textbf{AI-devs} and \textbf{!AI-devs} tend to shift toward the \textit{functional programming} paradigm. The proportion of solutions that introduce such constructs is 87\% and 76\%, respectively. Manual inspection shows that \textbf{AI-dev} submissions more frequently evolve existing usages of the Java Stream API, while \textbf{!AI-dev} submissions almost exclusively introduce new Stream APIs.

\paragraph{CodeScene Code Smells}
Table~\ref{tab:task1-code-smells} unpacks the CH of the base system and Task~1 solutions (see also Figure~\ref{fig:task1-ch} for the CH distributions). The code smells are described in Appendix~\ref{app:codescene}. The ``Base'' column depicts that the base RecipeFinder contains: 1) one Bumpy Road, 2) eleven Complex Methods, and 3) two Code Duplication smells. As shown in the last row, the average CH is similar across the three columns. Furthermore, Table~\ref{tab:smell-delta-combined} presents details about how participants worked with the base system's existing code smells. Code smells can be removed, improved (modified to be less severe), unchanged (moved from one location to another), degraded (more severe), and introduced (if a new instance appears) -- we refer to these five options as delta types. We observed differences between \textbf{AI-devs} and \textbf{!AI-devs}, which we discuss next. 

The \texttt{search} method in Figure~\ref{lst:code} shows the instance of the Bumpy Road -- and it also triggers one of the Complex Method smells. Since Task~1 participants had to modify this method, it serves as a natural refactoring target, which we investigate in detail. We found that 72\% of \textbf{AI-devs} resolved the Bumpy Road smell, compared to 46\% of \textbf{!AI-devs}. Interestingly, 14\% of \textbf{!AI-devs} even introduced another instance of this, whereas this never happened for \textbf{AI-devs}. 

These results suggest that AI assistants effectively resolved this instance of the Bumpy Road smell while \textbf{!AI-devs} appear to struggle with that specific structural cleanup. Furthermore, 38\% of the \textbf{!AI-devs} left the original code smell unchanged in the file. Manual inspection showed that this occurred when the problematic code block was merely extracted into a new method rather than being properly refactored to eliminate the underlying issue. Regarding effective refacorings, the most common resolution pattern was to rewrite the search method using a functional programming style -- which was slightly more common among \textbf{AI-devs}, as discussed under the regular expressions.

On the contrary, AI assistants appear to less effectively resolve Complex Method code smells. 18\% of the \textbf{AI-devs} (7 solutions) removed at least one Complex Method code smell compared to 32\% of the \textbf{!AI-devs} (12 solutions). On the other hand, 54\% of the \textbf{AI-devs} (21 solutions) improved smells in the Complex Method category, while 35\% of the \textbf{!AI-devs} (13 solutions) did. This indicates that AI-assisted developers are likely to improve existing Complex Methods, but less able to fully resolve them.


\begin{table}[ht]
\centering
\caption{Unpacking Task~1 CodeHealth into code smells, showing \textit{mean / std / median [min–max]}.}
\begin{tabular}{l|c|cc}
\toprule
\textbf{Code Smell} & \textbf{Base} & \textbf{AI-dev (n=39)} & \textbf{!AI-dev (n=37)} \\
\midrule
Bumpy Road & 1 & 0.28 / 0.46 / 0 [0–1] & 0.68 / 0.47 / 1 [0–1] \\
Code Duplication & 2 & 3.79 / 5.62 / 2 [2–36] & 3.08 / 3.94 / 2 [0–25] \\
Complex Conditionals & 0 & 0.15 / 0.43 / 0 [0–2] & 0.08 / 0.36 / 0 [0–2] \\
Complex Method & 11 & 11.56 / 1.94 / 11 [6–20] & 11.41 / 1.17 / 11 [6–14] \\
Constructor Over-Injection & 0 & 0.03 / 0.16 / 0 [0–1] & 0.00 / 0.00 / 0 [0–0] \\
Duplicated Assertion Blocks & 0 & 0.03 / 0.16 / 0 [0–1] & 0.00 / 0.00 / 0 [0–0] \\
Excess Function Args & 0 & 0.13 / 0.66 / 0 [0–4] & 0.05 / 0.33 / 0 [0–2] \\
Large Assertion Blocks & 0 & 0.23 / 1.44 / 0 [0–9] & 0.00 / 0.00 / 0 [0–0] \\
Large Method & 0 & 0.03 / 0.16 / 0 [0–1] & 0.00 / 0.00 / 0 [0–0] \\
Nested Complexity & 0 & 0.08 / 0.27 / 0 [0–1] & 0.00 / 0.00 / 0 [0–0] \\
Primitive Obsession & 0 & 0.13 / 0.41 / 0 [0–2] & 0.05 / 0.23 / 0 [0–1] \\
String-Heavy Args & 0 & 0.08 / 0.27 / 0 [0–1] & 0.14 / 0.42 / 0 [0–2] \\
\addlinespace
\textbf{Total code smells} & \textbf{14} & \textbf{16.51} & \textbf{15.49} \\
\textbf{Avg. CodeHealth} & \textbf{8.3} & \textbf{8.3} & \textbf{8.4} \\
\bottomrule
\end{tabular}
\label{tab:task1-code-smells}
\end{table}


\begin{table}[ht]
\centering
\caption{Overview of Task~1 changes in code smells grouped by delta types and code smells, showing \textit{mean / std / median [min–max]}.}
\label{tab:smell-delta-combined}
\begin{tabular}{ll|cc}
\toprule
\textbf{Delta Type} & \textbf{Code Smell} & \textbf{AI-dev (n=39)} & \textbf{!AI-dev (n=37)} \\
\midrule

\multirow{3}{*}{Removed}
  & Bumpy Road              & 0.72 / 0.46 / 1 [0-1] & 0.46 / 0.51 / 0 [0-1] \\
  & Code Duplication        & 0.00 / 0.00 / 0 [0-0] & 0.05 / 0.23 / 0 [0-1] \\
  & Complex Method          & 0.28 / 0.86 / 0 [0-5] & 0.46 / 1.04 / 0 [0-6] \\

\midrule
Improved
  & Complex Method          & 0.54 / 0.51 / 1 [0-1] & 0.41 / 0.60 / 0 [0-2] \\

\midrule
\multirow{3}{*}{Unchanged}
  & Bumpy Road              & 0.18 / 0.39 / 0 [0-1] & 0.38 / 0.49 / 0 [0-1] \\
  & Code Duplication        & 0.18 / 0.39 / 0 [0-1] & 0.27 / 0.45 / 0 [0-1] \\
  & Complex Method          & 0.03 / 0.16 / 0 [0-1] & 0.05 / 0.23 / 0 [0-1] \\

\midrule
Degraded
  & Complex Method          & 0.31 / 0.52 / 0 [0-2] & 0.43 / 0.50 / 0 [0-1] \\

\midrule
\multirow{5}{*}{Introduced}
  & Bumpy Road              & 0.00 / 0.00 / 0 [0-0] & 0.14 / 0.35 / 0 [0-1] \\
  & Code Duplication        & 0.41 / 1.07 / 0 [0-6] & 0.14 / 0.42 / 0 [0-2] \\
  & Complex Conditionals    & 0.15 / 0.43 / 0 [0-2] & 0.08 / 0.36 / 0 [0-2] \\
  & Complex Method          & 0.72 / 1.52 / 0 [0-7] & 0.62 / 0.83 / 0 [0-4] \\
  & Nested Complexity       & 0.08 / 0.27 / 0 [0-1] & 0.00 / 0.00 / 0 [0-0] \\
  & Primitive Obsession     & 0.13 / 0.41 / 0 [0-2] & 0.05 / 0.23 / 0 [0-1] \\

\bottomrule
\end{tabular}
\end{table}

\textbf{AI-devs} show a higher tendency to introduce certain new code smells not originally present in the base RecipeFinder project. Notably, at least one \textit{Complex Conditionals} code smell was introduced by 13\% of \textbf{AI-devs} (5 solutions) compared to 5\% of \textbf{!AI-devs} (2 solutions). Additionally, 8\% of \textbf{AI-devs} (3 solutions) also introduced the \textit{Nested Complexity} code smell while none of \textbf{!AI-devs} did. Furthermore, new \textit{Code Duplication} code smells are introduced by 23\% of \textbf{AI-devs} (9 solutions) while 11\% of \textbf{!AI-devs} (4 solutions). Lastly, 10\% of \textbf{AI-devs} (4 solutions) introduced the \textit{Primitive Obsession} when 5\% of \textbf{!AI-devs} (2 solutions) did. Although these differences are minor--showing notable differences in mean values but high variance and identical medians--they hint that AI assistants may be prone to introducing certain code smell patterns.

\citet{al_madi_how_2023} studied the readability of Copilot's generated code in a controlled experiment with students (n=21). Their results suggest that code written by a human pairing with an AI assistant is comparable in complexity and readability to code written by human pair programmers. Except for some outlier \textbf{AI-devs} who produced large volumes of code, our findings support their conclusion.

\paragraph{RefactoringMiner's Refactoring Operations}
Table~\ref{tab:task1-refactoringsminer} presents descriptive statistics of refactoring operations identified by RefactoringMiner in the \texttt{git diffs}. The data reveal notable differences between \textbf{AI-devs} and \textbf{!AI-devs}. First, we observe that \textbf{AI-Devs} are more inclined to apply simplification refactorings, such as \textit{Inline Variable}, \textit{Rename Variable}, and Spring Boot annotation improvements. In contrast, \textbf{!AI-Devs} more frequently perform structural refactorings, including \textit{Extract Class} and \textit{Move Method} (to another class). Moreover, \textit{Extract And Move Method}, another operation that requires deeper architectural considerations, was only recorded by \textbf{!AI-devs}. Despite these group-level patterns, we acknowledge considerable variation in how individuals refactor, no matter if they work with AI assistants or not.

\begin{table}[ht]
\centering
\caption{Subset of identified Task~1 refactoring operations, showing \textit{mean / std / median [min–max]}.}
\label{tab:task1-refactoringsminer}
\begin{tabular}{ll|cc}
\toprule
\textbf{Refactoring Operation} & & \textbf{AI-dev (n=39)} & \textbf{!AI-dev (n=37)} \\
\midrule

Add Method Annotation & & 0.36 / 1.35 / 0 [0–8] & 0.05 / 0.33 / 0 [0–2] \\ 
Add Param. Annotation  & & 0.15 / 0.67 / 0 [0–4] & 0.00 / 0.00 / 0 [0–0] \\
Change Attr. Access Modifier & & 0.26 / 0.79 / 0 [0–3] & 0.00 / 0.00 / 0 [0–0] \\
Extract And Move Method & & 0.00 / 0.00 / 0 [0–0] & 0.19 / 0.70 / 0 [0–3] \\
Extract Class & & 0.26 / 0.68 / 0 [0–3] & 0.43 / 0.83 / 0 [0–4] \\
Extract Method & & 1.56 / 3.65 / 0 [0–16] & 1.19 / 2.11 / 0 [0–8] \\
Inline Variable & & 0.64 / 1.74 / 0 [0–8] & 0.08 / 0.36 / 0 [0–2] \\
Modify Param Annotation & & 0.33 / 1.15 / 0 [0–6] & 0.00 / 0.00 / 0 [0–0] \\
Move Method & & 0.21 / 0.73 / 0 [0–4] & 1.89 / 4.21 / 0 [0–19] \\
Rename Variable & & 0.87 / 1.64 / 0 [0–7] & 0.30 / 0.57 / 0 [0–2] \\
Remove Method Modifier & & 0.00 / 0.00 / 0 [0–0] & 0.14 / 0.48 / 0 [0–2] \\

\addlinespace
\textbf{Total refactorings} & & 18.26 / 28.82 / 7 [1–126] & 15.43 / 19.23 / 5 [1–69] \\
\bottomrule
\end{tabular}
\end{table}

\paragraph{PMD Linting Results}
Table~\ref{tab:task1-pmd-rules} presents an overview of PMD violations (see Appendix~\ref{app:pmd}) organized per set rule, i.e., related rules grouped under a common theme. By far the most frequently violated rules for both \textbf{AI-devs} and \textbf{!AI-devs} belong to the \textit{Code Style} set, covering conventions such as naming, formatting, or method ordering. However, \textbf{AI-devs} display a much higher standard deviation (154 vs. 45) despite a lower median (258 vs. 267), confirming the impact of prolific outlier \textbf{AI-devs}. For all other rule sets, median violation counts barely differ between the groups.



Table~\ref{tab:task1-pmd-rules} further shows a sample of particularly interesting PMD rules. We find some interesting group-level patterns. First, \textbf{AI-devs} triggered significantly fewer violations related to algorithmic complexities, such as \textit{NPathComplexity} and \textit{CyclomaticComplexity}. Second, \textbf{AI-devs} also produced fewer violations related to documentation rules. Third, \textbf{AI-devs}' rule violations were more concentrated in surface-level issues, such as formatting and naming. A possible explanation is that the AI-generated code tends to be more verbose, including boilerplate code that may deviate from best-practice -- a pattern consistent with the \textit{Code Style} rule set in Table~\ref{tab:task1-pmd-rules}. 

Overall, it appears that contemporary use of AI assistants may reduce complexity but add stylistic clutter. This finding is in line with a study by \citet{siddiq_empirical_2022}, where they found that GitHub Copilot produced code with similar stylistic issues. We hypothesize that this can potentially increasing the cognitive burden of human code reviews and downstream maintainance tasks. This concludes our comparative analysis of \textbf{AI-dev} and \textbf{!AI-dev} solutions, and we are now ready to investigate Task~2's RCT results.

\begin{table}[ht]
\centering
\caption{Statistics for PMD set rules and a subset of individual rules for Task~1, showing \textit{mean / std / median [min–max]}.}
\label{tab:task1-pmd-rules}
\begin{tabular}{l|r|cc}
\toprule
\textbf{Rule} & \textbf{Base} & \textbf{AI-dev (n=39)} & \textbf{!AI-dev (n=37)} \\
\midrule
\textbf{Best Practices:} & 38  & 49.21 / 29.83 / 40 [32--207]     & 42.97 / 7.65 / 40 [31--62] \\
UnusedAssignment         & 1   & 0.21 / 0.41 / 0 [0--1]        & 0.41 / 0.50 / 0 [0--1] \\
\midrule
\textbf{Code Style:} & 253 & 294.97 / 153.68 / 258 [204--1037] & 279.30 / 44.54 / 267 [233--489] \\
AtLeastOneConstructor    & 11  & 12.62 / 6.52 / 11 [7--45]     & 12.08 / 2.03 / 11 [8--20] \\
ControlStatementBraces   & 23  & 19.67 / 3.28 / 18 [12--30]    & 21.73 / 3.49 / 23 [8--26] \\
LinguisticNaming         & 3   & 3.49 / 1.54 / 3 [3--9]        & 2.95 / 0.52 / 3 [0--4] \\
MethodArgCouldBeFinal    & 68  & 81.15 / 32.51 / 69 [53--232]  & 76.92 / 8.06 / 75 [69--99] \\
OnlyOneReturn            & 19  & 20.59 / 3.17 / 19 [15--33]    & 23.03 / 3.95 / 24 [12--30] \\
ShortVariable            & 25  & 25.79 / 4.36 / 25 [16--45]    & 26.03 / 2.75 / 25 [16--33] \\
UseExplicitTypes         & 0   & 0.08 / 0.35 / 0 [0--2]        & 3.35 / 12.75 / 0 [0--69] \\
\midrule
\textbf{Design:} & 27  & 26.21 / 5.84 / 25 [19--57]        & 26.51 / 3.43 / 27 [15--32] \\
AvoidCatchingGenExcep    & 0   & 0.33 / 1.46 / 0 [0--9]        & 0.00 / 0.00 / 0 [0--0] \\
CyclomaticComplexity     & 0   & 0.05 / 0.22 / 0 [0--1]        & 0.38 / 0.49 / 0 [0--1] \\
ImmutableField           & 4   & 3.05 / 1.45 / 4 [0--4]        & 3.54 / 1.50 / 4 [0--6] \\
NPathComplexity          & 1   & 0.31 / 0.47 / 0 [0--1]        & 0.70 / 0.46 / 1 [0--1] \\
\midrule
\textbf{Documentation:} & 114 & 122.38 / 23.56 / 115 [109--244]   & 124.19 / 12.68 / 118 [114--159] \\
CommentRequired          & 113 & 117.05 / 12.43 / 114 [102--181] & 122.68 / 12.12 / 117 [113--158] \\
\midrule
\textbf{Total Rule Violations}  & 450 & 512.18 / 222.17 / 456 [399–1632]  & 490.00 / 57.44 / 465 [438–705] \\
\bottomrule
\end{tabular}
\end{table}

\subsection{Analysis of Free-text Answers} \label{sec:res-freetext}
The qualitative insights provided in this section enrich our data, and we triangulate them with quantitative results in the reporting. Table~\ref{tab:coding} presents the coding scheme used for the qualitative analysis of responses to Q2-8 and, in some cases, follow-up emails. In total, we coded 177 pieces of free-text input, organized into three overarching groups.

\subsubsection{AI-assisted Development} \label{sec:freetext-ai}
Four Task~1 participants commented positively on using AI for refactoring, which lies at the core of the study. The two most positive reflections came from users of Cursor, both describing substantial improvements. For example:

\interviewquote{The AI did most of the work, really. I just had to approve its changes and ask it to solve error messages if encountered. It even provided some nice enhancement ideas when asked. I honestly felt more like a product manager than a programmer.}{\textit{anon148}, Task 1, AI-dev, Pro+Intermediate, Cursor, R4, 92 LoC, (3h, CH=8.31, TC=68\%, PP=4.6)}

\noindent The other two positive participants worked with GitHub Copilot and instead described assistance at the syntax level. This contrast reflects capability differences between the first and second generations of AI assistants. Cursor encourages more chat-based programming, while GitHub Copilot rather supports traditional code completion workflows. This is reflected below:

\interviewquote{Using Copilot made the task easy. I used it to find correct syntax and refactor code to better readability.}{\textit{anon103}, Task 1, AI-dev, Pro+Intermediate, GH Copilot, R4, 129 LoC, (1.5h, CH=8.42, TC=70\%, PP=4.8)}

Eight participants expressed how they found value in using AI assistants for learning and onboarding, demonstrating the relevance of this use case. Several seasoned developers had limited experience with Java and Spring Boot, and appreciated using AI assistance to get up to speed. Two of them even stated that they used AI solely during onboarding, to understand the stack and generate code examples, and then wrote all new code manually for Task~1. One explanation is that they found the onboarding challenge to be much greater than the actual coding task. A representative reflection is:

\interviewquote{I am an experienced software developer with other tech stacks, I have no professional experience with Java. This task would have been very difficult and time-consuming if not impossible for me without GitHub Copilot. It gave critical clues about where to start my debugging.}{\textit{anon110}, Task 1, AI-dev [HABITUAL], Hobbyist+Intermediate, GH Copilot, R3, 134 LoC, (5h, CH=8.49, TC=70\%, PP=4.4)}

However, regarding the initial debugging, we also observed contrasting comments. One participant reported that GitHub Copilot misguided them toward the wrong problem when doing initial debugging, which wasted time. A similar sentiment was shared by a student participant who highlighted that the assistant -- while good at generating new code -- was really bad at debugging:

\interviewquote{I tried using AI for all parts, I found that AI was good at new features, but was terrible at bug-fixing, it completely overlooked things and it took me some time to realize that the API call to retrieve the data doesn't overwrite the data, but appends it, so I would manually need to clear it after implementing the fix. AI didn't help with this at all.}{\textit{anon009}, Task1, AI-dev [HABITUAL], Student+Beginner, GH Copilot+ChatGPT, R3, 46 LoC, (3h, CH=8.22, TC=70\%, PP=2.5)}

Five participants mentioned how they experienced a productivity boost using AI, often in relation to onboarding into an unfamiliar codebase, as just discussed. Three of the participants, two of them using Cursor, explained how AI helped them with knowledge transfer, i.e., helping them apply their prior development experience to Java, which they had not worked with in years. For example:

\interviewquote{I have not used Java as a full-time professional language since 2011 [...] One thing where Copilot really accelerated my work was that once I understood the structure of the system I could make reference to constructs from languages and frameworks that I use on a daily basis and have it translate them into the Java and Spring [Boot] paradigm.}{\textit{anon133}, Task 1, AI-dev [HABITUAL], Pro+Intermediate, GH Copilot+Grok, R4, 90 LoC, (3h, CH=8.43, TC=70\%, PP=4.1)}

Obviously, not all experiences with AI assistants were positive. Ten participants mentioned limitations they found with the AI assistants during Task 1. Issues encountered include 1) \textit{``AI getting in the way,''} 2) ChatGPT being very sensitive to prompt formulation, i.e., finding the right prompt might be slower than using web search, and 3) AI generally focusing on syntax rather than program flow. As the last statement about syntax was mentioned by \textit{anon064}, who used GitHub Copilot, ChatGPT, JetBrains AI Assistant, and Gemini, it shows that this experience does not only hold for GitHub Copilot.

Regarding the trustworthiness of the output, two participants stressed the importance of double-checking what the AI assistants produced. For example:

\interviewquote{The interaction with the AI felt like pair programming with someone with a vast experience but lazy who sometimes gives hacky answers. I had to constantly remind the AI to use the most recent and best practice information to get better code.}{\textit{anon073}, Task 1, AI-dev, Pro+Advanced, Cursor, R4, 2635 LoC, (4.5h, CH=8.38, TC=67\%, PP=4.2)}

Finally, there was also an interesting reflection on AI-assisted development by \textit{anon063} who was assigned to the \textbf{!AI-dev} group: \textit{``Not relying on LLMs as I usually do these days made me reconsider their value. Honestly I'm not sure if they actually have that much value as I initially expected.'' } Still, we conclude that the general tone in the comments around AI assistants was moderately positive. We discuss this finding further in relation to prior empirical studies in Section~\ref{sec:disc}.

\begin{tcolorbox}[
  colback=lightblue,
  colframe=blue!75!black,
  coltitle=white,
  title=Key Takeaways – Task 1 Reflections on AI-assistance,
  fonttitle=\bfseries,
  enhanced,
  breakable,
  boxrule=0.5pt,
  leftrule=1pt,
  rightrule=1pt,
  toprule=1pt,
  bottomrule=1pt
]
\begin{itemize}
  \item Some users of second-generation AI assistants reported success with highly automated chat-based programming.
  \item Several participants described very positive experiences using AI assistants for onboarding.
  \item Others noted limitations, including poor debugging support and prompt sensitivity -- but the overall tone was moderately positive.
\end{itemize}
\end{tcolorbox}

\subsubsection{Development without AI}
This section presents reflections by participants who did not use any AI assistants. The quotes include both Task 1 and Task 2 concerns. 

Nine participants provided comments related to refactoring. Two expressed a sense of satisfaction or fulfilment from improving the code. This shows that developers in our study could have a positive experience regarding the tasks even without AI -- which is a reassuring observation, suggesting the tasks we designed were not just a boring chore.

Two participants raised insightful comments about the difference between refactoring in this study compared to a setting in a professional development team. \textit{anon098} explained: \textit{``I usually don't do deeper refactorings when starting with a new codebase, respecting the current conventions and asking more experienced devs what they think about my ideas before implementing them.’’} Another participant went a bit further down the refactoring path, but reflected critically in hindsight:

\interviewquote{I refactored the structure of part of the code, including tests, to make it easier for my brain, but I did not apply this change to the whole codebase. This will undoubtedly create problems for whoever comes next because part of the code is in the new format and part in the old. This is a typical legacy code problem though.}{\textit{anon027}, Task 2, Treatment, Pro+Advanced, R4, 606 LoC, (4h, CH=8.66, TC=72\%, PP=3.3)}

Seven participants mentioned how they approached the onboarding activity. As in the \textbf{AI-dev} group, several seasoned developers were either rather new to Java, or had not used it in a long time. Most comments focused on how they familiarized themselves with the tech stack, without using AI, as exemplified by the .Net and C++ developers below:

\interviewquote{I am not familiar [enough] with JPA and Java to effectively solve the coding challenge. In .NET I would have been able to finish the task in the suggested 2-4h. Most of my time was spent on discovering the framework for testing, Spring [Boot] annotations, Java standards and practice [...] I was reading one StackOverflow page after another.}{\textit{anon012}, Task 1, !AI-dev, Pro+Advanced, R5, 602 LoC, (3.5h, CH=8.42, TC=73\%, PP=3.7)}

\interviewquote{As a C++ developer for embedded systems, Java is not my first language. I had to do some functions earlier with it, and that is how far my knowledge [goes]. [...]  Figuring out the tools, environment and project structure took considerable time in the beginning. Not knowing language idioms and best practices slowed down the refactoring part and kept me from doing more.}{\textit{anon059}, Task 2, Control, Researcher+Beginner, R1, 161 LoC, (4h, CH=8.32, TC=66\%, PP=4.0)}

Five participants working without AI assistants were somewhat put off by the limited number of test cases available. Two explicitly expressed their preferences for test-driven development. Separately, \textit{anon116} complained about the mundane task of creating test data and felt slowed down. \textit{anon112} considered adding more tests, but found it would require refactoring the code for testability -- so they decided to finish the task without making such intrusive changes to the codebase. 

\begin{tcolorbox}[
  colback=lightblue,
  colframe=blue!75!black,
  coltitle=white,
  title=Key Takeaways – Task 1 and 2 Reflections on non-AI Development,
  fonttitle=\bfseries,
  enhanced,
  breakable,
  boxrule=0.5pt,
  leftrule=1pt,
  rightrule=1pt,
  toprule=1pt,
  bottomrule=1pt
]
\begin{itemize}
  \item Many participants had to ramp up on Java and Spring Boot using manual self-study, e.g., documentation and StackOverflow.
  \item Some participants avoided doing invasive refactoring either to respect existing conventions or due to the lack of test cases and testability.
\end{itemize}
\end{tcolorbox}

\subsubsection{Task Reflections} \label{sec:task-reflect}
Beyond tool usage, many participants shared reflections on the task itself -- feedback we actively elicited in Q2-8. This resulted in a total of 101 coded reflections, more than in the other two categories. 

The most recurring theme was the participants' unfamiliarity with the technology stack, i.e., modern Java, Spring Boot, and Jakarta Persistence (JPA). Twenty-eight participants mentioned this in some form. In the \textbf{AI-dev} group, it was common to mention how AI assistants helped them understand the new stack, as already discussed in Section~\ref{sec:freetext-ai}. Among the \textbf{!AI-devs}, it was instead common to mention which other stacks they typically used -- often pointing out that the task would have been faster to solve with their familiar tools. Again, this reinforces the value of controlling for Java proficiency in the causal analysis.

Almost as common, mentioned 25 times, were reflections on the effort required to set up the environment properly before starting the task. Notably, this was only mentioned once by \textbf{AI-devs}, suggesting that AI assistants may indeed help reduce setup friction. Among the \textbf{!AI-devs}, reported issues include cloning the repository to local environments, JDK versions, IDE settings, Spring Boot and Maven, setting up the test infrastructure, GitHub Single Sign-On (SSO) over HTTPS, git configurations, and database versions. The only \textbf{AI-dev} who mentioned setup friction proposed distributing the task as a containerized solution, which was rather a suggestion for improvement. We acknowledge that containerization could have saved several participants quite some time. Nevertheless, for Task 2, randomization should mitigate any systematic impact of setup time.

Twenty-one participants mentioned that it was unclear when a solution was ``good enough'' for submission. The vague instructions were intentional from our side, and the random assignment should again mitigate any systematic effects. Participants reported different strategies for managing quality, including: 1) spending hours on improving large parts of the codebase, 2) delivering only the bare minimum, 3) timeboxing a fixed amount of time for refactoring, 4) prioritizing improvements to the most critical parts, 5) focusing on product code rather than test code, and 6) refactoring only areas that were touched (in line with the ``Boy Scout rule''~\citep{martin_clean_2009}). One participant shared an insightful comment that also captured a broader sentiment of confusion, i.e., the need for discussions with the client before undertaking larger changes: 

\interviewquote{The biggest thing I was left feeling was an indecision on how far to take quality refactors/changes. If I were working on a real problem for a real client, I probably would've needed to have discussions about desirable performance characteristics, etc. that weren't possible here. That information could inform how far to take, and in what direction to take, additional changes.}{\textit{anon067}, Task 2, Control, Pro+Intermediate, R3, 404 LoC, (3h, CH=8.56, TC=65\%, PP=4.4)}

A related theme that recurred, often in combination with the above comments about quality expectations, was that the task was simple or even simplistic. Eighteen participants mentioned this in some way, but only four related to Task~1. It is evident that Task~2 was underwhelming to some participants:

\interviewquote{I was very confused about what I had to do during the entire experiment. Still unsure if I completed all the required tasks or not. The mentions of 3 hours and multiple sittings got me even more confused, since it probably took me less than 5 minutes for the actual code changes.}{\textit{anon136}, Task 2, Treatment, Pro+Intermediate, R4, 14 LoC, (1.5h, CH=8.42, TC=73\%, PP=3.0)}

Still, note that the median Task~2 completion time was 136 minutes (see Figure~\ref{fig:task2-time}) with a median of 56.5 added LoC for the Treatment group, and 173 minutes and 80 LoC for the Control group. The impression of Task~2 probably depended on how the preceding Task~1 was solved, as indicated by this quote mentioning a ``clear prior standard'':

\interviewquote{The task was extremely simple, only requiring less than 10 lines of code, and had a clear prior standard. I felt there was no room for creative thinking or really any problem solving at all.}{\textit{anon101}, Task 2, Control, Pro+Intermediate, R4, 19 LoC, (1h, CH=8.25, TC=71\%, PP=3.7)}

Finally, we identified nine mentions of uncertainties in the task instructions, apart from imprecise quality expectations. Three participants mentioned details in the \texttt{README.md} file: 1) a poorly described fact that the Backend-API was externally hosted and not part of the codebase, 2) the importance of using the Spring Boot framework's @Autowire annotation, and 3) unclear instructions regarding the acceptance tests. While it is possible that some participants dropped out because of bad instructions, we consider this threat minor, as the total number of comments related to instruction problems was low.

\begin{tcolorbox}[
  colback=lightblue,
  colframe=blue!75!black,
  coltitle=white,
  title=Key Takeaways – Reflections on Tasks 1 and 2,
  fonttitle=\bfseries,
  enhanced,
  breakable,
  boxrule=0.5pt,
  leftrule=1pt,
  rightrule=1pt,
  toprule=1pt,
  bottomrule=1pt
]
\begin{itemize}
  \item Many participants found the unfamiliar tech stack and setup requirements challenging, but AI assistance eased onboarding and reduced initial friction.
  \item Several participants were unsure how far to take refactoring and the vague quality expectations triggered a wide range of strategies.
  \item Although resembling previous development tasks, several participants found Task~2 simplistic -- especially when Task~1 had established a high baseline.

\end{itemize}
\end{tcolorbox}

\section{Discussion} \label{sec:disc}
Our study is unique in its two-stage design, in which the second phase constitutes an RCT. In this phase, developers manually evolve code written by an unknown predecessor -- who was either assisted by AI (treatment) or not (control). This section revisits the two RQs and discusses our findings in light of the closest related work -- for which we also share findings from Task~1. Finally, we discuss the implications for research and practice.

\subsection{RQ1: More Efficient Manual Evolution?}
Do developers manually evolve code that has been co-developed with AI assistants more efficiently? To answer this question, we analyzed two metrics collected in the RCT and enriched the picture through free-text answers. First, the completion time for the manual evolution task (Task~2) performed by a second developer. Second, the perceived productivity of that developer, guided by the SPACE framework~\citep{forsgren_space_2021}.  

We found substantial variability in Task~2 completion time, with no significant differences between the treatment and control groups. The median completion time was 136 minutes for the Treatment group and 173 minutes for the Control group, but the confidence intervals overlapped widely and the effect size was negligible. Free-text reflections from the participants provided several explanations for the variance, including large individual variations in 1) ramp-up time needed to become familiar with the Java tech stack, 2) setup friction in configuring the local development environment, and 3) personal ``definitions of done'' related to code quality. 

We designed the tasks around common Java technologies, but still observed a large variation in learning time, even among participants with the same self-reported Java proficiency level. Regarding setup friction, we underestimated the extent to which participants would invest in configuring their preferred local Java environment. The vague expectations on code quality were deliberate, however, and we counted on randomization to mitigate its effects -- but given the sources of variation, and an underpowered RCT, it was unlikely that frequentist statistics would detect systematic differences in completion time.

Bayesian analysis tends to be a suitable alternative in such situations, providing both an estimated effect and a transparent representation of uncertainty -- instead of dichotomous verdicts. The posterior mean suggests a small speedup (12.5\%), but the 95\% credible interval includes both zero and a small disadvantage for the treatment. In other words, at most a small and highly uncertain speed advantage emerged in our setting. Moreover, with an optimistic prior we observed a significant effect, which we interpret as a failed sensitivity analysis. We consider this another indication that the available data was insufficient to draw firm conclusions related to completion time.


As described in Section~\ref{sec:rw}, most previous work on AI-assisted development has focused on direct productivity gains. By examining the completion times in Task~1, we can position our findings to these studies. In our study, participants using AI assistants completed Task~1 with a 30.7\% shorter median completion time -- a statistically significant difference corresponding to a medium effect size. This result aligns with prior studies on speedups with GitHub Copilot. For example, \citet{peng_impact_2023} reported a 55.8\% speedup when implementing an HTTP server in JavaScript; \citet{paradis_how_2025} reported a 21\% speedup for C++ tasks at Google; and \citet{chatterjee_impact_2024} reported a 42.3\% speedup for algorithmic Python tasks at ANZ Bank.

Furthermore, our Bayesian analysis of Task~1 supports this trend, showing that habitual AI users completed the task about 59.6\% faster than the control group, matching the most optimistic results previously reported by~\citet{peng_impact_2023}. Thus, our study adds to the growing body of evidence that AI assistants can significantly accelerate development -- providing supporting data from a different task and a different programming language.

As recommended in the SPACE framework, we complement the direct measurements with participants' self-reported perceptions of productivity. The frequentist analysis found no significant difference between treatment and control. The Bayesian analysis indicates that the most likely effect is a modest disadvantage for those evolving AI-assisted code, but with wide uncertainty. A possible explanation for this tendency is that we observed a handful of AI-assisted Task~1 participants who reported feeling very productive and also produced substantial amounts of new code. We hypothesize that Task~2 participants assigned to evolve larger solutions experienced longer onboarding periods -- and thus felt less productive. While we find no clear patterns supporting this in the data, previous work highlights code size as a (or the) key maintainability challenge~\citep{sjoberg_quantifying_2013}. The risk that widespread adoption of AI-assistants will inflate codebases -- at an unprecedented scale -- is an important trend to closely monitor in the near future.

Several previous studies have examined the impact of AI assistants on perceived productivity. \citet{ziegler_measuring_2024} surveyed over 17,000 GitHub Copilot users and reported a large positive impact across different dimensions. Similarly, \citet{liang_large-scale_2024} and \citet{butler_dear_2025} found that developers generally feel productivity gains using the same assistant. \citet{weisz_examining_2025} contrasts the picture somewhat for the watsonx Code Assistant, showing that while most developers feel faster with the tool, a large fraction feel less effective. Our Task~1 results (see Figure~\ref{fig:task1-pp} reveal large individual variation. On the other hand, we found that AI-assisted developers felt more productive, and they are overrepresented among the participants who reported the highest scores. Free-text answers showed that some of them worked with more autonomous AI assistants, approaching the third-generation coding agents described in Section~\ref{sec:ai-gens}. Clearly, we need more empirical research on this emerging category of more capable tools.

\begin{tcolorbox}[
  colback=lightblue,
  colframe=blue!75!black,
  coltitle=white,
  title=RQ1: Is AI-assisted code more efficient to evolve manually?,
  fonttitle=\bfseries,
  enhanced,
  breakable,
  boxrule=0.5pt,
  leftrule=1pt,
  rightrule=1pt,
  toprule=1pt,
  bottomrule=1pt
]
Our RCT provides no reliable evidence that AI-assisted code differs in manual evolution efficiency. Both completion times and perceived productivity in Task~2 were consistent with a null effect, with at most small and highly uncertain differences between treatment and control.
\end{tcolorbox}

\subsection{RQ2: Higher Quality Upon Evolution?}
Does code co-developed with AI assistants result in higher quality upon manual evolution? Again, we addressed this question by analyzing two metrics from the RCT, complemented by free-text answers. First, the average CodeHealth of the resulting code after the manual evolution task (Task~2). Second, the test coverage of that same Task~2 solution.  

Using frequentist analysis, we found no evidence that Task~1 AI usage affected Task~2 average CodeHealth. One key source of quality variability, as discussed in Section~\ref{sec:task-reflect}, is that participants interpreted the deliberately vague task instructions differently, leading to a varied judgment about when the task was ``good enough'' to submit. The sample size was not large enough for randomization to fully mitigate this effect.

As with the completion time, the Bayesian analysis provides more nuanced insights. We found that the more AI-skilled the Task~1 participants were, the stronger the effect of their AI usage on Task~2 CodeHealth. For habitual AI users, the estimated mean effect was a significant absolute increase of +0.10 in CodeHealth. While the effect may appear small, we should note that this signal corresponds to half a standard deviation, detected in an average over about 50~files -- and the effect is robust across priors. Still, building on code developed by a proficient Java developer mattered even more.

The findings show that developers vary in their ability to work effectively with AI assistants. While this is not surprising, it suggests that training developers to properly use these tools remains important in the current AI era. Adding to that, we also found that the Task~1 developer's Java proficiency had a stronger influence on Task~2 outcomes than AI-usage, even among habitual AI users. Human development expertise clearly still matters when working with first- and second-generation AI assistants. How this finding translates to third-generation coding agents is an open question for future work.

Previous research has reported mixed findings on how developer seniority affects the impact of AI assistants. The early work by \citet{peng_impact_2023} reported that junior developers benefited the most from GitHub Copilot, while \citet{paradis_how_2025} reported greater speedups among senior developers at Google. Although our study includes a few juniors, we contribute a new perspective: habitual AI users using preferred tools refactor code in ways that leave lasting improvements -- effects that remain even after the code is manually evolved by someone else. This provides a strong argument for equipping capable developers with refactoring-oriented AI assistants to support long-term maintainability.

Finally, we complemented the CodeHealth measurements with test coverage. We found no significant differences in Task~2 test coverage between the treatment and control groups, with both achieving a median coverage of 70\%. Some AI-assisted Task~1 participants reached high coverage scores, but no free-text answers elaborated on the use of AI assistants to support testing, despite this being a realistic use case \citep{ouedraogo_large-scale_2024}. However, some participants mentioned issues related to testing, such as missing test cases or poor testability. The Bayesian analysis surprisingly revealed significantly lower test coverage in Task~2 solutions preceded by highly proficient Java developers in Task~1. Given the lack of a plausible causal explanation, we consider this to be a spurious result.


\begin{tcolorbox}[
  colback=lightblue,
  colframe=blue!75!black,
  coltitle=white,
  title=RQ2: Does prior AI use improve quality after manual evolution?,
  fonttitle=\bfseries,
  enhanced,
  breakable,
  boxrule=0.5pt,
  leftrule=1pt,
  rightrule=1pt,
  toprule=1pt,
  bottomrule=1pt
]
Our RCT provides no reliable frequentist evidence that prior AI-assisted development affects CodeHealth after continued manual evolution. Bayesian analysis suggests a small CodeHealth improvement when the original Task~1 solution was co-developed with AI by a habitual AI user. Test coverage was essentially identical between groups.
\end{tcolorbox}

\subsection{Implications for Industry Practice}
Our study has resulted in several novel insights backed by empirical data. But what does all this mean in practice? We conclude the section by discussing three interpretations for industry: one relieving finding, one risk that needs to be managed, and one challenge that remains open.

\paragraph{1) AI assistants tend to help with file-level maintainability issues}
From a maintainability perspective, our findings suggest that developers who prefer working with AI assistants should continue doing so. We observe no systematic maintainability issues for the next developer continuing to evolve the code down the line. On the contrary, habitual AI users appear to use these supporting tools in ways that benefit future manual evolution.

We speculate that AI assistants homogenize code into standard constructs. This is supported by a recent study on LLM creativity by~\citet{haase_has_2025}, which shows homogenization effects across LLMs, i.e., output from models lacks originality and clusters around similar solutions. This can be a major benefit for maintainability. Tornhill's notion of \textit{``Beauty in Code''}~\citep{tornhill_use_2024} defined high-quality code by its lack of surprises and the ease with which a developer can form a correct mental model of its behavior. If AI assistants successfully shave off surprises, they offer a substantial maintainability advantage by reinforcing idiomatic language patterns and avoiding novel solutions to known problems. While our work highlights the value of AI assistants among capable developers, it is reasonable that these ``surprise shaving'' benefits would manifest at least as much among less skilled developers.

\paragraph{2) Reckless use of AI assistants might quickly bloat codebases}
Generative AI is a new power tool in the developer's toolbox. But like any power tool, it can cause serious damage when misused. With the cost of creating new code nearing zero, the efforts to understand and maintain the code remain. This asymmetry risks bloating codebases with redundant logic, abandoned attempts, and code with questionable purpose. User discipline and project-wide retention policies will be critical in a time when we will inevitably see examples of ``firehose generation’’ of new code. 

Even if individual files become highly maintainable, the sheer code volume may undermine system-level maintainability. Code volume is a primary driver of complexity~\citep{sjoberg_quantifying_2013}, and AI-assisted development can accelerate growth at an unprecedented pace. One clear risk of unrestrained code synthesis is the unnoticed introduction of code clones -- a long-standing maintainability concern, even if empirical studies are not conclusive on their severity~\citep{rahman_clones_2012}. Regardless, the total amount of code on the planet will reach new magnitudes, potentially triggering unforeseen phenomena as tipping points in size are crossed.

We see two primary ways to mitigate the risks. First, organizations must continuously monitor the inflow of new code into the repositories. Given the expected adoption of AI-assisted generation, tools are essential since the output volume will surpass what humans can oversee. Beyond volume, tools should also assess various quality aspects such as clone rates and security vulnerabilities. Tool vendors are already exploring how to support organizations in scaling AI adoption reliably. Second, developers should receive training to ensure disciplined AI usage. Guardrails, both technical and procedural, will help embed best practices in the workflows of the AI era. Interestingly, as several participants in our study noted, AI assistants can also be helpful for onboarding and understanding new code -- suggesting that these tools may both contribute to the code size challenge and help us navigate its effects. Time will tell where the new equilibrium settles.

\paragraph{3) Over-reliance on AI assistants might erode skills and understanding}
Neither industry nor higher education institutions yet know how to proceed with knowledge management in the AI-assisted era \citep{franklin_generative_2025}. What skills will be central for future generations of ``LLM-native'' developers? How will sustained use of AI assistants affect software craftsmanship and deep understanding of complex software systems? More research along the lines of \citet{abrahao_software_2025} is clearly needed to carefully examine what AI assistance enhances, retrieves, reverses, and obsolesces. One thing is clear: AI assistance is here to stay. The alluring simplicity of working with the tools makes the path of least resistance difficult to resist. 

But as demonstrated in a recent electroencephalography experiment by~\citet{kosmyna_your_2025} (N=54), some resistance is vital for learning. In the context of essay writing, they found that participants assisted by ChatGPT showed significantly lower cognitive activity than those using web search or no tools at all. Moreover, the LLM-assisted participants underperformed in remembering details of their work. The authors refer to this as \textit{cognitive debt}, i.e., mental effort is deferred in the short term but accumulates long-term costs, such as diminished critical inquiry and decreased creativity. 

While AI-assisted development does not appear to introduce code-level technical debt, we warn about a future with build-up of cognitive debt in software organizations. The repercussions could be massive if developers end up with only a shallow understanding of codebases increasingly generated by machines. Over time, this might not only threaten maintainability but also impede innovation. To counter this risk, we should revisit decades-old research on ``innovative software engineering environments''~\citep{ambriola_towards_1991} and explore how to best support human inventiveness with the combinatorial creativity that LLMs excel at.

Finally, we recommend that organizations adopt a strategic approach to knowledge management in this new era. Tool support will be needed for provenance tracking, e.g., to monitor where only little human effort has been involved. Teams must make deliberate decisions about which components should remain under human cognitive control. Mitigating developer deskilling will become a priority when convenience drives LLM usage -- ``keepskilling'' may emerge as a complement to upskilling and reskilling. There is a new dawn for strategic HR in software organizations, where policies ensure that developer maintain their core skills. This is analogous to pilots who are encouraged to manually land aircraft from time to time to retain their flying skills. We find this to be a substantial challenge for the future of software engineering.




\section{Limitations and Threats to Validity} \label{sec:threats}
Selecting an empirical research method for a software engineering study always involves trade-offs. By opting for a controlled experiment, we prioritize control over realism. That said, we believe our Java development tasks are realistic and span development durations. The participants also largely agreed that they resembled prior development tasks (see Figure~\ref{fig:task-resemblance}). 

While we argue that our tasks are realistic, they represent only a slice of real-world Java development. Before discussing specific threats, we acknowledge four important limitations. First, our two-phase study only allows us to observe short-term maintainability effects. Maintainability is obviously part of the long game, but investigating long-term consequences requires longitudinal case studies. Second, our study focuses on individual developers. Although software engineering is inherently social, team dynamics falls outside our scope. Still, our design includes a rare controlled hand-off between developers. Third, the RecipeFinder system was kept small to allow reasonable onboarding times. Studying architectural considerations requires larger systems that deserve case study research. Fourth, we did not investigate security in this study, although previous research found that developers who work with AI assistants tend to write less secure code~\citep{perry_users_2023}. 

We organize the remaining discussion according to the categories proposed by the ACM SIGSOFT Empirical Standards~\citep{acm_sigsoft_experiments_2024}.

\subsection{Construct Validity}
A backbone construct in our study is \textit{AI-assisted development}. AI assistants vary in interaction mode and capability, and Task~1 participants might have interpreted the concept differently. We mitigated this by providing concrete tool examples in the Task~1 instructions and again in the exit questionnaire (Q2-3). We also asked participants to report how frequently they used their assistant (Q2-4). Moreover, the main target of this study is Task~2, in which participants were randomly assigned to evolve solutions that had been AI-assisted in a variety of ways. We consider this threat minor.

Section~\ref{sec:maint} explains how we define and measure the complex constructs of maintainability and productivity. We build on state-of-the-art research and operationalize the constructs using a combination of completion time, objective CodeHealth, and subjective perceived productivity guided by the SPACE framework. We consider both constructs valid, and we argue that the completeness of our measurement approach is sufficient for the scope of this study.

Still, we highlight two threats related to completion time and perceived productivity. First, as discussed in Section~\ref{sec:dataproc}, many participants did not complete the task in one uninterrupted session (Q2-1). To remedy this protocol violation, we asked them to provide their best time estimates. Since these participants tended to have longer completion times, we believe that they generally overestimated the time it took to complete the task. However, this risk is mitigated: in Task~1 the distribution of these time estimates (instead of git-based time stamps) was evenly distributed (19 \textbf{AI-devs} and 20 \textbf{!AI-devs}), and in Task~2, randomization should balance the bias, and we directly controlled for the effect in the Bayesian analysis. Second, we speculate that \textbf{AI-devs} may have been more inclined to evaluate their productivity positively due to personal beliefs in the value of AI assistance. We accept this second threat.

Finally, our qualitative analysis of free-text answers introduces construct threats. Coding activities involve interpretations and may thus reflect the researchers' expectations. To mitigate this threat, we employed two coders and iteratively developed the coding scheme. For the qualitative analysis of source code solutions, five of the authors were involved.

\subsection{Internal Validity}
As we conducted a controlled experiment, the internal validity is our main concern. RCTs are textbook examples of designs that maximize internal validity, although Bayesian causal analysis is equally potent. The core argument for high internal validity lies in the random assignment in Task~2, which guards against confounding threats and balances known and unknown factors between the treatment and control groups. Still, we elaborate on the most important threats below.

Most importantly, we identified substantial variation in participants' implicit definitions of ``good enough'' code quality for submission. As discussed in Section~\ref{sec:task-reflect}, participants differed in how much effort they chose to invest in both Task~1 and Task~2. The vague quality requirement was a deliberate design choice to provide realism, and we wanted participants to deliver according to their own quality bars. While randomization should mitigate these effects, we acknowledge that interpretations ranged from treating the assignment ``as if I were hired for a job'' to providing the bare minimum effort. In the worst case, randomization may have failed to distribute these differences evenly between the two experimental groups.

Participants completed tasks remotely, and we cannot guarantee protocol adherence. Regarding AI-assistance, Q2-2 caught one violation, i.e., an \textbf{!AI-dev} had to be moved to the \textbf{AI-dev} group. While we detected no indications of other violations, e.g., use of AI assistants in Task~2, we cannot entirely rule them out. Likewise, some participants may have received other types of assistance, such as pairing up with senior friends to solve the tasks collaboratively.

As reported in Section~\ref{sec:dataproc}, one \textbf{AI-dev} (\textit{anon126})  submitted two Task~1 solutions. The two solutions are very different in size and completion time, with the second being minimal in comparison. Thus, we accept to treat these as two separate inputs to Task~2. This design decision affects Task~1 analysis as there is obviously a learning effect at play that we choose to accept.

Given the home-based experiment setting, we expected some unusual variation points. Indeed, the free-text answers reveal some anomalies that surely had an impact on the outcome of the tasks. Self-reported anomalies include family business, such as taking care of babies, drinking a glass of wine while completing the task, and having the TV on in the background. Again, we trust that randomization can properly control these effects.

\subsection{External Validity}
The results from our study cannot be generalized to all software development contexts. Nonetheless, we argue that our tasks reflect a realistic scenario: progressing code previously developed by an unknown contributor. This is a common assignment in industry. Moreover, since 92\% of our participants are professional developers, and most of them between 30 and 49 years of age, we avoid the common debate about the validity of using students as subjects in controlled experiments~\citep{feldt_four_2018}. 

Still, the main threats to external validity relate to our system and task design. RecipeFinder is a small Java web application built on Spring Boot. We believe this codebase is the largest feasible size for participants to understand and extend within a reasonable time. However, the value of AI assistants may change with system size. In principle, AI assistants may be \textit{more valuable} as information navigation tools when a maintainer needs to onboard a vast codebase to add a new feature. On the other hand, very large codebases do not fit the context window of an LLM, so AI assistants only see partial views of the architecture and a small subset of files at a time. As a result, these tools may be \textit{less capable} at supporting cross-cutting, architecture-level maintenance in industrial-scale systems than in RecipeFinder.

In addition, our study does not provide any insight into what would happen if also the Task~2 developers used AI assistants. This might be standard practice in the near future and should be the focus of future work.
Another consideration is our choice of a tech stack. The results might differ in other languages or problem domains. For example, our results might not generalize to algorithmic challenges in C++ or Python. We note that many evaluations of LLM programming capabilities have targeted competitive programming, e.g., AlphaCode~\citep{li_competition-level_2022}, but we are instead happy to provide results for tasks that we co-designed with senior developers to ensure practically relevant research~\citep{garousi_practical_2020}.

Another threat to the external validity concerns the specific AI assistants used. We did not restrict participants in the \textbf{AI-dev} group to any particular assistants, as we prioritized allowing them to use their preferred tools. Table~\ref{tab:ai-tools} shows that a variety of tools were used, building on different underlying LLMs. Given this distribution, our findings primarily reflect the impact of mature commercial assistants. We cannot claim that the same results would hold for smaller, cheaper, or less sophisticated models running locally. Such models may  generate lower-quality code that could potentially challenge downstream maintainability.

Finally, we acknowledge sampling bias in the demographics of the participants. Our sample is heavily male (95\%), a stronger imbalance than in the general developer population. It is possible that societal roles make women less likely to participate in voluntary experiments of this nature. Geographic representation is also uneven. Large countries with substantial developer workforces are missing entirely, such as China, Japan, South Korea, and Iran. Our sample reflects the reach of the last author’s English-language YouTube channel. Future work should broaden demographic representation.

\subsection{Conclusion Validity} 
We used mixed-method research, combining objective and subjective measures, to assess the maintainability effects of AI-assisted development. We applied randomization, data cleaning, robust statistical methods, and multi-researcher coding for qualitative analysis.

A central limitation concerns the statistical power of our RCT. As quantified in Section~\ref{sec:freq}, our realized sample size of 75 Phase~2 participants yields 80\% power only for relatively large standardized mean differences. Hence, the non-significant results in Section~\ref{sec:res} rule out large effects of prior AI assistance on Task~2 outcomes, but remain inconclusive with respect to small or moderate effects. We therefore interpret our frequentist findings as evidence against large effects, rather than as evidence for the absence of any effect.

Beyond this, some additional risks remain. First, a fundamental assumption of statistical analysis is that observations are independent. We assume that the participants completed their tasks without communicating during or after the assignment. This cannot be verified, but the wide geographical spread and the variety of email domains suggest that the risk is minor. There is a group of participants from Equal Experts, but we trust that they did not discuss the tasks.

Second, several of our preregistered statistical tests assume normality of the underlying data. After running Shapiro-Wilk tests (see Section~\ref{sec:freq}), we found that the completion time and the test coverage violated this assumption. In these cases, we used non-parametric alternatives in the frequentist analysis. While most previous work reports mean completion times, which can be misleading for skewed distributions, we discuss medians. Moreover, our parallel Bayesian analysis relaxes the distributional assumptions and supports the robustness of our findings.

Third, our analysis of perceived productivity is based on a Likert scale. We verified high internal consistency (Cronbach’s $\alpha > 0.85$), but calculating means from ordinal data remains contested. We mitigated this threat by method triangulation, as the Bayesian analysis modeled a latent productivity variable based on the individual Likert items. 

Finally, the collection of free-text responses adds richness to our results but might introduce self-selection bias. Participants who were more motivated, or opinionated, probably submitted more detailed reflections. We consider this threat acceptable, as we use these responses to enrich our understanding, rather than to support any statistical inference.

\section{Conclusion and Future Work} \label{sec:conc}
We set out to investigate the impact of AI assistants on software maintainability. To this end, we formulated two research questions targeting evolution efficiency and code quality, respectively. We conducted a preregistered two-phase controlled experiment. In Phase~2, a randomized control trial (N=75), participants manually evolved a Phase~1 solution developed by someone else -- who had either worked with AI assistants such as GitHub Copilot, ChatGPT, and Cursor (treatment), or not (control). In total, we collected 151 solutions to our realistic development tasks, 95.4\% of which were completed by professional software developers.

Our results provide no clear evidence that code co-developed with AI assistants is more efficient to evolve manually. Task~2 completion times were highly variable -- largely due to variations in learning and setup times -- and any speed advantage for evolving AI-assisted code was small and statistically unreliable. For code quality, traditional frequentist analyses found no significant differences between treatment and control. Bayesian analysis, however, suggests a small CodeHealth improvement when the original Task~1 solution was co-developed with AI by a habitual AI user. These findings indicate that, in our setting, prior AI use neither clearly improves nor degrades downstream manual evolution, with at most a small positive signal for CodeHealth in this specific subgroup.

We found additional evidence that individual variation in AI proficiency matters: in Phase~1, the posterior mean effect on completion time for habitual AI users was a 55.9\% speedup, compared to a 30.7\% median decrease across the entire sample. We conclude that learning to work effectively with AI assistants is a valuable developer skill -- and that such proficiency may even benefit others working manually downstream. But being a proficient Java developer mattered even more in our study.

Some Phase~1 participants assisted by chat-based programming shared enthusiastic comments about their experience. Developers who reported the highest levels of perceived productivity all worked with AI assistants that went beyond code completion, most notably Cursor. This boost to perceived productivity did not carry over to Phase~2, where our results are consistent with a null effect. The Bayesian analysis shows a small and uncertain tendency toward lower perceived productivity when building on work by habitual AI users, but the evidence is too weak for firm conclusions. Future work should explore in what situations overly enthusiastic AI use might lead to downstream maintainability backlashes. We found no treatment effect on test coverage in Phase~2, but we observed that the few outstanding test suites in Phase~1 were all submitted by AI-assisted participants.

Our observational findings from Phase~1 adds to the growing body of evidence that AI assistants can effectively accelerate development. Moreover, we offer a cautiously reassuring signal for file-level maintainability. This is an encouraging finding, as we are convinced that AI assistants are here to stay. Within the scope of our tasks and measures, we found no indication that AI-assisted development harms maintainability at the code construct level. On the contrary, the large language model output is likely to shave off surprising or idiosyncratic code constructs that impede maintainability. The Bayesian analysis suggests that a positive maintainability effect is more probable than not, but any such effect would be at most small and highly uncertain. However, a greater concern is that reckless use of AI can bloat codebases. And large volumes of code, regardless of quality, are known to be hard for human developers to maintain.

Our concerns go beyond code volume. We also worry about the cognitive debt that over-reliance on AI assistants might introduce. If we increasingly generate code with minimal cognitive effort, developers may end up with only a shallow understanding of large, AI-generated codebases. Over time, this could erode core programming skills and probably stifle innovation. Whenever we automate an activity, something human is inevitably lost. How software organizations should navigate these neural tides in relation to knowledge management deserves substantial research attention in the next decade. And to what extent can the AI assistants help us cope with all the code they generate? This is not only an essential question for tool providers, but also for engineering managers, educators, and developers themselves.

These concerns point to several important directions for future work. The AI disruption is unfolding rapidly, and current AI assistants are already far more capable than the generation we studied half a year ago. Our study captures an early phase of AI-assisted development. Documenting the related downstream effects provides an empirical reference point for understanding how maintainability dynamics evolve as development practices shift toward more autonomous tooling. With coding agents entering the scene, human bottlenecks will increasingly be sidelined. What will be the direct and indirect effects of this shift? Longitudinal case studies are needed to track these changes as they unfold, ideally complemented by action research that explores possible interventions in real time.

We also see a potential need for further controlled studies. We encourage replication of our work to explore several new angles. First, we did not study evolution by developers who were themselves AI-assisted. This is a very relevant scenario going forward. Second, future experiments should evaluate third-generation AI assistants, such as Claude Code or Sonargraph's AMP, which radically change developer workflows. Third, our analysis should be extended to consider security implications. This will be particularly important for the agentic era, since agent ensembles will emerge. Combining agentic workforces, e.g., through model context protocols, opens up cans of new security worms where autonomous agents may unintentionally expose vulnerabilities or be exploited by antagonistic actors. However, such security research is an endeavor quite different from our focus on maintainability.

\section*{Acknowledgment}
We thank all participants of the study. Their willingness to voluntarily invest time and effort in completing the tasks was essential to this research.

\section*{Declarations}

\subsection*{Funding}
This work was partly funded by the NextG2Com Competence Centre – Next-Generation Communication and Computing Infrastructures and Applications – under the Vinnova grant number 2023-00541. Nadim Hagatulah was supported by the Wallenberg AI, Autonomous Systems and Software Program (WASP) funded by the Knut and Alice Wallenberg Foundation. Emma S\"oderberg was supported by the Swedish strategic research environment ELLIIT and the Swedish Foundation for Strategic Research (grant nbr. FFL18-0231).

\subsection*{Ethical Approval}
The design of our study adheres to the essential attributes of the ACM SIGSOFT Empirical Standard ``Ethics (Studies with Human Participants)''~\citep{acm_sigsoft_ethics_2024}. Additionally, the peer review of the registered report~\citep{borg_does_2024} acted as an independent assessment that the study design meets the ethical standards of the empirical software engineering community. 

There were no anticipated risks of harm to participants. The task required an estimated time investment of 2–4 hours, similar in nature and duration to programming assessments used in recruitment. All participants who completed that task received a signed copy of Dave Farley's book ``Modern Software Engineering: Doing What Works to Build Better Software Faster''~\citep{farley_modern_2022} as an incentive to complete the task.

\subsection*{Informed Consent}
This study involved voluntary participation by developers who completed programming tasks and questionnaires. Before assigning tasks, we ensured that participants understood their privacy was protected, that the study was intended for academic publication, and that they could withdraw at any time without consequence. Several participants chose to do so. All participants gave informed consent prior to participation.

\subsection*{Author Contributions}
We specify our contributions according to the CRediT taxonomy:
Markus Borg: Conceptualization, Methodology, Supervision, Investigation, Data Curation, Formal Analysis, Writing -- Original Draft Preparation, Writing -- Review \& Editing.
Dave Hewett: Conceptualization, Investigation, Resources, Writing -- Review \& Editing.
Nadim Hagatulah: Formal Analysis, Writing -- Original Draft Preparation, Writing -- Review \& Editing.
Noric Couderc: Methodology, Formal Analysis, Writing -- Original Draft Preparation, Writing -- Review \& Editing.
Emma Söderberg: Methodology, Supervision, Formal Analysis, Writing -- Review \& Editing.
Donald Graham: Conceptualization, Software, Writing -- Review \& Editing.
Uttam Kini: Software, Resources, Writing -- Review \& Editing.
Dave Farley: Resources, Writing -- Review \& Editing.

\subsection*{Data Availability Statement}
All materials required to replicate this study, e.g., task instructions, cleaned and anonymized participant data, and the causal graph, are available on Zenodo~\citep{equal_experts_does_2025} under the Creative Commons Attribution 4.0 International license. A GitHub repository containing the full code base, scripts for data cleaning and analysis, and notebooks to conveniently reproduce the figures is available at \url{https://github.com/codescene-research/echoes-of-ai-emse-2025}. The contents of the GitHub repository are versioned and mirrored in the Zenodo archive. CodeScene was used to calculate the Code Health metric as an indicator of maintainability. While CodeScene is a commercial tool, academic researchers can request a free license for research purposes.

\subsection*{Conflict of Interest}
Markus Borg is employed by CodeScene, the company behind the CodeScene analysis tool used in this study, including the CodeHealth metric. Dave Hewett, Donald Graham, and Uttam Kini are consultants at Equal Experts, a company offering AI-accelerated software delivery services among other offerings. Dave Farley is an independent consultant and author who runs the YouTube channel Modern Software Engineering. The authors declare that no commercial interests compromised the scientific rigor of the study design, data collection, or analysis.

\subsection*{Clinical Trial Number}
Clinical trial number: not applicable.

\bibliography{llm-prog}

\begin{thebibliography}{63}
\providecommand{\natexlab}[1]{#1}
\providecommand{\url}[1]{{#1}}
\providecommand{\urlprefix}{URL }
\providecommand{\doi}[1]{\url{https://doi.org/#1}}
\providecommand{\eprint}[2][]{\url{#2}}
 \bibcommenthead

\bibitem[{Abrahao et~al(2025)Abrahao, Grundy, Pezze, Storey, and
  Tamburri}]{abrahao_software_2025}
Abrahao S, Grundy J, Pezze M, et~al (2025) Software {Engineering} by and for
  {Humans} in an {AI} {Era}. ACM Trans Softw Eng Methodol 34(5):129:1--129:46.
  \doi{10.1145/3715111}

\bibitem[{{ACM SIGSOFT}(2024{\natexlab{a}})}]{acm_sigsoft_ethics_2024}
{ACM SIGSOFT} (2024{\natexlab{a}}) Ethics ({Studies} with {Human}
  {Participants}).
  \urlprefix\url{https://github.com/acmsigsoft/EmpiricalStandards/blob/master/docs/supplements/EthicsHumanParticipants.md},
  commit f82836c

\bibitem[{{ACM SIGSOFT}(2024{\natexlab{b}})}]{acm_sigsoft_experiments_2024}
{ACM SIGSOFT} (2024{\natexlab{b}}) Experiments (with {Human} {Participants}).
  \urlprefix\url{https://github.com/acmsigsoft/EmpiricalStandards/blob/master/docs/standards/Experiments.md},
  commit c4dbe93

\bibitem[{Al~Madi(2023)}]{al_madi_how_2023}
Al~Madi N (2023) How {Readable} is {Model}-generated {Code}? {Examining}
  {Readability} and {Visual} {Inspection} of {GitHub} {Copilot}. In: Proc. of
  the 37th {International} {Conference} on {Automated} {Software}
  {Engineering}, pp 1--5, \doi{10.1145/3551349.3560438}

\bibitem[{Ambriola et~al(1991)Ambriola, Ciancarini, Corradini, and
  DeFrancesco}]{ambriola_towards_1991}
Ambriola V, Ciancarini P, Corradini A, et~al (1991) Towards {Innovative}
  {Software} {Engineering} {Environments}. Journal of Systems and Software
  14(1):17--29. \doi{10.1016/0164-1212(91)90085-K}

\bibitem[{Ani et~al(2024)Ani, Hamid, and Zhamri}]{ani_recent_2024}
Ani ZC, Hamid ZA, Zhamri NN (2024) The {Recent} {Trends} of {Research} on
  {GitHub} {Copilot}: {A} {Systematic} {Review}. In: Zakaria NH, Mansor NS,
  Husni H, et~al (eds) Computing and {Informatics}. Springer Nature, Singapore,
  pp 355--366, \doi{10.1007/978-981-99-9589-9_27}

\bibitem[{Anthropic(2024)}]{anthropic_model_2024}
Anthropic (2024) Model {Context} {Protocol}: {Official} {Specification}. Tech.
  rep., \urlprefix\url{https://modelcontextprotocol.io}

\bibitem[{Avgeriou et~al(2016)Avgeriou, Kruchten, Ozkaya, and
  Seaman}]{avgeriou_managing_2016}
Avgeriou P, Kruchten P, Ozkaya I, et~al (2016) Managing {Technical} {Debt} in
  {Software} {Engineering} ({Dagstuhl} {Seminar} 16162). Dagstuhl Reports
  6(4):110--138. \doi{10.4230/DagRep.6.4.110}

\bibitem[{Barke et~al(2023)Barke, James, and Polikarpova}]{barke_grounded_2023}
Barke S, James MB, Polikarpova N (2023) Grounded {Copilot}: {How} {Programmers}
  {Interact} with {Code}-{Generating} {Models}. In: Proc. of the {ACM} on
  {Prgramming} {Languages}, pp 85--111, \doi{10.1145/3586030}

\bibitem[{Basili et~al(1994)Basili, Caldiera, and Rombach}]{basili_goal_1994}
Basili V, Caldiera G, Rombach D (1994) The {Goal} {Question} {Metric}
  {Approach}. In: Encyclopedia of {Software} {Engineering}. p 528--532

\bibitem[{Borg et~al(2023)Borg, Tornhill, and Mones}]{borg_u_2023}
Borg M, Tornhill A, Mones E (2023) U {Owns} the {Code} {That} {Changes} and
  {How} {Marginal} {Owners} {Resolve} {Issues} {Slower} in {Low}-{Quality}
  {Source} {Code}. In: Proc. of the 27th {International} {Conference} on
  {Evaluation} and {Assessment} in {Software} {Engineering}, pp 368--377,
  \doi{10.1145/3593434.3593480}

\bibitem[{Borg et~al(2024{\natexlab{a}})Borg, Ezzouhri, and
  Tornhill}]{borg_ghost_2024}
Borg M, Ezzouhri M, Tornhill A (2024{\natexlab{a}}) Ghost {Echoes} {Revealed}:
  {Benchmarking} {Maintainability} {Metrics} and {Machine} {Learning}
  {Predictions} {Against} {Human} {Assessments}. In: Proc. of the 40th
  {International} {Conference} on {Software} {Maintenance} and {Evolution}, pp
  678--688

\bibitem[{Borg et~al(2024{\natexlab{b}})Borg, Hewett, Graham, Couderc,
  Soderberg, Church, and Farley}]{borg_does_2024}
Borg M, Hewett D, Graham D, et~al (2024{\natexlab{b}}) Does {Co}-{Development}
  with {AI} {Assistants} {Lead} to {More} {Maintainable} {Code}? {A}
  {Registered} {Report}. \doi{10.48550/arXiv.2408.10758}

\bibitem[{Borg et~al(2024{\natexlab{c}})Borg, Pruvost, Mones, and
  Tornhill}]{borg_increasing_2024}
Borg M, Pruvost I, Mones E, et~al (2024{\natexlab{c}}) Increasing, {Not}
  {Diminishing}: {Investigating} the {Returns} of {Highly} {Maintainable}
  {Code}. In: Proc. of the 7th {International} {Conference} on {Technical}
  {Debt}, pp 21--30

\bibitem[{Butler et~al(2025)Butler, Suh, Haniyur, and
  Hadley}]{butler_dear_2025}
Butler J, Suh J, Haniyur S, et~al (2025) Dear {Diary}: {A} {Randomized}
  {Controlled} {Trial} of {Generative} {AI} {Coding} {Tools} in the
  {Workplace}. In: Proc. of the {International} {Conference} on {Software}
  {Engineering}. arXiv

\bibitem[{Bürkner and Charpentier(2020)}]{burkner_modelling_2020}
Bürkner PC, Charpentier E (2020) Modelling {Monotonic} {Effects} of {Ordinal}
  {Predictors} in {Bayesian} {Regression} {Models}. The British Journal of
  Mathematical and Statistical Psychology 73(3):420--451.
  \doi{10.1111/bmsp.12195}

\bibitem[{Chatterjee et~al(2024)Chatterjee, Liu, Rowland, and
  Hogarth}]{chatterjee_impact_2024}
Chatterjee S, Liu CL, Rowland G, et~al (2024) The {Impact} of {AI} {Tool} on
  {Engineering} at {ANZ} {Bank} {An} {Empirical} {Study} on {GitHub} {Copilot}
  within {Corporate} {Environment}. In: Proc. of the 10th {International}
  {Conference} on {Software} {Engineering} ({SEC}),
  \doi{10.48550/arXiv.2402.05636}

\bibitem[{Cook et~al(2003)Cook, Ji, and Harrison}]{cook_software_2003}
Cook S, Ji H, Harrison R (2003) Software {Evolution} and {Software}
  {Evolvability}. In: Proc. of the {Workshop} on {Software} {Analysis} and
  {Maintenance}

\bibitem[{Cui et~al(2025)Cui, Demirer, Jaffe, Musolff, Peng, and
  Salz}]{cui_effects_2025}
Cui ZK, Demirer M, Jaffe S, et~al (2025) The {Effects} of {Generative} {AI} on
  {High} {Skilled} {Work}: {Evidence} from {Three} {Field} {Experiments} with
  {Software} {Developers}. \doi{10.2139/ssrn.4945566}

\bibitem[{{Equal Experts} et~al(2025){Equal Experts}, Borg, and
  Couderc}]{equal_experts_does_2025}
{Equal Experts}, Borg M, Couderc N (2025) Does {Co}-{Development} with {AI}
  {Assistants} {Lead} to {More} {Maintainable} {Code}? {Replication} {Package}.
  \doi{10.5281/zenodo.17977295}

\bibitem[{Farley(2022)}]{farley_modern_2022}
Farley D (2022) Modern {Software} {Engineering}: {Doing} {What} {Works} to
  {Build} {Better} {Software} {Faster}. Addison Wesley, Boston, MA, USA

\bibitem[{Feldt et~al(2018)Feldt, Zimmermann, Bergersen, Falessi, Jedlitschka,
  Juristo, Munch, Oivo, Runeson, Shepperd, Sjoberg, and
  Turhan}]{feldt_four_2018}
Feldt R, Zimmermann T, Bergersen GR, et~al (2018) Four {Commentaries} on the
  {Use} of {Students} and {Professionals} in {Empirical} {Software}
  {Engineering} {Experiments}. Empirical Software Engineering 23(6):3801--3820.
  \doi{10.1007/s10664-018-9655-0}

\bibitem[{Fenton(1994)}]{fenton_software_1994}
Fenton N (1994) Software {Measurement}: {A} {Necessary} {Scientific} {Basis}.
  IEEE Transactions on Software Engineering 20(3):199--206

\bibitem[{Forsgren et~al(2021)Forsgren, Storey, Maddila, Zimmermann, Houck, and
  Butler}]{forsgren_space_2021}
Forsgren N, Storey MA, Maddila C, et~al (2021) The {SPACE} of {Developer}
  {Productivity}: {There}'s more to it than you think. Queue
  19(1):10:20--10:48. \doi{10.1145/3454122.3454124}

\bibitem[{Franklin et~al(2025)Franklin, Denny, Gonzalez-Maldonado, and
  Tran}]{franklin_generative_2025}
Franklin D, Denny P, Gonzalez-Maldonado DA, et~al (2025) Generative {AI} in
  {Computer} {Science} {Education}: {Challenges} and {Opportunities}. Cambridge
  University Press, Cambridge, MA, USA

\bibitem[{Garousi et~al(2020)Garousi, Borg, and Oivo}]{garousi_practical_2020}
Garousi V, Borg M, Oivo M (2020) Practical {Relevance} of {Software}
  {Engineering} {Research}: {Synthesizing} the {Community}’s {Voice}.
  Empirical Software Engineering 25(3):1687--1754.
  \doi{10.1007/s10664-020-09803-0}

\bibitem[{Gelman et~al(2021)Gelman, Hill, and Vehtari}]{gelman_regression_2021}
Gelman A, Hill J, Vehtari A (2021) Regression and {Other} {Stories}. Cambridge

\bibitem[{Gorla et~al(2025)Gorla, Kumar, Lorenzini, and
  Alipourfaz}]{gorla_cubetesterai_2025}
Gorla D, Kumar S, Lorenzini PNR, et~al (2025) {CUBETESTERAI}: {Automated}
  {JUnit} {Test} {Generation} using the {LLaMA} {Model}. In: Proc. of the 18th
  {International} {Conference} on {Software} {Testing}, {Verification} and
  {Validation}

\bibitem[{Haase et~al(2025)Haase, Hanel, and Pokutta}]{haase_has_2025}
Haase J, Hanel PHP, Pokutta S (2025) Has the {Creativity} of {Large}-{Language}
  {Models} {Peaked}? {An} {Analysis} of inter- and {Intra}-{LLM} {Variability}.
  \doi{10.48550/arXiv.2504.12320},
  \urlprefix\url{http://arxiv.org/abs/2504.12320}

\bibitem[{He et~al(2026)He, Miller, Agarwal, Kästner, and
  Vasilescu}]{he_speed_2026}
He H, Miller C, Agarwal S, et~al (2026) Speed at the {Cost} of {Quality}: {How}
  {Cursor} {AI} {Increases} {Short}-{Term} {Velocity} and {Long}-{Term}
  {Complexity} in {Open}-{Source} {Projects}. In: Proc. of the 23rd
  {International} {Conference} on {Mining} {Software} {Repositories},
  \doi{10.1145/3793302.3793349}

\bibitem[{Husein et~al(2025)Husein, Aburajouh, and Catal}]{husein_large_2025}
Husein RA, Aburajouh H, Catal C (2025) Large language models for code
  completion: {A} systematic literature review. Computer Standards \&
  Interfaces 92:103917. \doi{10.1016/j.csi.2024.103917}

\bibitem[{{International Organization for
  Standardization}(2011)}]{international_organization_for_standardization_systems_2011}
{International Organization for Standardization} (2011) {ISO/IEC 25010:2011
  S}ystems and {Software} {Engineering} - {Systems} and {Software} {Quality}
  {Requirements} and {Evaluation} ({SquaRE}) - {System} and {Software}
  {Quality} {Models}

\bibitem[{Jaspan and Sadowski(2019)}]{jaspan_no_2019}
Jaspan C, Sadowski C (2019) No {Single} {Metric} {Captures} {Productivity}. In:
  Sadowski C, Zimmermann T (eds) Rethinking {Productivity} in {Software}
  {Engineering}. Apress, Berkeley, CA, p 13--20

\bibitem[{JetBrains(2024)}]{jetbrains_state_2024}
JetBrains (2024) The {State} of {Developer} {Ecosystem} 2024. Tech. rep.,
  JetBrains s.r.o.,
  \urlprefix\url{https://www.jetbrains.com/lp/devecosystem-2024/}

\bibitem[{Kim and Yegge(2025)}]{kim_vibe_2025}
Kim G, Yegge S (2025) Vibe {Coding}: {Building} {Production}-{Grade} {Software}
  {With} {GenAI}, {Chat}, {Agents}, and {Beyond}. IT Revolution, to appear

\bibitem[{Kosmyna et~al(2025)Kosmyna, Hauptmann, Yuan, Situ, Liao, Beresnitzky,
  Braunstein, and Maes}]{kosmyna_your_2025}
Kosmyna N, Hauptmann E, Yuan YT, et~al (2025) Your {Brain} on {ChatGPT}:
  {Accumulation} of {Cognitive} {Debt} when {Using} an {AI} {Assistant} for
  {Essay} {Writing} {Task}. \doi{10.48550/arXiv.2506.08872}

\bibitem[{Lacerda et~al(2020)Lacerda, Petrillo, Pimenta, and
  Gueheneuc}]{lacerda_code_2020}
Lacerda G, Petrillo F, Pimenta M, et~al (2020) Code {Smells} and {Refactoring}:
  {A} {Tertiary} {Systematic} {Review} of {Challenges} and {Observations}.
  Journal of Systems and Software 167:110610

\bibitem[{Li et~al(2022)Li, Choi, Chung, Kushman, Schrittwieser, Leblond,
  Eccles, Keeling, Gimeno, Dal~Lago, Hubert, Choy, de~Masson~d’Autume,
  Babuschkin, Chen, Huang, Welbl, Gowal, Cherepanov, Molloy, Mankowitz,
  Sutherland~Robson, Kohli, de~Freitas, Kavukcuoglu, and
  Vinyals}]{li_competition-level_2022}
Li Y, Choi D, Chung J, et~al (2022) Competition-{Level} {Code} {Generation}
  {With} {AlphaCode}. Science 378(6624):1092--1097.
  \doi{10.1126/science.abq1158}

\bibitem[{Liang et~al(2024)Liang, Yang, and Myers}]{liang_large-scale_2024}
Liang JT, Yang C, Myers BA (2024) A {Large}-{Scale} {Survey} on the {Usability}
  of {AI} {Programming} {Assistants}: {Successes} and {Challenges}. In: Proc.
  of the {IEEE}/{ACM} 46th {International} {Conference} on {Software}
  {Engineering}, pp 1--13, \doi{10.1145/3597503.3608128}

\bibitem[{Mantyla and Lassenius(2006)}]{mantyla_subjective_2006}
Mantyla MV, Lassenius C (2006) Subjective {Evaluation} of {Software}
  {Evolvability} {Using} {Code} {Smells}: {An} {Empirical} {Study}. Empirical
  Software Engineering 11(3):395--431. \doi{10.1007/s10664-006-9002-8}

\bibitem[{Martin(2009)}]{martin_clean_2009}
Martin R (2009) Clean {Code}: {A} {Handbook} of {Agile} {Software}
  {Craftsmanship}. Upper Saddle River, NJ

\bibitem[{McElreath(2020)}]{mcelreath_statistical_2020}
McElreath R (2020) Statistical {Rethinking}: {A} {Bayesian} {Course} with
  {Examples} in {R} and {STAN}, 2nd edn. Boca Raton, FL, USA

\bibitem[{Ouedraogo et~al(2024)Ouedraogo, Kaboré, Tian, Song, Koyuncu, Klein,
  Lo, and Bissyande}]{ouedraogo_large-scale_2024}
Ouedraogo WC, Kaboré K, Tian H, et~al (2024) Large-scale, {Independent} and
  {Comprehensive} {Study} of the {Power} of {LLMs} for {Test} {Case}
  {Generation}. \doi{10.48550/arXiv.2407.00225}

\bibitem[{Papke and Wooldridge(1996)}]{papke_econometric_1996}
Papke LE, Wooldridge JM (1996) Econometric {Methods} for {Fractional}
  {Response} {Variables} {With} an {Application} to 401(k) {Plan}
  {Participation} {Rates}. Journal of Applied Econometrics 11(6):619--632

\bibitem[{Paradis et~al(2025)Paradis, Grey, Madison, Nam, Macvean, Zhang,
  Ferrari-Church, and Chandra}]{paradis_how_2025}
Paradis E, Grey K, Madison Q, et~al (2025) How {Much} {Does} {AI} {Impact}
  {Development} {Speed}? {An} {Enterprise}-based {Randomized} {Controlled}
  {Trial}. In: Proc. of the 47th {International} {Conference} on {Software}
  {Engineering}. arXiv, \doi{10.1109/ICSE-SEIP66354.2025.00060}

\bibitem[{Peng et~al(2023)Peng, Kalliamvakou, Cihon, and
  Demirer}]{peng_impact_2023}
Peng S, Kalliamvakou E, Cihon P, et~al (2023) The {Impact} of {AI} on
  {Developer} {Productivity}: {Evidence} from {GitHub} {Copilot}.
  \doi{10.48550/arXiv.2302.06590}

\bibitem[{Perry et~al(2023)Perry, Srivastava, Kumar, and
  Boneh}]{perry_users_2023}
Perry N, Srivastava M, Kumar D, et~al (2023) Do {Users} {Write} {More}
  {Insecure} {Code} with {AI} {Assistants}? In: Proc. of the 2023 {ACM}
  {SIGSAC} {Conference} on {Computer} and {Communications} {Security}, pp
  2785--2799, \doi{10.1145/3576915.3623157}

\bibitem[{Prosser(2025)}]{prosser_worried_2025}
Prosser D (2025) Worried {About} {AI}-{Generated} {Code}? {Ask} {AI} {To}
  {Review} {It}. Forbes
  \urlprefix\url{https://www.forbes.com/sites/davidprosser/2025/05/07/worried-about-ai-generated-code-ask-ai-to-review-it/}

\bibitem[{Rahman et~al(2012)Rahman, Bird, and Devanbu}]{rahman_clones_2012}
Rahman F, Bird C, Devanbu P (2012) Clones: {What} {Is} {That} {Smell}?
  Empirical Software Engineering 17(4):503--530.
  \doi{10.1007/s10664-011-9195-3}

\bibitem[{Sadowski and Zimmermann(2019)}]{sadowski_rethinking_2019}
Sadowski C, Zimmermann T (eds)  (2019) Rethinking {Productivity} in {Software}
  {Engineering}. Apress, Berkeley, CA,
  \urlprefix\url{https://link.springer.com/10.1007/978-1-4842-4221-6}

\bibitem[{Schnappinger et~al(2020)Schnappinger, Fietzke, and
  Pretschner}]{schnappinger_defining_2020}
Schnappinger M, Fietzke A, Pretschner A (2020) Defining a {Software}
  {Maintainability} {Dataset}: {Collecting}, {Aggregating} and {Analysing}
  {Expert} {Evaluations} of {Software} {Maintainability}. In: Proc. of the 36th
  {International} {Conference} on {Software} {Maintenance} and {Evolution}, pp
  278--289

\bibitem[{Siddiq et~al(2022)Siddiq, Majumder, Mim, Jajodia, and
  Santos}]{siddiq_empirical_2022}
Siddiq ML, Majumder SH, Mim MR, et~al (2022) An {Empirical} {Study} of {Code}
  {Smells} in {Transformer}-based {Code} {Generation} {Techniques}. In: Proc.
  of the {International} {Working} {Conference} on {Source} {Code} {Analysis}
  and {Manipulation}, pp 71--82, \doi{10.1109/SCAM55253.2022.00014}

\bibitem[{Silva et~al(2024)Silva, Saavedra, and
  Monperrus}]{silva_gitbug-java_2024}
Silva A, Saavedra N, Monperrus M (2024) {GitBug}-{Java}: {A} {Reproducible}
  {Benchmark} of {Recent} {Java} {Bugs}. In: Proc. of the 21st {International}
  {Conference} on {Mining} {Software} {Repositories},
  \doi{10.48550/arXiv.2402.02961}

\bibitem[{Sjoberg et~al(2013)Sjoberg, Yamashita, Anda, Mockus, and
  Dybå}]{sjoberg_quantifying_2013}
Sjoberg D, Yamashita A, Anda B, et~al (2013) Quantifying the {Effect} of {Code}
  {Smells} on {Maintenance} {Effort}. IEEE Transactions on Software Engineering
  39(8):1144--1156. \doi{10.1109/TSE.2012.89}

\bibitem[{Storey et~al(2025)Storey, Hoda, Milani, and
  Baldassarre}]{storey_guiding_2025}
Storey MA, Hoda R, Milani AMP, et~al (2025) Guiding {Principles} for {Using}
  {Mixed} {Methods} {Research} in {Software} {Engineering}.
  \doi{10.48550/arXiv.2404.06011}

\bibitem[{Tornhill(2024)}]{tornhill_use_2024}
Tornhill A (2024) Use {Beauty} as a {Guiding} {Principle}. In: Your {Code} as a
  {Crime} {Scene}: {Use} {Forensic} {Techniques} to {Arrest} {Defects},
  {Bottlenecks}, and {Bad} {Design} in {Your} {Programs}, 2nd edn. The
  Pragmatic Programmers, Raleigh, NC, USA

\bibitem[{Tornhill and Borg(2022)}]{tornhill_code_2022}
Tornhill A, Borg M (2022) Code {Red}: {The} {Business} {Impact} of {Code}
  {Quality} - {A} {Quantitative} {Study} of 39 {Proprietary} {Production}
  {Codebases}. In: Proc. of the 5th {International} {Conference} on {Technical}
  {Debt}, pp 11--20

\bibitem[{Treude and Storey(2025)}]{treude_generative_2025}
Treude C, Storey MA (2025) Generative {AI} and {Empirical} {Software}
  {Engineering}: {A} {Paradigm} {Shift}. In: Proc. of the 2nd {International}
  {Conference} on {AI}-powered {Software}. arXiv,
  \doi{10.48550/arXiv.2502.08108}

\bibitem[{Tsantalis et~al(2022)Tsantalis, Ketkar, and
  Dig}]{tsantalis_refactoringminer_2022}
Tsantalis N, Ketkar A, Dig D (2022) {RefactoringMiner} 2.0. IEEE Transactions
  on Software Engineering 48(3):930--950. \doi{10.1109/TSE.2020.3007722}

\bibitem[{Weber et~al(2024)Weber, Brandmaier, Schmidt, and
  Mayer}]{weber_significant_2024}
Weber T, Brandmaier M, Schmidt A, et~al (2024) Significant {Productivity}
  {Gains} through {Programming} with {Large} {Language} {Models}. Proc of the
  ACM on Hum-Comput Interact 8(EICS):256:1--256:29. \doi{10.1145/3661145}

\bibitem[{Weisz et~al(2025)Weisz, Kumar, Muller, Browne, Goldberg, Heintze, and
  Bajpai}]{weisz_examining_2025}
Weisz JD, Kumar SV, Muller M, et~al (2025) Examining the {Use} and {Impact} of
  an {AI} {Code} {Assistant} on {Developer} {Productivity} and {Experience} in
  the {Enterprise}. In: Proc. of the {Extended} {Abstracts} of the {CHI}
  {Conference} on {Human} {Factors} in {Computing} {Systems}, pp 1--13,
  \doi{10.1145/3706599.3706670}

\bibitem[{Ziegler et~al(2022)Ziegler, Kalliamvakou, Li, Rice, Rifkin, Simister,
  Sittampalam, and Aftandilian}]{ziegler_productivity_2022}
Ziegler A, Kalliamvakou E, Li XA, et~al (2022) Productivity {Assessment} of
  {Neural} {Code} {Completion}. In: Proc. of the 6th {ACM} {SIGPLAN}
  {International} {Symposium} on {Machine} {Programming}, {MAPS} 2022, pp
  21--29, \doi{10.1145/3520312.3534864}

\bibitem[{Ziegler et~al(2024)Ziegler, Kalliamvakou, Li, Rice, Rifkin, Simister,
  Sittampalam, and Aftandilian}]{ziegler_measuring_2024}
Ziegler A, Kalliamvakou E, Li XA, et~al (2024) Measuring {GitHub} {Copilot}'s
  {Impact} on {Productivity}. Communications of the ACM 67(3):54--63.
  \doi{10.1145/3633453}

\end{thebibliography}

\appendix

\section{System and Tasks Under Study} \label{app:system_tasks}
Controlled experiments on software maintainability require both a realistic system and carefully designed tasks. The system must be large enough to simulate real-world project conditions, and the tasks must reflect typical software development work. At the same time, the tasks must be possible to solve within a reasonable time. This section first describes how we designed the experimental artifacts, then outlines the two maintainability tasks, and finally provides a technical overview of the system under study.

\subsection{Task Development and Context} \label{app:taskdevcon}
We initiated the development of the experimental artifacts with a brainstorming workshop with four senior consultants from Equal Experts and the first author. Together, we defined the design goal as specifying a typical real-world problem that should be recognizable by any professional developer. This approach helps mitigate threats to internal validity related to task familiarity. Furthermore, we specified that the system under study should exhibit the following attributes: 

\begin{itemize}
    \item code spread across multiple files;
    \item subpar code quality;
    \item unit tests present, but not complete coverage;
    \item API and database integration;
    \item involving a well-known framework;
    \item easily understandable domain;
    \item problem statement should be fun -- or at least interesting;
    \item possible to complete in 2--4 hours -- to reduce dropouts.
\end{itemize}

Guided by the above, we introduced participants to the experiment through a lightweight role-playing scenario. They took on the role of consultants hired by a hypothetical new business, Recipes4Success (R4S), whose mission is to ignite a passion for cooking among young people. Participants were told that R4S had previously engaged a software consultancy to develop a recipe service: RecipeFinder. Unfortunately, the collaboration did not meet expectations -- a working web application was eventually delivered, but with poor software quality. R4S now seeks to enhance this service with new features and to establish a fruitful partnership with another consultancy, setting the stage for the participant's involvement.

\begin{wrapfigure}{r}{0.4\textwidth}
    \centering
    \includegraphics[width=0.4\textwidth]{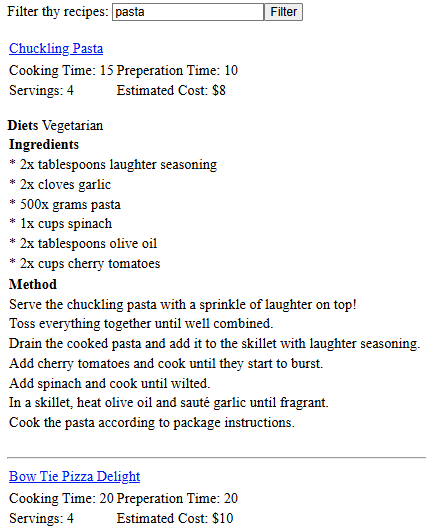}
    \caption{Screenshot of the web application.}
    \label{fig:screenshot}
\end{wrapfigure}

RecipeFinder consists of a rudimentary web application (see Figure~\ref{fig:screenshot}) and a supporting API. The base system offers the following features: 1) filtering the recipe list by a search term, 2) listing all recipes when the search term is blank, and 3) filtering recipes by total time (preparation time + cooking time). The service is built around a back-end API that exposes two endpoints: 1) listing all recipe IDs (with names and descriptions) and 2) getting recipe details by ID. 

\newpage

We stress that the task instructions and technical details presented next were refined over several iterations together with developers from Equal Experts. Senior software engineering consultants from Equal Experts co-designed and implemented the RecipeFinder system, focusing on realism: a small-scale web service with minor quality issues and only partial test coverage. Keeping the system small and scoping the maintainability tasks were necessary to ensure completion within the target time frame. We also intentionally kept the task descriptions somewhat imperfect to reflect how feature requests are often phrased in practice: R4S is not a software expert and customer assignments are rarely fully specified.

\subsection{Maintainability Tasks} \label{app:tasks}
Participants will complete one of two tasks, both presented as high-level feature requests rather than detailed technical specifications. In Task~1, they were tasked with enhancing the existing search feature to include filtering recipes by the total time required to prepare a meal. Task~1 participants were explicitly told that changing the API was not expected. Finally, we explained to the participants that there are several design problems in the existing codebase, as well as a known bug related to the presentation of preparation and cooking times. We concluded the feature request by stating: \textit{``We would like you to build this feature for us, and we expect a high level of quality. In particular, it should be easy to read and maintain.''} We intentionally kept this quality requirement vague to reflect realistic stakeholder communication, even though this inherently leaves room for participant interpretation and an open definition of done.

Furthermore, we deliberately introduced the bug to increase realism and the flavor of subpar code quality listed as a design goal in Appendix~\ref{app:taskdevcon}. The small defect resides in the controller code that participants must modify to implement the requested feature. Moreover, the acceptance tests for Task~1 require this defect to be fixed, and we only consider submissions that pass all tests. The injected bug thus adds a minor element of corrective maintenance and realism without biasing our comparisons, since all participants start from the same buggy baseline and only passing solutions proceed to the next phase.

In Task~2, participants were asked to further develop the Task~1 search feature by incorporating a filter for the cost per serving. Completing this task required the participant to build on the solution provided by the Task~1 participant. Note that we used the same R4S role-playing scenario for Task~2, no matter the quality of the Task~1 solution. Again, the feature request concluded with the same statement about our quality expectations. Both tasks included three acceptance tests to be executed using GitHub Actions, which target the expected behavior of the completed solution -- and ensure that the Task~1 defect has been resolved.

To validate the setup, five consultants from Equal Experts pilot-tested either Task~1 or Task~2. These pilots tested the clarity of the instructions, the estimated time budget, and the supporting infrastructure (especially \textit{snapcode.review}, described in Section~\ref{sec:datacol}). The exact task descriptions are available in the replication package~\citep{equal_experts_does_2025}.

\subsection{RecipeFinder Codebase: Technical Details} \label{app:recipefinder}
The base application, RecipeFinder, is a deliberately substandard Java/Spring Boot application used as the starting point for Task 1. We chose to rely on Spring Boot as it is widely regarded as a standard framework in the contemporary Java stack~\citep{gorla_cubetesterai_2025}. Features related to our tasks include Spring Data JPA for database access, Spring MVC for REST endpoints, and annotation-based dependency injection.

The total size of the base system, i.e., the codebase sent to the Task~1 participants, is 2 KLoC across 47 files. We analyzed the maintainability of the base system using CodeScene and complemented the results with low-level style checks using the PMD\footnote{\url{https://pmd.github.io/}} linting tool (v7.13.0) using all Java rules\footnote{\url{https://github.com/pmd/pmd/blob/e5516daad869a4895ca14e3257af15f528993bc7/pmd-core/src/main/resources/rulesets/internal/all-java.xml}}. We present these analyses when comparing how developers working with or without AI assistants evolved the Task~1 code in Section~\ref{sec:code-comparison}.

Figure~\ref{lst:code} shows a Java method from \texttt{RecipeFinderController.java}, which is at the core of both tasks. In Task~1, participants modified this method to implement the new feature -- and in Task~2, other participants changed it again. The code displays subpar characteristics and violations of standard Java conventions. The method contains two CodeScene code smells, i.e., Bumpy Road and Complex Method, which will be further discussed in Section~\ref{sec:code-comparison}.

\begin{figure}[t]
\centering
\caption{Original search handler in \texttt{RecipeFinderController.java} used in Task~1. The red dashes indicate the Bumpy Road code smell (bumps on lines 13, 20, and 27). The Complex Method smell is triggered due to a cyclomatic complexity of 13.}
\label{lst:code}
\begin{lstlisting}[language=Java, numbers=left, frame=single, basicstyle=\ttfamily\tiny]
@GetMapping("/recipes")
public String search(@RequestParam(name="query", required=false) String query, Model model) {
@r@ @List<Recipe> recipes = recipeRepository.findAll();
@r@ @model.addAttribute("recipes", recipes.stream().filter(recipe -> {
@r@ @ @r@ @if (query == null) {
@r@ @ @r@ @ @r@ @return true;
@r@ @ @r@ @}
@r@ @ @r@ @if (recipe.getName().toLowerCase().contains(query.toLowerCase())) return true;
@r@ @ @r@ @if (recipe.getDescription().toLowerCase().contains(query.toLowerCase())) return true;
@r@ @ @r@ @long count = 0L;
@r@ @ @r@ @for (Ingredient x : recipe.getIngredients()) {
@r@ @ @r@ @ @r@ @if (x.getName().toLowerCase().contains(query.toLowerCase())) {
@r@ @ @r@ @ @r@ @ @r@ @count++;
@r@ @ @r@ @ @r@ @}
@r@ @ @r@ @}
@r@ @ @r@ @ @r@ @if (count > 0) return true;
@r@ @ @r@ @ @r@ @count = 0;
@r@ @ @r@ @ @r@ @for (Method m : recipe.getMethod()) {
@r@ @ @r@ @ @r@ @if (m.getDescription().toLowerCase().contains(query.toLowerCase())) {
@r@ @ @r@ @ @r@ @ @r@ @count++;
@r@ @ @r@ @ @r@ @}
@r@ @ @r@ @}
@r@ @ @r@ @if (count > 0) return true;
@r@ @ @r@ @count = 0;
@r@ @ @r@ @for (Diet diet : recipe.getDiets()) {
@r@ @ @r@ @ @r@ @if (diet.getName().toLowerCase().contains(query.toLowerCase())) {
@r@ @ @r@ @ @r@ @ @r@ @count++;
@r@ @ @r@ @ @r@ @}
@r@ @ @r@ @}
@r@ @ @r@ @if (count > 0) return true;
@r@ @ @r@ @count = 0;
@r@ @ @r@ @
@r@ @ @r@ @return false;
@r@ @}).collect(toList()));
@r@ @
@r@ @return "recipe-search";
}
\end{lstlisting}
\end{figure}

The codebase contains a unit test suite that helps participants understand the behavior of key Java methods. Moreover, it provides a starting point for those who prefer test-driven development. Participants could add or modify the unit tests as they added new features. Note that the unit tests did not expose the injected defect.

Finally, to preserve the integrity of the study, i.e., preventing any leakage to AI assistants' training data~\citep{silva_gitbug-java_2024}, we chose not to host the code in a public git repository. Instead, all relevant documents and code are shared as PDF documents in the replication package~\citep{equal_experts_does_2025}.

\section{Bayesian Statistical Models} \label{app:bayes}
This appendix provides an introduction to Bayesian analysis and presents details for our modeling of the four dependent variables.

\subsection{Preamble}

In the models described below, we describe a generative model of the data. Each model contains a \textit{likelihood}, which describes how we think the observed data was generated. This likelihood usually has parameters that we do not know and want to \textit{infer}. For each of these parameters, we provide a \textit{prior} (another distribution). Bayesian inference software (in our case, the Turing Julia library\footnote{\url{https://turing.ml/}}) then estimates a \textit{posterior distribution} of the likely values of the parameters. Posterior distributions are a compromise between the likelihood and the prior distributions.
For example, the model below estimates the mean and variances ($\mu$ and $\sigma$) of a dataset $y$. 

\begin{align*}
    \mu &\sim \text{Normal}(0, 1) \\
    \sigma &\sim \text{Exponential}(1) \\
    y_i &\sim \text{Normal}(\mu, \sigma)
\end{align*}

We follow the same notation for all our models, $\alpha \sim d$ denotes that parameter $\alpha$ has distribution $d$. The last line is the likelihood, and uses parameters $\mu$ and $\sigma$, which we want to estimate, we therefore give them priors. In the following models, we use the normal, exponential, Dirichlet, and ordered logsitic distributions.
The normal distribution is the familiar ``bell curve'' distribution, with the difference that we denote $\text{Normal}(\mu, \sigma) = \mathcal{N}(\mu, \sigma^2)$, for simplicity. The exponential distribution covers a continuous range of positive values, we use it for variance terms, which are strictly positive.

\begin{figure}
    \centering
    \includegraphics[width=0.75\linewidth]{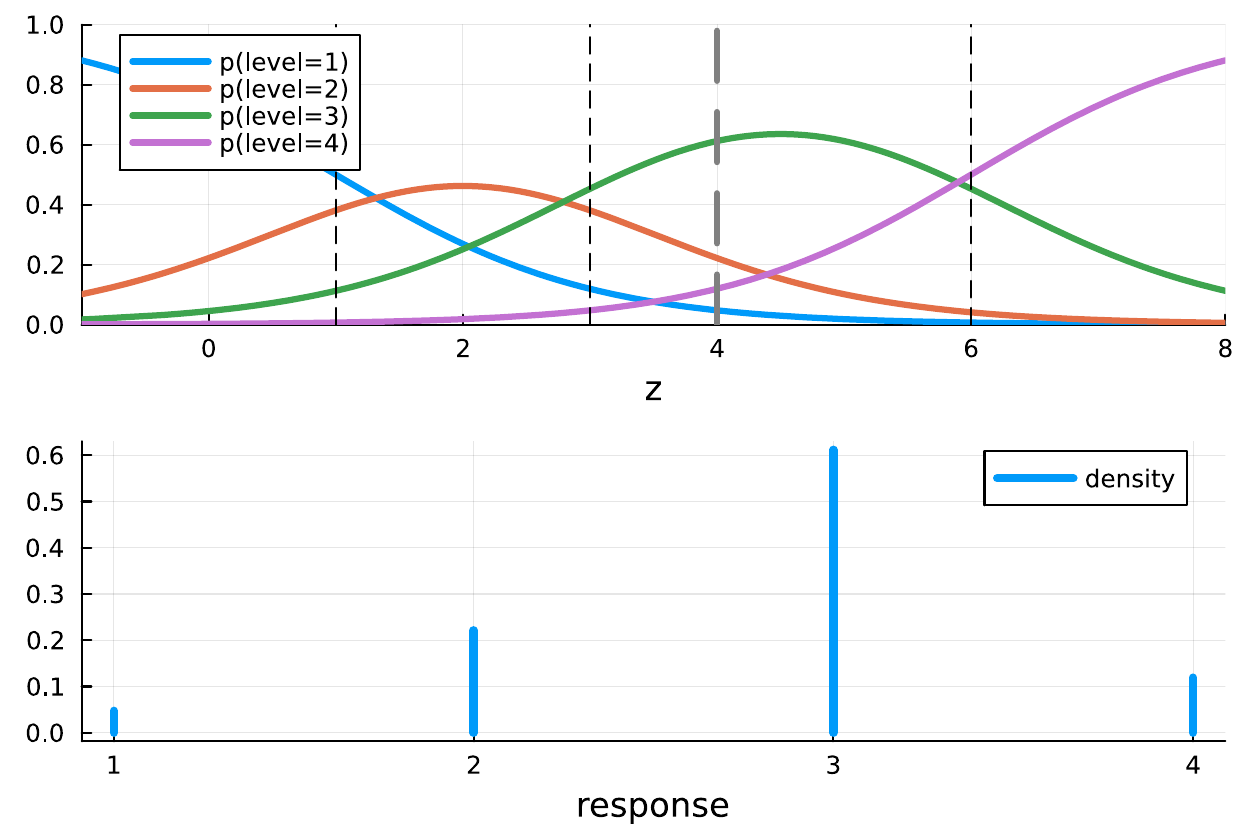}
    \caption{Two plots, showing the relationship between the latent variable $z$ and an ordered logistic distribution with cutoffs $[1, 3, 6]$. 
    The top plot shows the probability of each of the 4 levels as a function of $z$, where the cutoffs are shown using black dashed lines. 
    The bottom plot shows the probability of each response level for $z = 4$. The bottom plot is a slice indicated on the top plot by a gray dashed line. As $z$ increases, higher response levels become more probable, but there is no assumption that cutoff levels are evenly separated.}
    \label{fig:ordered-logistic-dist}
\end{figure}

The Dirichlet distribution is a distribution of vectors, where the sum of the vectors is equal to 1. We use it to represent monotonic effects, which we clarify with an example. Suppose we want to quantify the effect of education (BSc, MSc, or PhD) on another variable (e.g., yearly income). Modeling this with independent effects does not necessarily make sense, because people who have a master's degree usually have a bachelor's degree too. Instead, we represent the effect of education as a \emph{cumulative sum} of effects, with two components. First, we have a parameter for the \emph{maximum} effect of the variable (e.g., for education, the effect of having a PhD). And second, we have a vector of reals between 0 and 1 that sum to 1. This vector represents the effect of each level as a fraction of the maximum.

In our models, we usually use this distribution to represent the effect of 1) experience in Java and 2) experience in AI-assisted development. 
Consider the following equations:

\begin{align*}
\beta &\sim \text{Normal}(0, 1),~\boldsymbol{\lambda}_{\text{xp}} \sim \text{Dirichlet}(5, 1.0) \\
\theta_{\text{xp}}[j] &= \left( \sum_{k=1}^{j} \lambda_{\text{xp}}[k] \right) \cdot \beta \quad \text{for } j = 1,2,3,4,5 \\
\end{align*}

$\theta_{\text{xp}}[j]$ represents the effect of having the experience level $j$ on the outcome (not shown here). 
In our analyses, 1 corresponds to ``Minimal experience'', and 5 to ``High experience''. 
$\beta$ represents the effect of the maximum level of experience in AI (``High''). 
$\boldsymbol{\lambda}_{\text{xp}}$ represents the ``ladder'' of effects. The example shows a vector of 5 values, which sum to 1, and each element represents the difference with the previous level.

The ordered logistic distribution is useful for modeling variables on a discrete scale on $K$ levels, where the levels are ordered but we do not know how distant they are from each other. It takes two parameters: $z$ is real number and a vector of ``cutoffs'' $c_1, c_2, \dots, c_{K-1}$. It describes the probability of obtaining a discrete response level $r \in [1, 2,\dots,K]$. In our analysis, we assume that when $z < 0$, the most probable level is the minimum, and as $z$ increases, the probabilities of observing each of the $K$ levels behave like displayed in Figure \ref{fig:ordered-logistic-dist}. 

Next, we present the details related to the four dependent variables under study.

\subsection{Completion Time}

\begin{itemize}
  \item \( a_i \): AI usage (binary)
  \item \( x_i \in \{1,2,3,4,5\} \): Degree of habitual AI assistant use (Q1-7a)
  \item \( s_i \in \{1,2,3\} \): Developer 1 Java skill level (Q1-5)
  \item \( u_i \in \{1,2,3\} \): Developer 2 interruption level (Q2-1)
  \item \( y_i \): Observed completion time
  \item \( \mu_\beta \) and \( \sigma_\beta \) : Mean and variance of the prior we select.
\end{itemize}

In the model below, the likelihood of the logarithm of completion time $\log(y_i)$ is modeled with a normal distribution. The mean $\mu_i$ is the sum of an intercept $\alpha$, the effect of Java proficiency of the developer $\theta_{\text{skill}}[s_i]$, and the effect of their experience in AI $a_i \cdot \theta_{\text{xp}}[x_i]$ (multiplied by $a_i$ so that this effect disappears if AI is not used ($a_i = 0$)), and the effect of interruptions $\theta_{\text{int}}[u_i]$. The notation $v[i]$ refers to vector indexing, starting at 1. 

\begin{align}
\alpha &\sim \text{Normal}(0, 1) \\
\beta &\sim \text{Normal}(\mu_\beta, \sigma_\beta),~\boldsymbol{\lambda}_{\text{xp}} \sim \text{Dirichlet}(5, 1.0) \\
\theta_{\text{xp}}[j] &= \left( \sum_{k=1}^{j} \lambda_{\text{xp}}[k] \right) \cdot \beta \quad \text{for } j = 1,2,3,4,5 \\
\gamma &\sim \text{Normal}(0, 1),~\boldsymbol{\lambda}_{\text{skill}} \sim \text{Dirichlet}(3, 1.0) \\
\theta_{\text{skill}}[j] &= \left( \sum_{k=1}^{j} \lambda_{\text{skill}}[k] \right) \cdot \gamma \\
\delta &\sim \text{Normal}(0, 1),~\boldsymbol{\lambda}_{\text{int}} \sim \text{Dirichlet}(3, 1.0) \\
\theta_{\text{int}}[j] &= \left( \sum_{k=1}^{j} \lambda_{\text{int}}[k] \right) \cdot \delta \\
\sigma_j &\sim \text{Exponential}(1) \quad \text{for } j = 1,2,3 \\
\mu_i &= \alpha + \theta_{\text{skill}}[s_i] + a_i \cdot \theta_{\text{xp}}[x_i] + \theta_{\text{int}}[u_i] \\
\log(y_i) &\sim \text{Normal}(\mu_i, \sigma_{u_i})
\end{align}

\subsection{CodeHealth}

\begin{itemize}
  \item \( a_i \): AI usage (binary)
  \item \( x_i \in \{1,2,3,4,5\} \): Degree of habitual AI assistant use (Q1-7a)
  \item \( s_i \in \{1,2,3\} \): Developer 1 Java skill level (Q1-5)
  \item \( y_i \): Observed CodeHealth score
  \item \( \mu_\beta \) and \( \sigma_\beta \) : Mean and variance of the prior we select.
\end{itemize}

\textbf{Model:}
\begin{align}
\alpha &\sim \text{Normal}(0, 1) \\
\beta &\sim \text{Normal}(\mu_\beta, \sigma_\beta) \\
\boldsymbol{\lambda}_{\text{xp}} &\sim \text{Dirichlet}(5, 1.0) \\
\theta_{\text{xp}}[j] &= \left( \sum_{k=1}^{j} \lambda_{\text{xp}}[k] \right) \cdot \beta \quad \text{for } j = 1,2,3,4,5 \\
\gamma &\sim \text{Normal}() \\
\boldsymbol{\lambda}_{\text{skill}} &\sim \text{Dirichlet}(3, 1.0) \\
\theta_{\text{skill}}[j] &= \left( \sum_{k=1}^{j} \lambda_{\text{skill}}[k] \right) \cdot \gamma \quad \text{for } j = 1,2,3 \\
\sigma &\sim \text{Exponential}(1) \\
\mu_i &= \alpha + a_i \cdot \theta_{\text{xp}}[x_i] + \theta_{\text{skill}}[s_i] \\
y_i &\sim \text{Normal}(\mu_i, \sigma)
\end{align}

\subsection{Test Coverage}

\begin{itemize}
  \item \( a_i \): AI usage (binary)
  \item \( x_i \in \{1,2,3,4,5\} \): Degree of habitual AI assistant use (Q1-7a)
  \item \( s_i \in \{1,2,3\} \): Developer 1 Java skill level (Q1-5)
  \item \( y_i \): Observed logit-transformed test coverage
  \item \( \mu_\beta \) and \( \sigma_\beta \) : Mean and variance of the prior we select.
\end{itemize}

\begin{align}
\alpha &\sim \text{Normal}(0, 1) \\
\beta &\sim \text{Normal}(\mu_\beta, \sigma_\beta) \\
\boldsymbol{\lambda}_{\text{xp}} &\sim \text{Dirichlet}(5, 1.0) \\
\theta_{\text{xp}}[j] &= \left( \sum_{k=1}^{j} \lambda_{\text{xp}}[k] \right) \cdot \beta \quad \text{for } j = 1,2,3,4,5 \\
\gamma &\sim \text{Normal}() \\
\boldsymbol{\lambda}_{\text{skill}} &\sim \text{Dirichlet}(3, 1.0) \\
\theta_{\text{skill}}[j] &= \left( \sum_{k=1}^{j} \lambda_{\text{skill}}[k] \right) \cdot \gamma \quad \text{for } j = 1,2,3 \\
\sigma &\sim \text{Exponential}(1) \\
\mu_i &= \alpha + a_i \cdot \theta_{\text{xp}}[x_i] + \theta_{\text{skill}}[s_i] \\
\text{logit}(y_i) &\sim \text{Normal}(\mu_i, \sigma)
\end{align}

\subsection{Perceived Productivity}

\begin{itemize}
  \item \( a_i \): AI usage (binary)
  \item \( x_i \in \{1,2,3,4,5\} \): Degree of habitual AI assistant use (Q1-7a)
  \item \( s_i \in \{1,2,3\} \): Developer 1 Java skill level (Q1-5)
  \item \( y_{i,q} \in \{1, \dots, L\} \): Ordinal response for person \( i \) on question \( q \)
  \item \( L \): Number of ordinal levels
  \item \( Q \): Number of questions
  \item \( \mu_\beta \) and \( \sigma_\beta \) : Mean and variance of the prior we select.
\end{itemize}

\begin{align}
\alpha &\sim \text{Normal}(0, 1) \\
\beta &\sim \text{Normal}(\mu_\beta, \sigma_\beta) \\
\boldsymbol{\lambda}_{\text{xp}} &\sim \text{Dirichlet}(5, 1.0) \\
\theta_{\text{xp}}[j] &= \left( \sum_{k=1}^{j} \lambda_{\text{xp}}[k] \right) \cdot \beta \quad \text{for } j = 1,2,3,4,5 \\
\gamma &\sim \text{Normal}() \\
\boldsymbol{\lambda}_{\text{skill}} &\sim \text{Dirichlet}(3, 1.0) \\
\theta_{\text{skill}}[j] &= \left( \sum_{k=1}^{j} \lambda_{\text{skill}}[k] \right) \cdot \gamma \quad \text{for } j = 1,2,3 \\
\pi_i &= \alpha + a_i \cdot \theta_{\text{xp}}[x_i] + \theta_{\text{skill}}[s_i] \\
\delta_{q,\ell} &\sim \text{Exponential}(1) \quad \text{for } q = 1,\dots,Q,\ \ell = 1,\dots,L-2 \\
c_{q,0} &= 0 \\
c_{q,\ell} &= \sum_{m=1}^{\ell} \delta_{q,m} \quad \text{for } \ell = 1,\dots,L-2 \\
y_{i,q} &\sim \text{OrderedLogistic}(\pi_i,\ [c_{q,0}, c_{q,1}, \dots, c_{q,L-2}])
\end{align}

\section{CodeScene Code Smells} \label{app:codescene}

This appendix describes the CodeScene code smells observed and referenced in the article. These smells were identified using the \texttt{CodeScene CLI} tool through file-level analysis. The following commands were used: \texttt{cs check \$\{FILE\}} for analyzing individual files, and \texttt{cs delta \$\{FIRST\_USER\_COMMIT\} HEAD} for analyzing the diff of individual files between commits.

\begin{description}
    \item[\textbf{Bumpy Road}] (or Bumpy Road Ahead). Assigned to a function that contains multiple chunks of nested control structures. The deeper the nesting, the more bumps, and the more severe the Bumpy Road smell is. These bumps in the code represent missing abstractions and make the function harder to understand and maintain.
    \item[\textbf{Complex Conditionals}] indicates branch expressions that combine multiple logical operations, such as conjunctions and disjunctions, within a single condition. These statements are harder to read and reason about than a well-named boolean identifier that could represent the behavior and logic of the conditions. Complex conditionals obscure the intent of the code and contribute to the overall complexity of the method in which they appear.
    \item[\textbf{Complex Method}] is assigned to functions with high cyclomatic complexity, which means that they contain many independent logical paths. Such methods are more difficult to test thoroughly and are more prone to bugs. They often try to do too much and lack a clear separation of concerns. Splitting complex methods into smaller, more focused units improves both readability and testability.
    \item[\textbf{Nested Complexity}] (or Deep, Nested Complexity) is a smell that arises when control structures, such as loops and conditionals, are nested within each other to multiple levels. This kind of nesting increases the cognitive effort required to understand the flow of execution.
    \item[\textbf{Excessive Function Arguments}] is assigned when functions take too many parameters. This can indicate that the function is doing too much or that it lacks a proper abstraction to group related data. Functions with many arguments are harder to call correctly and more difficult to refactor.
    \item[\textbf{Large Assertion Blocks}] is a test code smell that is a sign of poor structure. When many assertions are grouped together without clear separation, it can become harder to understand what each test is verifying, which reduces the effectiveness of the test suite and makes failures harder to diagnose.
    \item[\textbf{Large Method}] indicates a method with many LoC, this typically indicates multiple responsibilities, which can make the method harder to read and should be decomposed into smaller, more cohesive units.
    \item[\textbf{Code Duplication}] is a file-level smell that potentially makes code evolution harder. Any change must be replicated across all instances, increasing the risk of inconsistencies and bugs. Duplication also inflates the codebase, making it potentially harder to navigate and understand.
    \item[\textbf{Duplicated Assertion Blocks}] is a test code smell similar to code duplication, making code harder to maintain. Duplicated test criteria indicate missing abstractions or a test suite that attempts to test too many things inside the same module.
    \item[\textbf{Constructor Over-Injection}] is assigned when a constructor has many arguments. This indicates that either a unit has low cohesion or an injection of dependencies on the wrong abstraction level.
    \item[\textbf{Primitive Obsession}] is assigned to code that uses a high degree of built-in primitives such as integers, strings, and floats. This shows a lack of domain language that encapsulates the validation and semantics of function arguments.
    \item[\textbf{String-heavy Arguments}] is related to primitive obsession, where the heavy usage of strings could indicate a missing domain language. String is a generic type that often fails to capture the constraints of the domain object it represents.
\end{description}

\section{PMD Rules} \label{app:pmd}

This appendix describes the significant PMD rules violations observed and referenced in the article\footnote{\url{https://pmd.github.io/pmd/pmd_rules_java.html}}.

\begin{description}
    \item[\textbf{\textit{UnusedAssignment}}]  
    Triggers when a variable is assigned a value that is never used. This may indicate leftover code or incomplete logic.
    \item[\textbf{\textit{AtLeastOneConstructor}}]  
    Flags non-static classes that do not declare an explicit constructor. Adding one makes the class definition clearer and more consistent.
    \item[\textbf{\textit{ControlStatementBraces}}]  
    Enforces the use of curly braces in control structures such as \texttt{if} and \texttt{for} to reduce ambiguity and prevent logic errors.
    \item[\textbf{\textit{LinguisticNaming}}]  
    Detects mismatches between identifier names and their types, such as a non-boolean variable named like a boolean. This improves naming consistency and code readability.
    \item[\textbf{\textit{MethodArgCouldBeFinal}}]    
    \item[\textbf{\textit{OnlyOneReturn}}]  
    Encourages methods to have a single return statement at the end, simplifying control flow and enhancing maintainability.
    \item[\textbf{\textit{ShortVariable}}]  
    Flags variable names shorter than three characters unless used in conventional contexts like loop counters. Short names reduce clarity and hinder comprehension.
    \item[\textbf{\textit{UseExplicitTypes}}]  
    Warns when the \texttt{var} keyword is used. Requiring explicit types improves code clarity and reduces cognitive overhead for readers.
    \item[\textbf{\textit{AvoidCatchingGenExcep}}]  
    Warns against catching overly generic exceptions such as \texttt{Exception} or `Throwable`, which can mask programming errors and complicate debugging. 
    \item[\textbf{\textit{CyclomaticComplexity}}]  
    Flags methods with high cyclomatic complexity (threshold: 10). Complex methods are harder to test and understand and often benefit from refactoring.
    \item[\textbf{\textit{ImmutableField}}]  
    Identifies fields that can be marked \texttt{final} but are not. Immutability strengthens code reliability and reduces potential side effects.
    \item[\textbf{\textit{NPathComplexity}}]  
    Calculates the number of possible execution paths through a method. A high value (threshold: 200) suggests the method may be overly complex and hard to test.
    \item[\textbf{\textit{CommentRequired}}]  
    Ensures that key elements like public classes and methods include Javadoc comments. Proper documentation supports better understanding and long-term maintainability.
\end{description}

\end{document}